\documentclass[12pt]{article}
\pdfoutput=1

\usepackage{pdflscape}
\usepackage{graphicx}
\usepackage{algorithmic}
\usepackage{algorithm}
\usepackage{listings}
\usepackage{amssymb}
\usepackage{amsmath}
\usepackage{amsfonts}
\usepackage{natbib}
\usepackage{lineno}
\usepackage{bm}
\usepackage{xr}

\usepackage{pdfpages}

\usepackage{fixltx2e}
\usepackage{textcomp}
\usepackage{fullpage}
\usepackage{amsfonts}
\usepackage{verbatim}
\usepackage[english]{babel}
\usepackage{pifont}
\usepackage{color}
\usepackage{setspace}
\usepackage{lscape}
\usepackage{indentfirst}
\usepackage[normalem]{ulem}
\usepackage{booktabs}
\usepackage{nag}
\usepackage{float}
\usepackage{latexsym}
\usepackage{url}
\usepackage{hyperref}
\usepackage{epsfig}
\usepackage{array}
\usepackage{ifthen}
\usepackage{amsthm}
\usepackage{amstext}

\usepackage{longtable}

\usepackage{footnote}
\usepackage[para]{footmisc}
\usepackage{tikz}
\usepackage{fp}   
\usetikzlibrary{fpu}
\usepackage{nameref}
\usepackage{siunitx}
\usepackage{boxhandler}

\newcommand*{\MyNum}[1]{%
\pgfmathprintnumber[
        fixed,
        precision=3,
        fixed zerofill=false,
        ]{#1}}%
\usepackage{xr}

\newcommand{\be}{\begin{equation}} \newcommand{\ee}{\end{equation}}
\newcommand{\bd}{\begin{displaymath}} \newcommand{\ed}{\end{displaymath}}
\newcommand{\ba}{\begin{align}} \newcommand{\ea}{\end{align}}
\newcommand{\baa}{\begin{align*}} \newcommand{\eaa}{\end{align*}}
\newcommand{\ben}{\begin{enumerate}} \newcommand{\een}{\end{enumerate}}
\newcommand{\bi}{\begin{itemize}} \newcommand{\ei}{\end{itemize}}
\newcommand{\supp}[1]{\operatorname{supp}\left( #1 \right)}

\newcommand{\ud}{\mathrm{d}}
\newcommand{\E}[1]{\operatorname{E}\left[ #1 \right]}
\newcommand{\Var}[1]{\operatorname{Var}\left[ #1 \right]}
\newcommand{\cov}[2]{\operatorname{Cov}\left[ #1,#2 \right]}

\newtheorem{lemma}{Lemma}
\newtheorem{definition}{Definition}

\begin{document}

\title{Phylogenetic effective sample size}
\author{{\sc Krzysztof Bartoszek}} 

\maketitle

\begin{abstract}
In this paper I address the question --- \textit{how large is a phylogenetic sample?}
I propose a definition of a phylogenetic effective sample size for Brownian motion
and Ornstein--Uhlenbeck processes --- the \textit{regression effective sample size}.
I discuss how mutual information can be used to define an effective sample size
in the non--normal process case and compare these two definitions to an
already present concept of effective sample size (the mean effective sample size).
Through a simulation study I find that the AIC$_{c}$ is robust if one corrects for the
number of species or effective number of species. Lastly I discuss how 
the concept of the phylogenetic effective sample size can be useful for 
biodiversity quantification, identification of interesting clades and 
deciding on the importance of phylogenetic correlations.
\end{abstract}


\section{Introduction}\label{secIntro}
One of the reasons to introduce phylogenetic comparative methods (PCMs) in the words of \citet{EMarTHan1996AB},
was to address the problem of statistical dependence. They called the issue the ``degrees of freedom''
or ``effective sample size'' problem. If we have $n$ species related by a phylogenetic tree,
unless it is a star phylogeny, then our effective sample size is less than $n$ 
(in extreme cases even one). Taking into consideration the number of 
independent observations
is important in evaluating the accuracy of parameter
estimation or hypothesis tests. The performance of such statistical procedures 
depends on the number of independent
data points and not on the observed number of data points \citep{EMarTHan1996AB}. 
Ignoring the correlations (and hence inflating the sample size) results in
too narrow confidence intervals, inflated
p--values and power. All of this leads to type I and II errors
of which the user may be oblivious of. 

In a phylogenetic context the calculation of the effective number of observations has not 
been often addressed directly. In statistical literature effective sample
size (ESS) is usually parameter specific, it can be understood as ``the 
number of independent measurements one would need to reach the same 
amount of information about a parameter as in the original data''
\citep{CFaeetal2009} --- in other words how many independent points do we
have for estimating a particular parameter. 
\citet[p. 145][]{CNun2011} points out that often phylogenetic comparative
methods have been viewed in a restricted manner as a ``degrees of freedom''
correction procedure that ``reduce the number of data points'', due to the nonindependence.
Most phylogenetic comparative methods
work in the following way --- one assumes a model and maximizes the likelihood under
that model. Hence, the issue of ESS, as mentioned above, has been taken care
of but only for the estimation problem. In other situations, as \citet{CNun2011} following
\citet{MPag1993} reminds, the ``degrees of freedom analogy can be misleading''. It 
is more important how the variance is partitioned among species. In 
fact in the case of model selection, or when
one wants to know how many ``independent'' taxa one has e.g. for conservation purposes
the situation becomes much more complex. As we will see it is more important
how the covariance is structured.

\citet{RSmi1994} directly approached the problem of effective sample size. 
He studied interspecies phenotypic data by a nested ANOVA and 
``\textit{Determination of the taxonomic levels that account for most of the variation 
can be used to select a single level at which it is most reasonable to consider
the data points as independent}''. From the perspective of modern phylogenetic
comparative methods this is a ``hack'', as \citet{RSmi1994} himself wrote
``the method improves the nonindependence problem but does not eliminate it''.
From our perspective his work is important, as from the nested ANOVA setup,
he partitioned the variance into components from different levels of the 
phylogeny and then defined the effective sample size as

\be
\begin{array}{rcl}
n_{e} & = & (\#\mathrm{of~superfamilies})(\mathrm{PVC~for~superfamilies})
+ (\#\mathrm{of~families})(\mathrm{PVC~for~families}) \\&&
+ (\#\mathrm{of~genera})(\mathrm{PVC~for~genera})
(\#\mathrm{of~species})(\mathrm{PVC~for~species})
\end{array}
\ee
where PVC is percentage of variance component. 
\citet{RSmi1994} importantly notices, that in principle 
``\textit{The method does not require that levels of the nested hierarchy are defined by
taxonomic categories.}''
In this work I develop the idea described in 
\citet{RSmi1994}'s own words: to 
``\textit{consider each
species as some fraction of a free observation varying between $0$ and $1.0$,
a value could be computed $\ldots$ that would reflect the balance 
between constraint and independent evolution. This value is defined
as the effective sample size (effective $N$) for the data set and trait,
as opposed to the traditionally used observed sample size (observed $N$).}''
Building up on the modern development of stochastic models for phylogenetic comparative
methods, I do not have to restrict myself to partitioning the data into hierarchical levels
containing different fractions of the variance, but rather look holistically at the dependence 
pattern induced by the tree and model of evolution.
This might make it impossible (but maybe not always)
to assign to each species (or taxonomic level) its fraction of free observations
but as we shall see it will allow me to calculate the sum of fractions of free observations.

An analysis of phylogenetically structured phenotypic
data often has as its goal to identify the mode of evolution, i.e. 
is the trait(s) adapting (and if so to what trait/phenotype) or rather exhibiting
neutral evolution. Information criteria like the Akaike Information Criterion \citep[AIC][]{HAka1974}, 
Akaike Information Criterion corrected for small sample size  \citep[AIC$_{c}$][]{CHurCTsa1989} 
or Bayesian Information Criterion \citep[BIC][]{GSch1978} are commonly used to identify
the model better supported by the data.
However, 
if one goes back to the derivation of the AIC$_{c}$ \citep{CHurCTsa1989} 
and BIC \citep{GSch1978} 
one can see that the $n$ observations are assumed independent. Therefore 
a phylogenetic comparative model seems to violate this assumption, in the best
case by inflating the sample size. 
In a way such an inflation corresponds to not penalizing 
enough for additional parameters. 
However in their original paper \citet{CHurCTsa1989} derive
the same AIC$_{c}$ formula for autoregressive models so this 
warrants further study in the phylogenetic setting 
where the covariance structure is hierarchical. 

Therefore, using the number of species (unless the phylogeny is a star)
results in a risk of overfitting for small phylogenies or those with most
speciation events near the tips. In this work I
propose a way of taking into
account the \textit{effective number of species} during the model selection procedure.
The newest version of mvSLOUCH (available from \url{http://cran.r-project.org/web/packages/mvSLOUCH/index.html}) 
allows for automatic model selection if one treats $n$ as the true sample size and also
if one corrects for the dependencies using an
effective sample size. 
Importantly mvSLOUCH allows for an arbitrary pattern of missing data ---
no observation is removed and the likelihood is based on all provided information.
Using this new version of mvSLOUCH, I include in this work a simulation study and analyze a number of data sets 
to see how much a difference does it make whether, one uses the observed or effective number 
of species for model selection. In most cases, the two ways of counting species
lead to the same conclusion. However,
for small samples (see Tab. \ref{tabDataRes}) using the effective number of species 
can result in a different outcome.
In fact we should expect this to be so, a good correction method should be robust --- 
with enough observations the data (or rather likelihood) should decide no matter how one corrects.
It is only with few observations (and hence little power) that correction 
methods should play a role by pointing to different possibilities of interpreting the observed data.

\section{Effective sample size}

Effective sample size is intuitively meant to represent the number of independent particles
of data in the sample. 
If the sample is correlated, then each observation will only have a certain fraction of the
information it carries particular to itself.
The rest of the information will be shared 
with one/some/all other points in the sample. We would like to quantify what 
proportion of the whole sample is made up of these independent bits of information.
If this proportion is $p$, then our phylogenetic effective sample size (pESS) will be $n_{e}=pn$. 
However our situation is a bit different. It is reasonable to assume that 
we have a least one observation --- at least one species described by at 
least a single trait. One way is to define $p$ to be between $1$ and $1/n$.
Alternatively we can define as

\be
n_{e} = 1+ p(n-1),\label{eqphylESS}
\ee
where $p\in [0,1]$. I will call this $p$ of Eq. \eqref{eqphylESS} the phylogenetic ESS factor.
The value $n_{e}/n$ is useful in practice to compare between different sized
phylogenies and I will call it the relative phylogenetic ESS.

\citet{EMarTHan1996AB}
point out, that in the discrete trait case, the ESS cannot be greater
than the number of independent evolutionary changes regardless of the number of observed species.
\citet{WMadRFit2015} very recently remind us of this again. Phylogenetic comparative 
methods are there to take care of ``pseudoreplicates'' due to the tree induced correlations. However,
especially in the discrete case, tests of significance might have inflated power as one uses
the number of species instead of the (unknown) number of independent evolutionary changes. 
Unfortunately, at the moment, there does not seem to be any solution for this problem \citep{WMadRFit2015}.
Hopefully the phylogenetic effective sample size concept presented here could 
indicate a direction for finding one. An alternative potential approach in the discrete
case, is phylogenetic informativeness based on the number of mutations (i.e. changes) shared by
tip taxa under the Poisson process \citep{WMulFCra2015,JTow2007}. It however, remains to study
the probabilistic properties of phylogenetic informativeness in order to understand
whether and how it may be applied in the pESS context.

Statistical definitions of effective sample size are commonly introduced
in the context of parameter estimations --- what is the ESS for a given
parameter/set of parameters. I am in a different situation
--- I want to quantify how many independent particles do I observe.
In this situation one has to propose
one's own definition of effective sample size that will be 
useful from a practical point of view.
This is not an obvious task 
in the situation of $n$ dependent observations. 
The case of 
multivariate observations, where individual components are dependent between each other and correlations
between traits can be negative, will be even more complicated. Below I will discuss a couple of possible approaches 
for defining an effective sample size and in the next section discuss how they can 
be applied in the phylogenetic comparative methods field.

\citet{CAne2008} defined an effective sample
size for estimating the root state under a Brownian motion (BM) model of evolution.
She noticed that it can be very small --- $6$ for a phylogeny of $49$ species 
\citep[mammal phylogeny of][]{TGarADicCJanJJon1993}. In fact 
my simulations and reanalysis of this data (Tab. \ref{tabDataRes})
give very similar numbers. She defined the effective sample size as

\be
n_{e}^{\mathrm{E}} := \mathbf{1}^{T} \mathbf{R}^{-1} \mathbf{1},
\ee
where $\mathbf{R}$ is the between species correlation matrix. 
I call $n_{e}^{\mathrm{E}}$ the \emph{mean effective sample size} (mESS), as 
$n_{e}^{\mathrm{E}}$ is the number of independent random variables
that result in the same precision for estimating the 
mean value (intercept) of a linear with $n$ 
correlated, by $\mathbf{R}$, observations \citep{CAne2008}.
It is important for the reader to notice that $n_{e}^{\mathrm{E}}$ is not connected
to any average of sample sizes. The word ``mean'' in the name refers to the fact
that $n_{e}^{\mathrm{E}}$ quantifies the information available on the mean value in a linear model.

For our purpose the mean effective sample size is not 
completely satisfactory. 
The $n_{e}^{\mathrm{E}}$ value does not say how much independent signal there is 
in the sample, but
only how much information we have about the expected value. In the scope
of this work we are more interested in the former and not the latter. 
In fact we can observe (Fig. \ref{figphylESS} and Tab. \ref{tabDataRes}),
that in a phylogenetic sample $n_{e}^{\mathrm{E}}$ is usually rather low. 
Such small numbers are due to the high variance of the sample average
\citep[an estimator of the mean value][]{CAne2008,KBarSSag2015,SSagKBar2012},
resulting in low precision for the mean value. I therefore consider 
alternative approaches to define a phylogenetic effective sample size.

Currently Ornstein--Uhlenbeck (OU) process are the state of the art in modelling
trait evolution \citep{KBaretal2012,JBeauetal2012,JClaGEscGMer2015,CCreMButAKin2015,TIngDMah2013,JUyeLHar2014}.
This OU process on a phylogeny is multivariate normal.
Therefore all the information will be contained in the mean vector and covariance matrix. 
In fact we have a natural multiple regression approach and each species, $y_{i}$, can be
represented as

$$
y_{i} = \E{y_{i} \vert \vec{y}_{-i}} + \epsilon_{i},
$$
where $\vec{y}_{-i}$ is the vector of measurements without the $i$--th entry.
The above equation will be of course of the form

$$
y_{i} = a_{i} + \vec{b}_{i} \cdot \vec{y}_{-i} + \epsilon_{i},
$$
where $\epsilon_{i}$ will be independent of $\vec{y}_{-i}$. The residual 
$\epsilon_{i}$ is mean $0$, normally distributed with variance

$$\sigma_{i}^{2} - \mathbf{V}_{i,-i}\mathbf{V}_{-i,-i}^{-1}\mathbf{V}_{-i,i},$$
where $-i$ notation again means removing the appropriate rows and/or columns.
As the variance of $y_{i}$ is $\sigma_{i}^{2}$, then the independent of the other
species part of this variance equals 
$(1-  \mathbf{V}_{i,-i}\mathbf{V}_{-i,-i}^{-1}\mathbf{V}_{-i,i}/\sigma_{i}^{2})\sigma_{i}^{2}$.
Standardizing every species to variance $1$ will mean that
each species carries 
$1- \mathbf{V}_{i,-i}\mathbf{V}_{-i,-i}^{-1}\mathbf{V}_{-i,i}/\sigma_{i}^{2}$
signal specific to itself. Therefore I propose to define a phylogenetic effective
sample size, called \emph{regression effective sample size} (rESS), in the following way.
Let

$$
v_{R} = \sum\limits_{i=1}^{n}(1- \mathbf{V}_{i,-i}\frac{\mathbf{V}_{-i,-i}^{-1}}{\sigma_{i}^{2}}\mathbf{V}_{-i,i}).
$$
be the total independent signal. 
The sum $v_{R}$ can be can be easily lesser than one.
We therefore consider

$$
\frac{1}{n}\sum\limits_{j=1}^{n}\sum\limits_{1=i\neq j}^{n}(1- \mathbf{V}_{i,-i}\frac{\mathbf{V}_{-i,-i}^{-1}}{\sigma_{i}^{2}}\mathbf{V}_{-i,i})
=\frac{n-1}{n}v_{R}.
$$
In the above one averages over all species, for each one considering the amount of distinct signal from it.
As we know that there is at least $1$ species I now define the rESS as

\be\label{defESSmvReg}
n_{e}^{R} = 1+ \frac{n-1}{n}v_{R}.
\ee
It can be easily checked that $n_{e}^{R}\in [1,n]$, equalling $n$ when the species are independent.
Taking the pseudoinverse instead of the inverse gives the value of $1$ when all $n$
species are identical. 

The rESS, just as the mESS, can be calculated for any process evolving 
on a phylogenetic tree. However, just as the mESS does not catch everything about a normal process,
the rESS will not catch everything in the non--normal process situation.
In the non--normal process, e.g. heavy tailed distributions \citep{MEllAMoo2014}, 
situation it is necessary to reach for more
complicated mathematical tools. The motivation behind the multiple regression
approach is to measure how much signal is contained about each species in 
other species and how much is specific to that species. Another way of formulating
the problem is to ask: how much information is contained in the joint distribution
of all of the species, when compared with only the marginal distributions.
The natural mathematical framework for this is information theory and the concept
of mutual information.

As the name itself suggests mutual information quantifies how much information do different probabilistic objects contain
about each other. I will briefly introduce a few concepts from information theory pointing the reader
to e.g. \citet{IKoc2014} for a more detailed discussion. 
\begin{definition}\label{defMIE}\citep{IKoc2014}
Let $\vec{X}\in \mathbb{R}^{n}$ be a random vector with density $f$ such that it has mean $\vec{\mu}$ and covariance $\mathbf{V}$. 
Further let $f_{j}$ $(j=1,\ldots,n)$ be the marginal densities of $f$ and 
$f_{G}$ be a Gaussian density with the same mean $\vec{\mu}$ and covariance $\mathbf{V}$, i.e. for $x \in \mathbb{R}^{n}$

$$
f_{G}(\vec{x}) = \left(\sqrt{(2\pi)^{n}\det\left(\mathbf{V}\right)}\right)^{-1}\exp\left(-\frac{1}{2}\left(\vec{x} - \vec{\mu} \right)^{T}\mathbf{V}^{-1}\left(\vec{x} - \vec{\mu} \right) \right).
$$
We then define the following.
\begin{enumerate}
\item The entropy of $f$ as 

$$
\mathcal{H}(f) = - \int\limits_{\supp{f}} f(\vec{x}) \log f(\vec{x}) \ud \vec{x},
$$
where $\supp{f} = \{ \vec{x}\in \mathbb{R}^{n} : f(\vec{x}) > 0 \}$ is the support of $f$.
\item The negentropy of $f$ as

$$
\mathcal{J}(f) = \mathcal{H}(f_{G}) -  \mathcal{H}(f).
$$
\item The mutual information of $f$ as

$$
\mathcal{I}(f) = \sum\limits_{j=1}^{n} \mathcal{H}(f_{j}) - \mathcal{H}(f).
$$
\end{enumerate}
\end{definition}
Intuitively the entropy of a density (or rather random variable behaving according to its law) is the 
measure of uncertainty about the value of this random variable prior to observation. The negentropy
from our perspective is more of a technical term however, the mutual information between
two densities (or random variables) will be very important in proposing an effective sample size
definition. 

The maximum sample size attained is $n$, when all species are independent of each other (we have a star phylogeny).
In this situation the density function of our $n$ dimensional vector of observations will be 
the product of the marginal $n$ densities. 
No observation
contains relatively more information about any other one observation than any other does.
Therefore, to quantify how much information do sample points contain about each other, we will
consider in Lemma \ref{lemMIEprops} the mutual information between the sample's $n$--dimensional density
and the density defined as the product of the marginal densities. If we recall that all the
considered evolutionary models here (Brownian motion, Ornstein--Uhlenbeck) are multivariate
normal, then we should expect that the entropy based measures be dependent only
on the covariance matrix and marginal variances. In the Gaussian case,
all shared knowledge is coded in the covariance structure, see Lemma 
\ref{lemMIEprops}.
\begin{lemma}\label{lemMIEprops}\citep{IKoc2014}
Using the notation of Definition \ref{defMIE} the entropy, negentropy and mutual information posses the below properties and relationships between them.
\begin{enumerate}
\item The negentropy $\mathcal{J}\ge 0$ and $\mathcal{J}(f)=0$ iff $f$ is Gaussian.
\item The mutual information $\mathcal{I}\ge 0$ and $\mathcal{I}=0$ iff $f=\prod\limits_{j=1}^{d}f_{j}$.
\item If $f$ is Gaussian, then it has entropy
\be
\mathcal{H}(f) = \frac{1}{2}\left(n\left(1 + \log\left( 2\pi\right) \right)+ \log\det \mathbf{V} \right).
\ee
\item If $\mathbf{V}$ is invertible, then
\be
\mathcal{I}(f) = \mathcal{J}(f) - \sum\limits_{j=1}^{n}\mathcal{J}(f_{j}) + \frac{1}{2} \log \left(\frac{\prod\limits_{j=1}^{n}\sigma_{j}^{2}}{\det \mathbf{V}} \right)
\ee
where $\sigma_{j}^{2}$ are the diagonal elements of $\mathbf{V}$ --- the marginal variances.
If $f$ is Gaussian  this simplifies to
\be
\mathcal{I}(f) = \frac{1}{2} \log \left(\frac{\prod\limits_{j=1}^{n}\sigma_{j}^{2}}{\det \mathbf{V}} \right).
\ee
\end{enumerate}
\end{lemma}
It would be tempting to propose mutual information effective sample size as something like

\bd
\left(1-\frac{\mathcal{I}(f)}{\sum\limits_{j=1}^{n}\mathcal{H}_{j}}\right)n.
\ed
However, $\mathcal{H}_{j}$ can easily be negative. We therefore have
to find some other way of using the entropy. 
\citet{XLinJPitBCla2007} used a similar motivation to define 
an effective sample size in order to obtain correct standard errors for parameter estimates. 
Theirs was a Bayesian setting and they define the effective sample
size as a minimizer of a relative entropy. The relative entropy is between the posterior parameter 
distribution under the true model and the the posterior parameter distribution under the effective 
sample. 
However, their approach does not allow for fractional sample sizes and could require, 
in the phylogenetic case, optimizing over the power set of species. 
Therefore, I propose to define the mutual information ESS (miESS) as

\be\label{defESSMI}
n_{e}^{MI} = 1 + \frac{1}{e(\mathcal{I}(f))}(n-1),
\ee
where $e(\cdot)$ is a strictly increasing function such that $e(0)=1$ and $e(\infty)=\infty$.
One example of such a function is the logarithm of $\mathcal{I}(f)$ increased by $\exp(1)$, considered in this work.
I choose such a function as compared to other formulae, e.g. $\exp(\cdot)$, it 
resulted in phylogenetic ESSs similar to those defined by the two other formulae. However, 
the proposed formula for $e(\cdot)$ should only be treated as a temporary definition. Further work
is needed to appropriately define it so that e.g. in the case of normal processes
(like BM or OU ones) it agrees with the rESS. In order to calculate miESS one needs
knowledge of the joint distribution of the tip species, or at least posses a numerical
procedure for obtaining it. Both could be unfortunately difficult to obtain in the non--normal
case, but \citep{MEllAMoo2014} present a family of heavy--tailed stable distributions for which
the joint likelihood is calculable.

The ESS, defined as such, has the desirable properties of being between $1$ and $n$.
In the Gaussian the formula for the miESS will equal

\be\label{defESSMIGauss}
n_{e}^{MI} = 1 + 
\log \left(
\exp(1)+\frac{1}{2}  \left(\sum\limits_{j=1}^{n}\log\sigma_{j}^{2} - \log \det \mathbf{V} \right)
\right)^{-1} (n-1).
\ee

It is important to notice, that the three proposed concepts of effective 
sample sizes are not compatible with each other. Firstly the mESS is meant
to quantify only information about the expected value of the sample, not
about independent signal. The motivations behind miESS and rESS are the same,
but it remains for a further study to define an appropriate transformation
$e(\cdot)$ that will make miESS equal to rESS in the normal process case.
In Sections \ref{secpESS}, \ref{secphilIC}, \ref{SecPDcons} and \ref{BiolpESS} I study their 
behaviour 
for simulated and real data.

\subsection{Multivariate extension}\label{secESSMvdef}

All of the above three definitions assumed that the each of the sample points is univariate. 
However,
methods for studying multiple co--evolving traits on the phylogeny are being developed
\citep[see e.g.][]{KBaretal2012,JBeauetal2012,JClaGEscGMer2015,THanJPieSOrz2008}
and all three considered ESS concepts are immediately generalizable to higher dimensions.  
Assume now that we have 
a $d$ dimensional trait. 
Each of our $n$ points 
a $d$ dimensional observation,
our sample is of size $d\cdot n$ correlated points instead of $n$ 
and $\mathbf{V} \in \mathbb{R}^{nd \times nd}$
instead of $\mathbb{R}^{n \times n}$. Hence, for 
model selection purposes we can use the above described procedures replacing $n$ with $d\cdot n$ 
inside all formulae, as most software packages do. 

The miESS and rESS can be elegantly generalized to quantify how many $d$--dimensional observations
we have effectively, i.e. how many
effectively independent species do we have amongst our $n$ species, regardless
of the dimensionality of each species. Notice that Eq. \eqref{defESSMI} does not depend
on the dimensionality of the species and can be used nearly without change

\be\label{defESSMImv}
n_{e}^{MI} = 1+ \frac{1}{e(\mathcal{I}(f))}(n-1).
\ee
The only difference is that here $\mathcal{H}_{j}$ is the entropy not of a univariate
random variable, but of the $d_{j}$--dimensional random vector of the $j$--th species. 
In the Gaussian case, we obtain 

\be\label{defESSMIGaussmv}
n_{e}^{MI} = 
1+\log\left(\exp(1)+\frac{1}{2}\left(
\sum\limits_{j=1}^{n}\log\det \mathbf{V}_{j} - \log \det \mathbf{V}
\right) \right)^{-1}(n-1),
\ee
where $\mathbf{V}_{j}$ is the $j$--th $d_{j}$--dimensional diagonal block of $\mathbf{V}$, i.e. the marginal
covariance matrix of the $j$--th $d_{j}$--dimensional observation. 

In a similar fashion we can adapt the $n_{e}^{R}$ to count the number of effective species in the multitrait case.
We sum the conditional total variances i.e.,

\be\label{defESSmvRegmv}
n_{e}^{R} = 1+ \frac{1}{n}\sum\limits_{j=1}^{n}\sum\limits_{1=i\neq j}^{n}\det \left(\mathbf{I}- \mathbf{V}^{-1}_{i,i}\mathbf{V}_{i,-i}\mathbf{V}_{-i,-i}^{-1}\mathbf{V}_{-i,i}\right),
\ee
where $\mathbf{I}$ is the unit matrix of dimension equalling the number of traits. 
Here $-i$ notation means removing rows/columns corresponding to the $i$th species.
Notice that in no case
is it required that all species are of the same dimension, allowing for proper handling of 
missing data. 

\section{Phylogenetic effective sample size} \label{secpESS}

Effective sample size calculation is very important in the phylogenetic context
but it seems to have received little attention. 
Phylogenetic comparative methods have taken care of the inflated sample size phenomena
for the most important inference issues.
We obtain the correct likelihood value and may in principle obtain
correct confidence intervals, and p--values. However, 
further development is needed for problems that 
actually depend on the sample size. 
Effective (and not observed) sample sizes 
are important when quantifying the biodiversity
of a clade to e.g. develop conservation strategies
or when doing model selection. 

It would seem desirable, to be able to calculate the effective sample 
size directly from the phylogeny and base any further estimates on this value 
of $n_{e}$. In fact, this seems to be the postulated approach by
\citet[Ch. 11][]{CNun2011}, 
that one should use the tree's phylogenetic diversity
to obtain the amount of information in the sample.
\citet{CNun2011} does not formulate it exactly in this way but this
is how mathematically it should be understood. In Section \ref{SecPDcons},
on phylogenetic diversity and conservation
I discuss this in detail. However, using phylogenetic diversity
to obtain an effective sample size for a trait (or suite of them) will be akin
to assuming a Brownian motion (neutral drift) model of evolution. 
Phylogenetic diversity is the sum of all branch lengths on a tree and this
is proportional to the sum of the variances of independent changes on the tree.

However, as \citet{THanSOrz2005} pointed out Brownian change is not appropriate
for traits under stabilizing selection.
I discussed earlier, that all considered 
definitions of effective sample
size will depend on $\mathbf{V}$, the between--species covariance matrix,
and how it differs from a diagonal matrix. 
Therefore, we need to calculate $n_{e}$ based on $\mathbf{V}$ and not just
the phylogeny. The between species covariance matrix depends not only on the phylogeny,
but also on the model of evolution. 
We denote by $\mathbf{T}=[t_{ij}]_{1\le i,j,\le n}$
the matrix of speciation times, where $t_{ij}$ is the speciation time of
species $i$ and $j$ and $t_{i}$ the time of species $i$ 
(these will be all equal to the tree height if the tree is ultrametric).
\citet{KBaretal2012} report the form of $\mathbf{V}$ for various models of evolution.
\begin{itemize}
\item Unconstrained evolutionary model --- univariate Brownian motion defined by the 
stochastic differential equation (SDE):
$\ud X_{t} = \sigma \ud B_{t}$ 

\be
\mathbf{V} = \sigma^{2}\mathbf{T},
\ee
where $B_{t}$ is the standard Wiener process.
\item Constrained evolutionary model --- univariate Ornstein--Uhlenbeck process defined by the 
SDE $\ud X_{t} = -\alpha(X_{t} - \theta_{t}) \ud t +\sigma \ud B_{t}$: 

\be
\mathbf{V}[i,j] = \frac{\sigma^{2}}{2\alpha}\left(e^{-\alpha (t_{i}+t_{j}-2t_{i,j})} - e^{-\alpha(t_{i}+t_{j})} \right).
\ee
\item Multitrait unconstrained evolutionary model --- multivariate Brownian motion defined by the SDE
$\ud \vec{X}_{t} = \mathbf{\Sigma} \ud \vec{B}_{t}$:

\be
\mathbf{V} = \mathbf{T} \otimes \left(\mathbf{\Sigma}\mathbf{\Sigma}^{T}\right),
\ee
where $\otimes$ is the Kronecker product.
\item Multitrait constrained evolutionary model, traits adapting to constrained traits --- multivariate Ornstein--Uhlenbeck process
defined by the SDE $\ud \vec{X}_{t} = -\mathbf{A}(\vec{X}_{t} - \vec{\theta}_{t}) \ud t +\mathbf{\Sigma} \ud \vec{B}_{t}$:

\be
\begin{array}{rcl}
\mathbf{V}_{ij} & = &
e^{-\mathbf{A}(t_{i}-t_{i,j})}
\int\limits_{0}^{t_{i,j}} e^{-\mathbf{A}v}  \mathbf{\Sigma}\mathbf{\Sigma}^{T} e^{-\mathbf{A}^{T}v} \ud v
e^{-\mathbf{A}^{T}(t_{j}-t_{i,j})}
\\ & = &
\mathbf{P} e^{-\mathbf{\Lambda}(t_{i}-t_{i,j})}
\left(
\left[\frac{1}{\lambda_{k}+\lambda_{r}}\left(1-e^{-(\lambda_{k}+\lambda_{r})t_{i,j}} \right) \right]_{1 \le r,k \le d} 
\odot \mathbf{P}^{-1}\mathbf{\Sigma}\mathbf{\Sigma}^{T}\mathbf{P}^{-T}
\right) e^{-\mathbf{\Lambda}(t_{i}-t_{i,j})} \mathbf{P}^{T},
\end{array}
\ee
where $\odot$ is the Hadamard product, $\mathbf{P}$, $\mathbf{\Lambda} = \mathrm{diag}(\lambda_{1},\ldots,\lambda_{d})$
are the eigenvectors and eigenvalues of $\mathbf{A}$
and $\mathbf{V}_{ij}$ is the block $i$, $j$ of dimension $d\times d$ of $\mathbf{V}$, i.e. the intersection
of the rows $( (i-1)d,\ldots,id)$ and columns $( (j-1)d,\ldots,jd)$.
\item Multitrait constrained evolutionary model, traits adapting to unconstrained traits --- 
multivariate Ornstein--Uhlenbeck process
defined by the SDE system

$$
\begin{array}{rcl}
\ud \vec{Y}_{t} & = & -\mathbf{A}\left(\vec{Y}_{t} -\left( \vec{\theta}_{t} + \mathbf{B}\vec{X}_{t}\right) \right) \ud t +\mathbf{\Sigma}_{y} \ud \vec{B}^{y}_{t} \\
\ud \vec{X}_{t} & = & \mathbf{\Sigma}_{x} \ud \vec{B}^{x}_{t}:
\end{array}
$$

\be
\mathbf{V}_{ij} = 
\left[
\begin{array}{c|c}
\begin{array}{l}
e^{-\mathbf{A}(t_{i}-t_{i,j})}
\left(
\int\limits_{0}^{t_{i,j}} e^{-\mathbf{A}v}  \mathbf{\Sigma}_{y}\mathbf{\Sigma}_{y}^{T} e^{-\mathbf{A}^{T}v} \ud v
\right. \\ \left.
+ \int\limits_{0}^{t_{i,j}} e^{-\mathbf{A}v} \mathbf{B} \mathbf{\Sigma}_{x}\mathbf{\Sigma}_{x}^{T}\mathbf{B}^{T} e^{-\mathbf{A}^{T}v} \ud v
\right)
e^{-\mathbf{A}^{T}(t_{j}-t_{i,j})}
\\
+e^{-\mathbf{A}(t_{i}-t_{i,j})}(\mathbf{I} - e^{-\mathbf{A}t_{i,j}})\mathbf{A}^{-1}\mathbf{B}\mathbf{\Sigma}_{x}\mathbf{\Sigma}_{x}^{T}\mathbf{B}^{T}
\\
+
\mathbf{B}\mathbf{\Sigma}_{x}\mathbf{\Sigma}_{x}^{T}\mathbf{B}^{T}\mathbf{A}^{-T}(\mathbf{I} - e^{-\mathbf{A}^{T}t_{i,j}})e^{-\mathbf{A}^{T}(t_{j}-t_{i,j})}
\\
+t_{i,j} \mathbf{B} \mathbf{\Sigma}_{x}\mathbf{\Sigma}_{x}^{T}\mathbf{B}^{T} 
\end{array}
& 
\begin{array}{l}
t_{i,j} \mathbf{B} \mathbf{\Sigma}_{x}\mathbf{\Sigma}_{x}^{T}
\\
-e^{-\mathbf{A}(t_{i}-t_{i,j})}(\mathbf{I} - e^{-\mathbf{A}t_{i,j}})\mathbf{A}^{-1}\mathbf{B}\mathbf{\Sigma}_{x}\mathbf{\Sigma}_{x}^{T}
\end{array}
\\
\hline
\begin{array}{l}
t_{i,j} \mathbf{\Sigma}_{x}\mathbf{\Sigma}_{x}^{T}\mathbf{B}^{T} 
\\
-\mathbf{\Sigma}_{x}\mathbf{\Sigma}_{x}^{T}\mathbf{B}^{T}\mathbf{A}^{-T}(\mathbf{I} - e^{-\mathbf{A}^{T}t_{i,j}})e^{-\mathbf{A}^{T}(t_{j}-t_{i,j})}
\end{array}
&
t_{i,j}\mathbf{\Sigma}_{x}\mathbf{\Sigma}_{x}^{T}
\end{array}
\right]
\ee
where $\mathbf{I}$ is the identity matrix of dimensions $d\times d$.
\end{itemize}
Hence, before reporting 
an effective sample size for a clade one has to estimate the parameters
of the evolutionary model. It would be also interesting to consider
more complex Gaussian setups, like function--valued traits. \citet{NJonJMor2013}
consider such a setup: for each species they observe measurements at a vector
of coordinates. As they assume normality, jointly the data is multivariate
normal, indicating the usefulness of all three proposed pESSs.

Given a phylogenetic tree and model of evolution, we can easily calculate the
effective sample size by plugging in the appropriate formula.
Below I present the values of the different definitions of ESS 
for the BM model of evolution. 
Formulae for OU based models would 
be too lengthy to be readable. We assume the tree is ultrametric with height $T$.

\be\label{eqBMne}
\begin{array}{rcl}
n_{e}^{MI} & = & 1 + \left(\sqrt{\frac{\det \mathbf{T}}{T^{n}}} \right)(n-1) \\
n_{e}^{R} & = & \sum\limits_{i=1}^{n}\left( 1 - \frac{1}{T}\mathbf{T}_{i,-i}\mathbf{T}^{-1}_{-i,-i}\mathbf{T}_{-i,i}\right)\\
n_{e}^{E} & = & T\vec{1}^{T}\mathbf{T}^{-1}\vec{1}.
\end{array}
\ee

In the phylogenetic context it would be tempting to take for the ESS factor, $p$, the interspecies
correlation coefficient \citep{SSagKBar2012}

$$
\rho_{n} := \binom{n}{2}^{-1}\sum\limits_{i<j}^{n}\frac{\cov{X_{i}}{X_{j}}}{\sqrt{\Var{X_{i}}\Var{X_{j}}}},
$$
where the sum is over all pairs of tip species. The above random variable is very well studied for 
the pure birth tree. The expectation of $\rho_{n}$ was derived for the BM and OU process
\citep[also with jumps][]{KBar2014,KBarSSag2015,SSagKBar2012}. 
Recently \citet{WMulFCra2015} calculated the distribution under the above modes of evolution.
However, in all the considered models $\E{\rho_{n}} \to 0$. Furthermore, for BM on a tree
with extinction one can consider death coefficients such that $\E{\rho_{n}} \to 1$.
As $0 \le \rho_{n} \le 1$, by the dominated convergence theorem, we have $\rho_{n} \to 0$ (alternatively $\to 1$)
almost surely. Such almost sure $0$ or $1$ asymptotic behaviour is not consistent with the motivation
behind studying a pESS, where the sample should be somewhere between $1$ and $n$, not exactly $1$ or $n$.

\begin{figure}
\begin{center}
\includegraphics[width=0.3\textwidth]{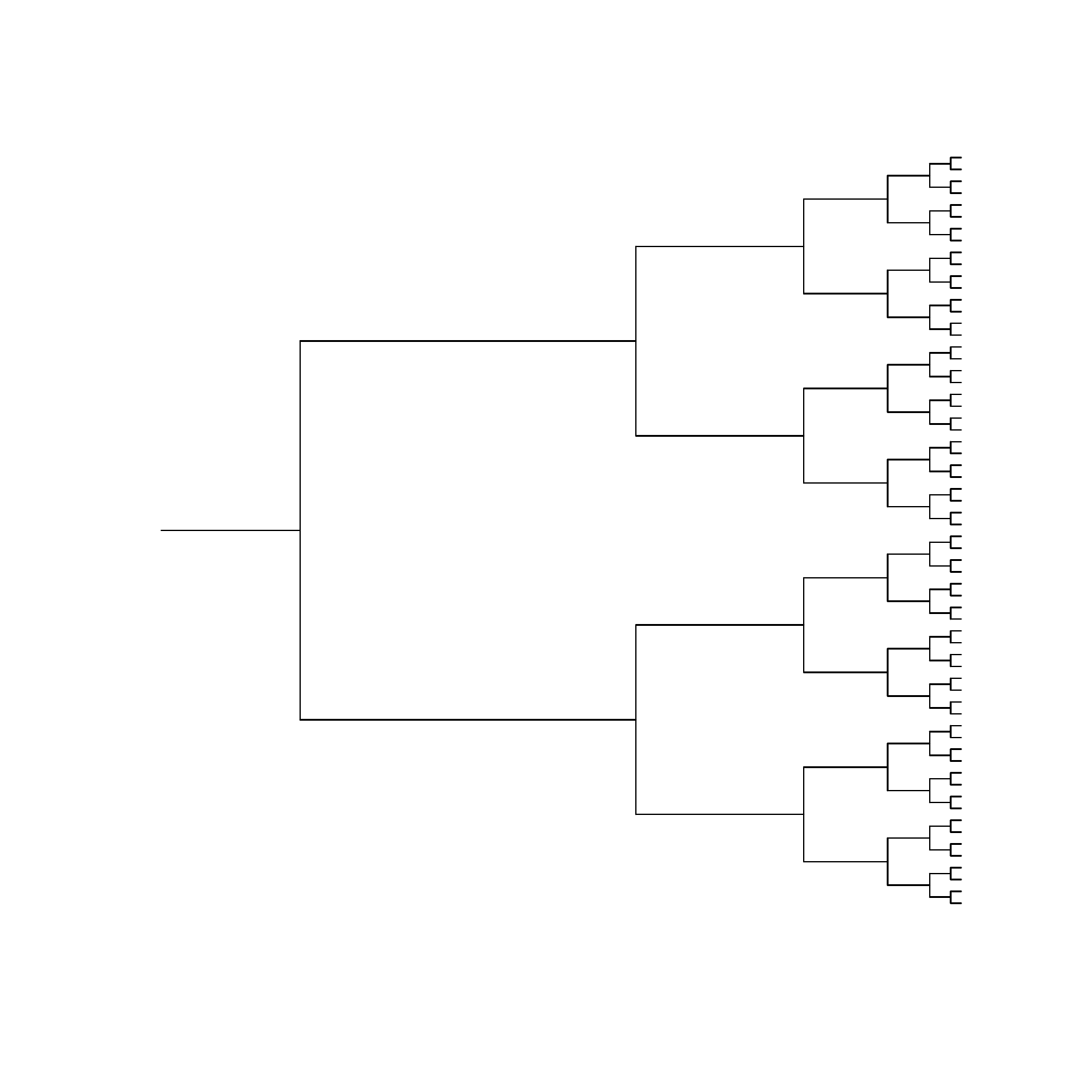} 
\includegraphics[width=0.3\textwidth]{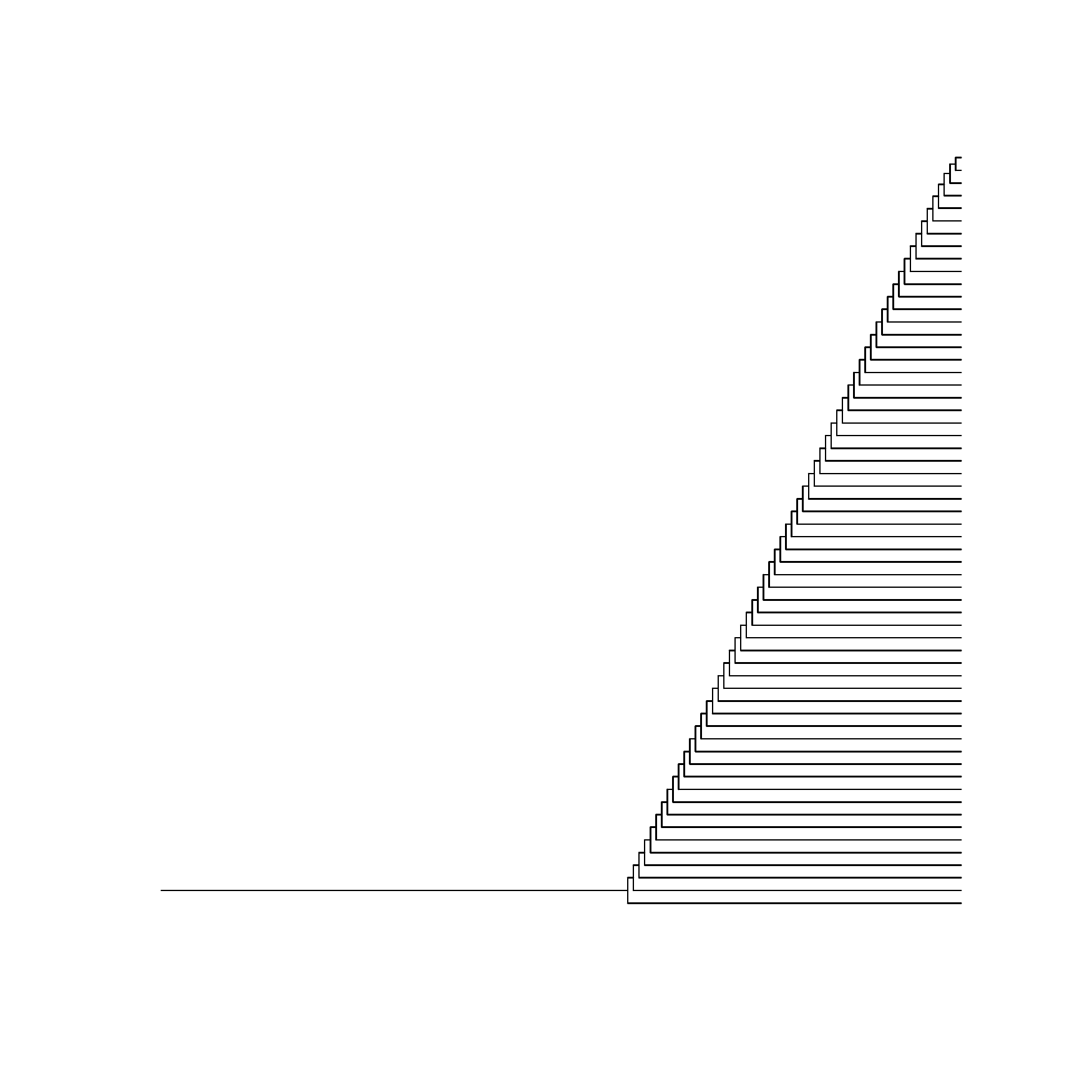} 
\includegraphics[width=0.3\textwidth]{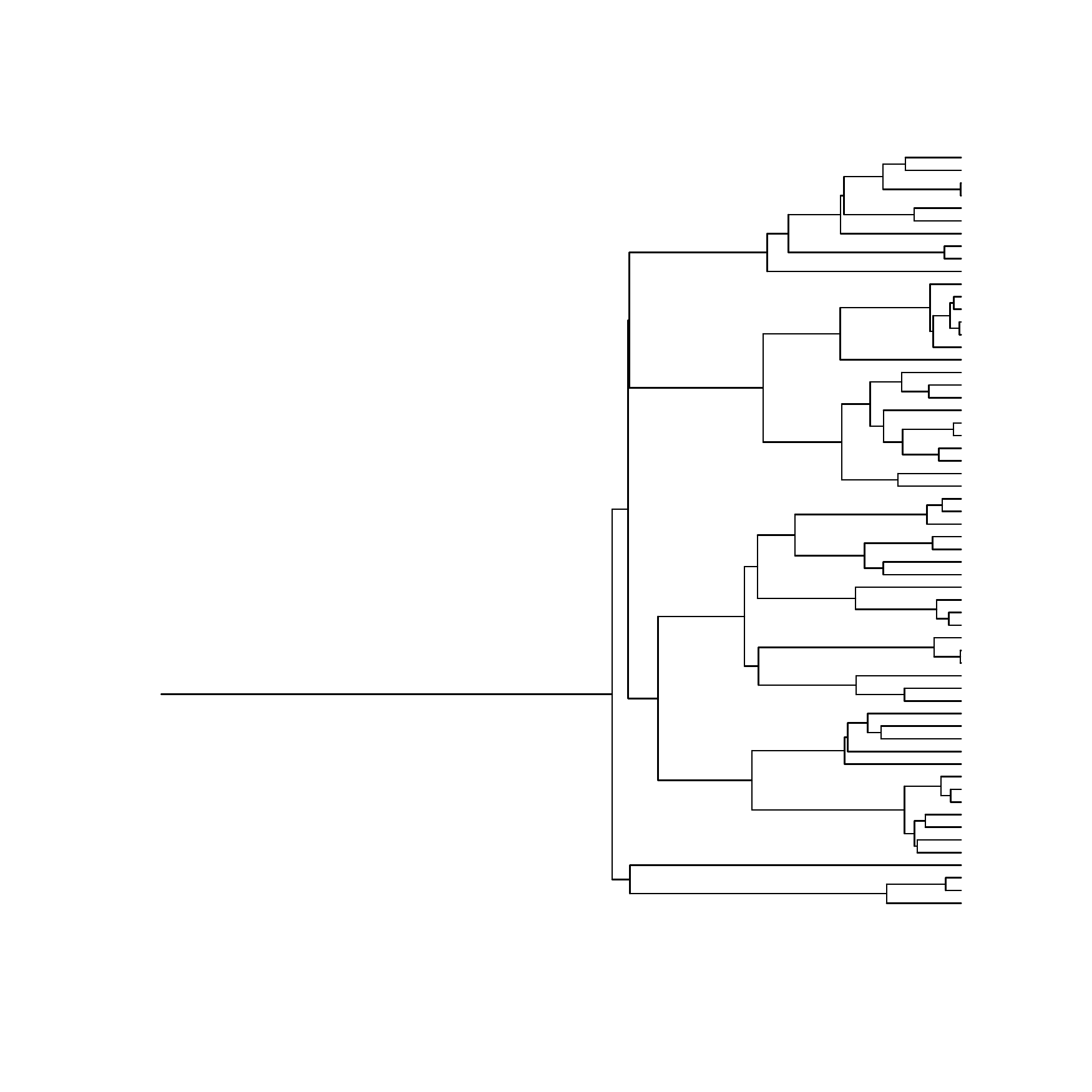} 
\caption{
Different binary phylogenetic tree setups used in the simulation studies.
Left: fully balanced tree, centre fully unbalanced tree, right:
single realization of a pure birth tree. 
The balanced tree has $64$ tips, the other two $60$.\label{figTreeTypes}
}
\end{center}
\end{figure}

\begin{figure}
\begin{center}
\includegraphics[width=0.4\textwidth]{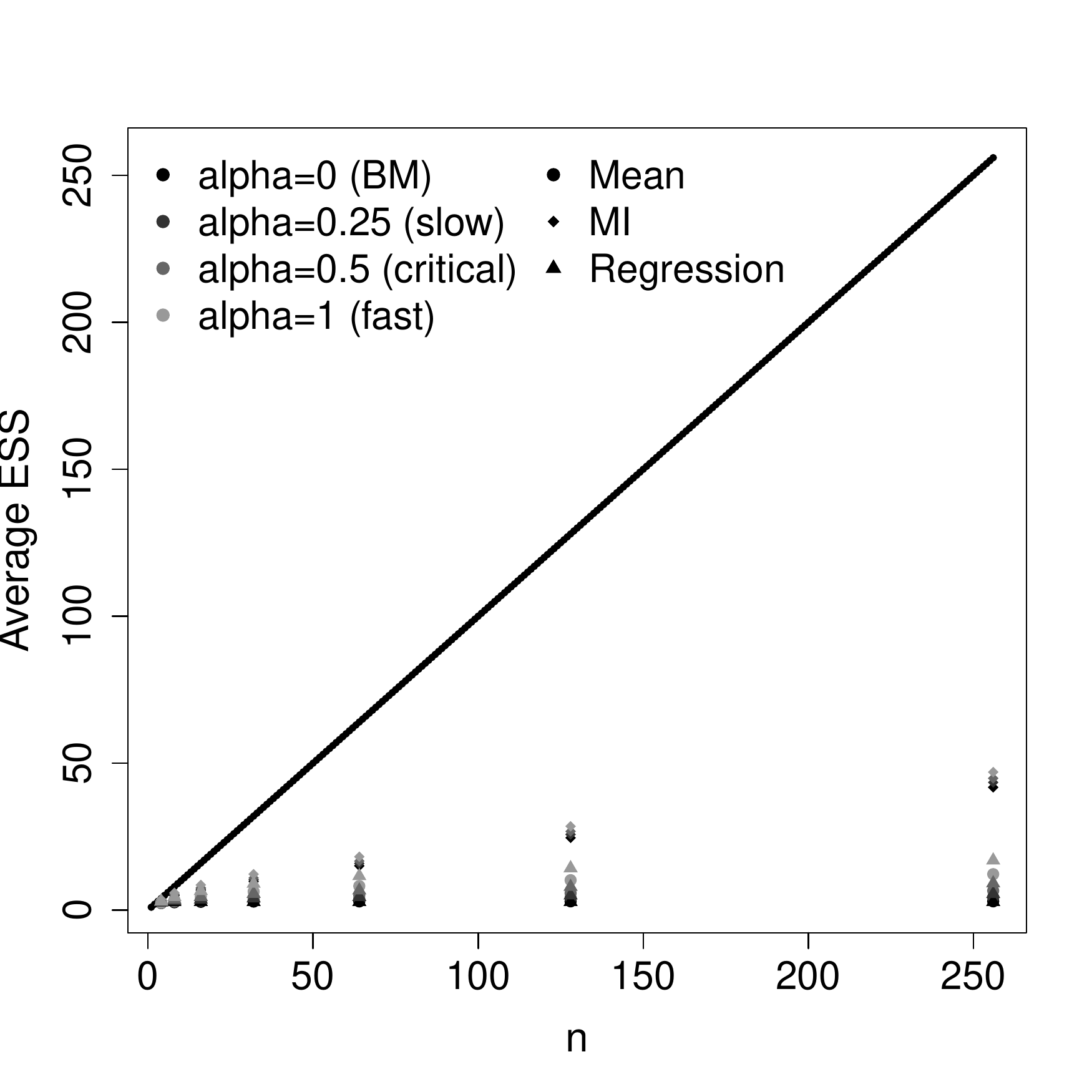} 
\includegraphics[width=0.4\textwidth]{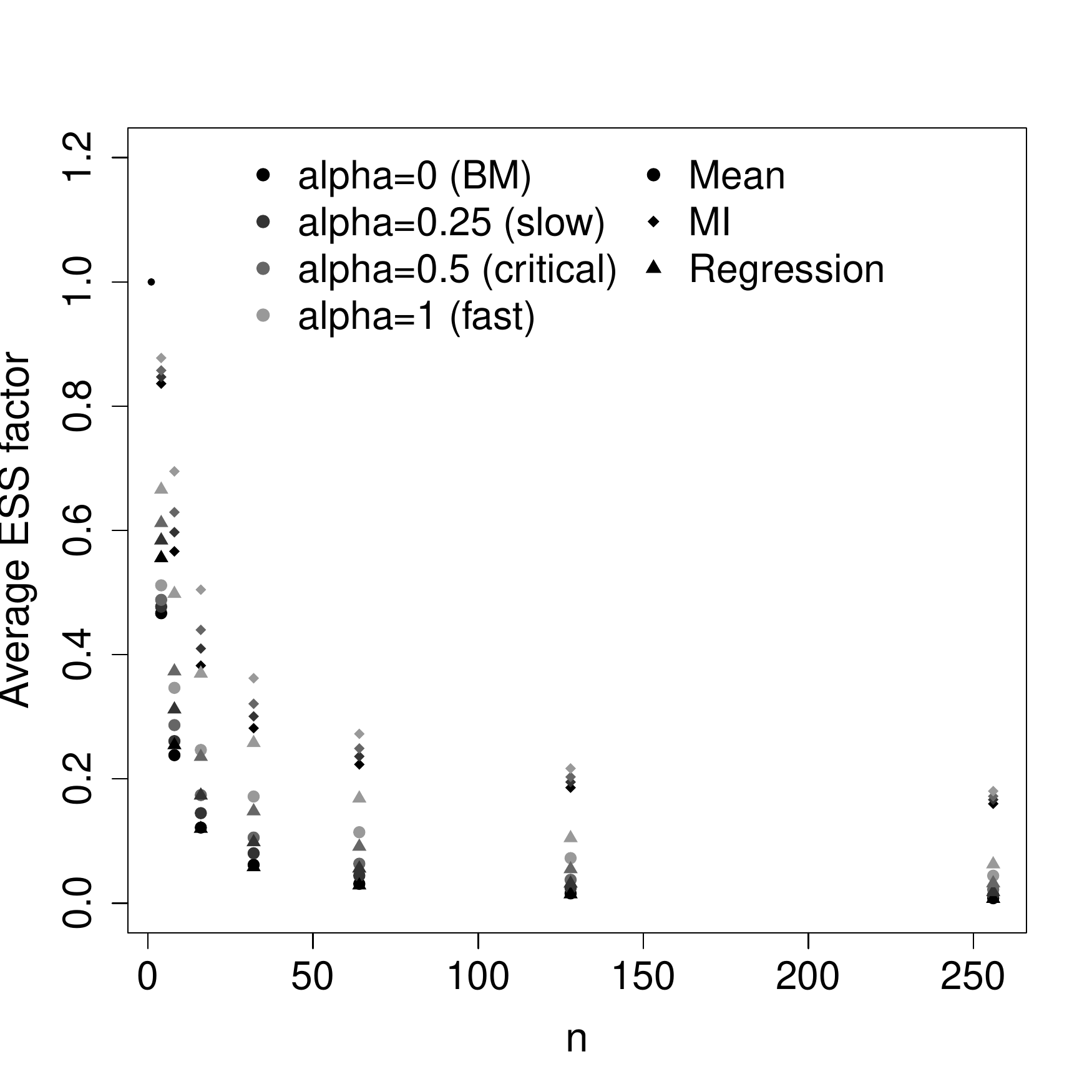} \\
\includegraphics[width=0.4\textwidth]{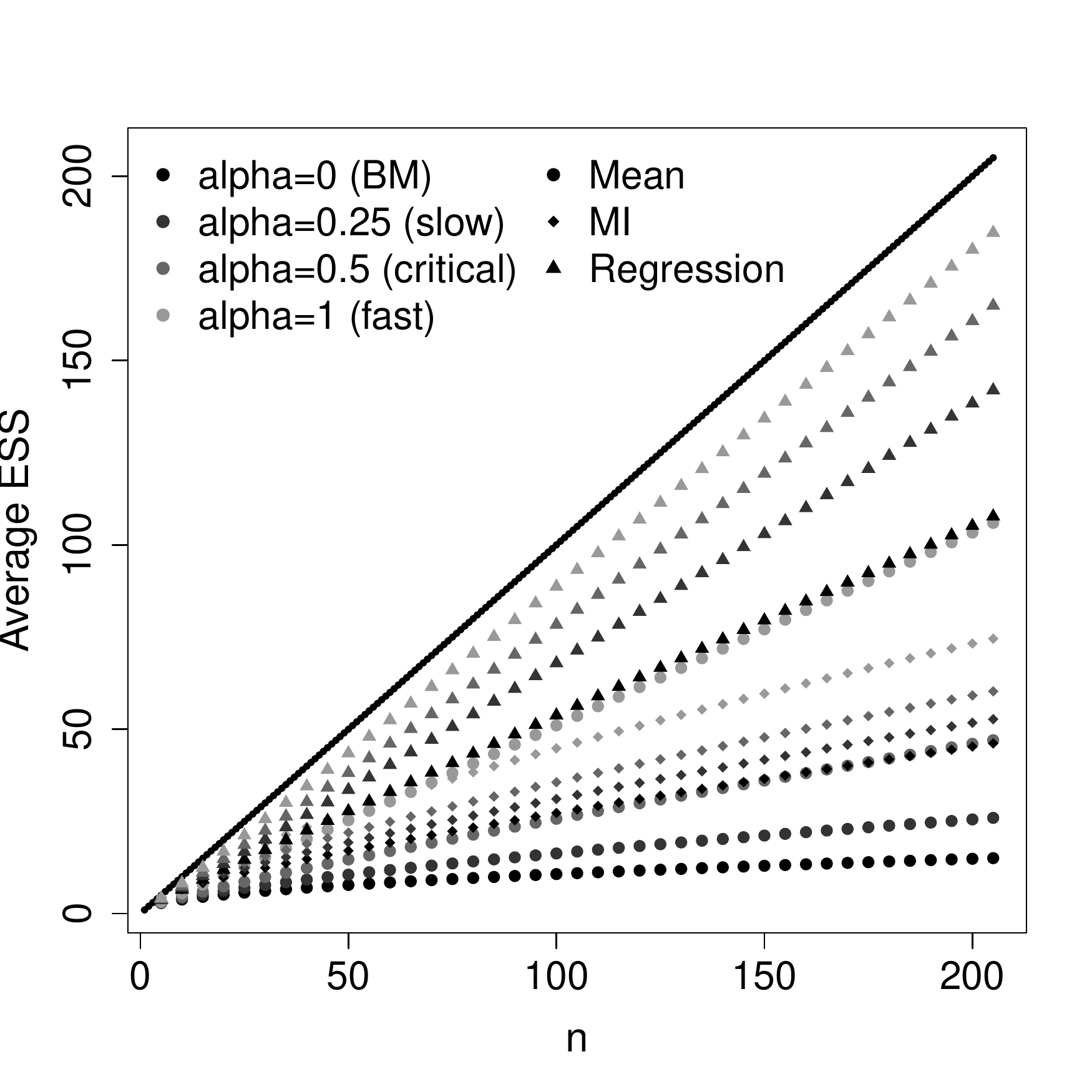} 
\includegraphics[width=0.4\textwidth]{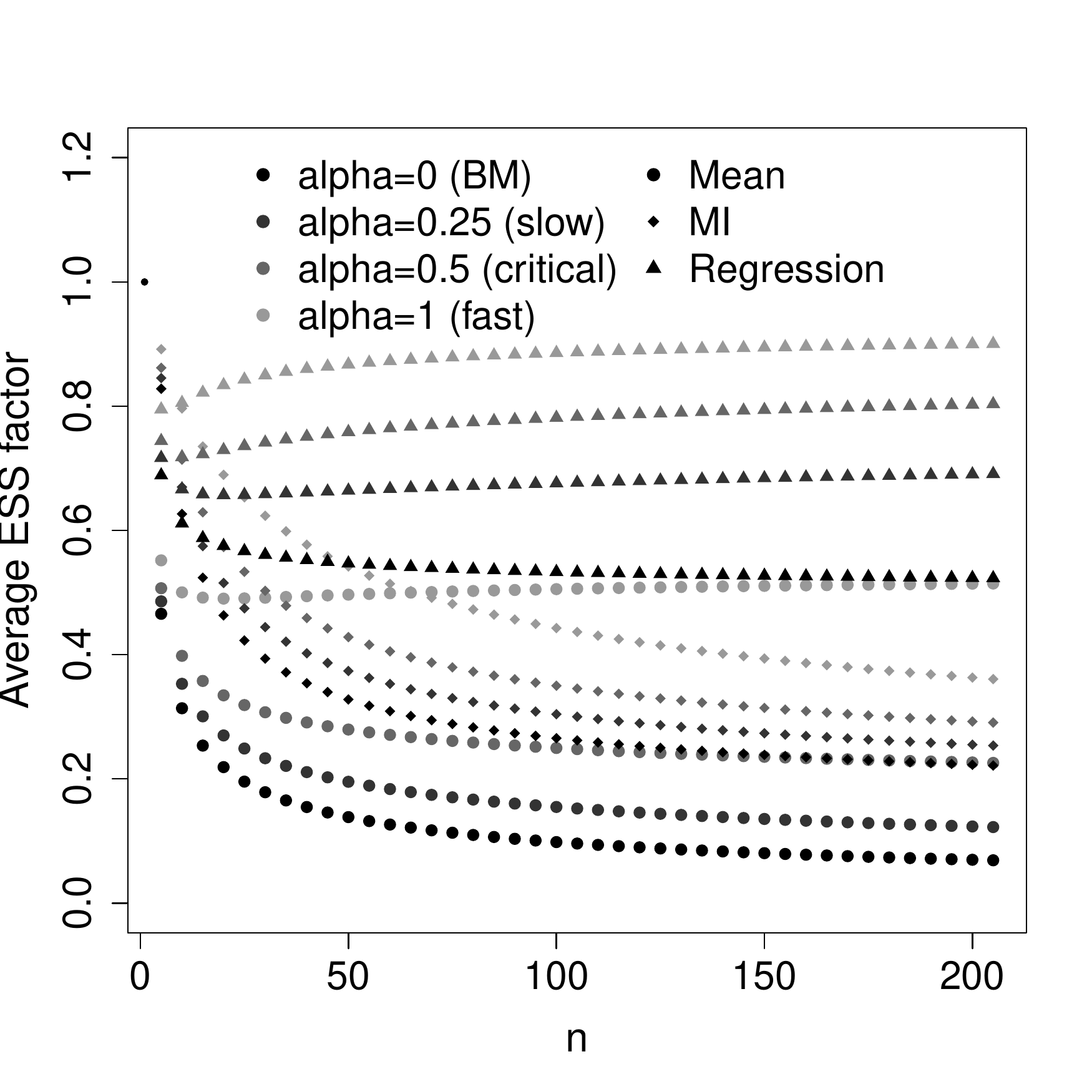} \\
\includegraphics[width=0.4\textwidth]{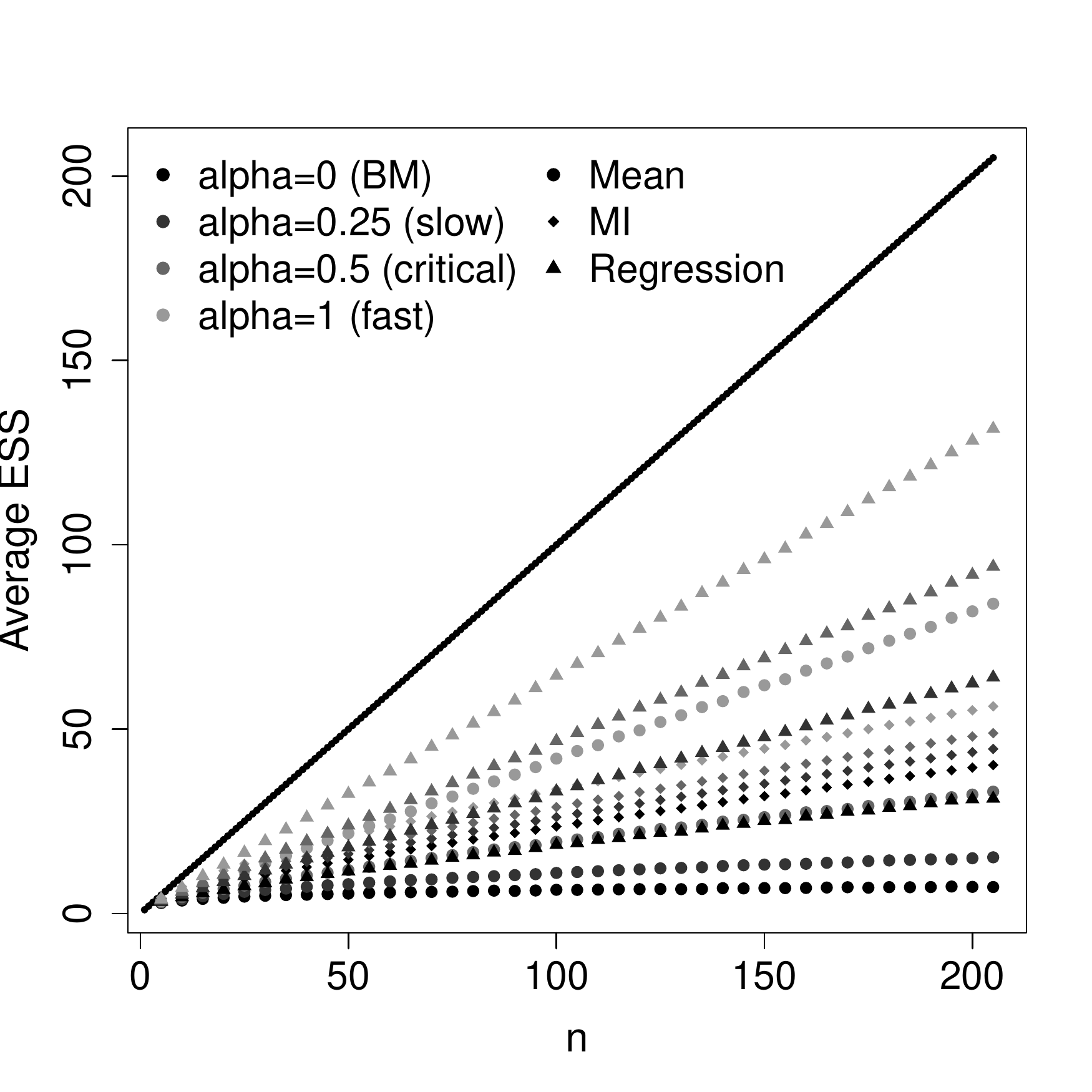}  
\includegraphics[width=0.4\textwidth]{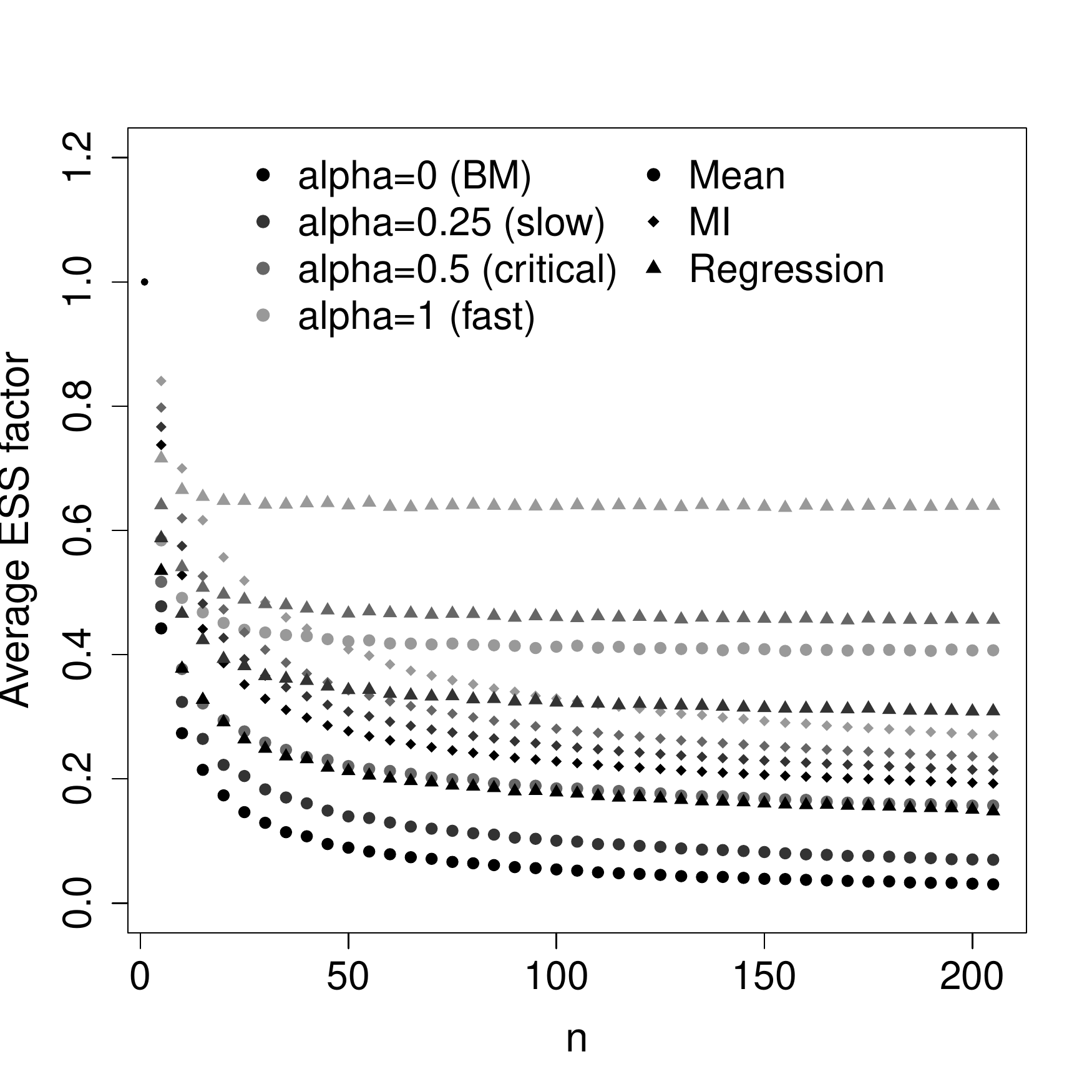}
\caption{Phylogenetic effective sample sizes for different types of trees and evolutionary processes.
First row: balanced tree, second row: left unbalanced tree, third row: average of $1000$ pure--birth Yule trees 
($\lambda=1$).
The balanced trees and unbalanced trees were generated using the function
stree() of the R \citep{R} ape package \citep{EPar2012}, the
Yule trees by the TreeSim R package \citep{TreeSim1,TreeSim2}.
First column: phylogenetic effective sample sizes, $n_{e}$ second column:
phylogenetic effective sample size factors, $p$.
The parameters of the processes are 
Brownian motion ($X_{0}=0$, $\sigma^{2}=1$),
second row: Ornstein--Uhlenbeck process ($\alpha=0.25$, $\sigma^{2}=1$, $X_{0}=0$, $\theta=0$),
third row: Ornstein--Uhlenbeck process ($\alpha=0.5$, $\sigma^{2}=1$, $X_{0}=0$,  $\theta=0$),
fourth row: Ornstein--Uhlenbeck process ($\alpha=1$, $\sigma^{2}=1$, $X_{0}=0$, $\theta=0$).
The straight black line is the observed number of taxa $n$. \label{figphylESS}}
\end{center}
\end{figure}

I illustrate Eq. \eqref{eqBMne} in Fig. \ref{figphylESS}. I also include the effective 
sample sizes for Ornstein--Uhlenbeck models. 
The considered evolutionary scenarios are a Brownian motion and Ornstein--Uhlenbeck process.
We fix the initial state $X_{0}=0$ and $\sigma^{2}=1$. For the OU 
process
we also fix the optimum $\theta=1$. We vary the adaptation rate $\alpha=0,0.25,0.5,1$.
We consider three binary phylogenetic tree setups (see Fig. \ref{figTreeTypes}). 
Two are deterministic trees: 
a completely unbalanced tree, a completely balanced tree (number of tips is a 
power of two). The third type is a random one ---
a conditioned on the number of tip species  Yule (pure birth) tree
\citep{KBarSSag2015,TGer2008a,TGer2008b,SSagKBar2012}. 
The rate of speciation is taken at $\lambda=1$. 
I take the number of tip
species to be from $5$ to $200$. Of course in the balanced tree only those that are
powers of two are allowed, hence there were significantly fewer trees. Each point
is the average over $1000$ simulations.

To make the simulations comparable 
the heights of the two deterministic tree types were scaled to $\log n$, the expected
height of the Yule tree. Also for these topologies randomness was added by
drawing the length of the root branch from the exponential with rate $1$ distribution.
In the case of the OU model, it allows the process to approach stationarity/stasis before 
speciation starts to take effect.

\section{Phylogenetic information criteria}\label{secphilIC}

My main motivation for studying the effective sample size in the phylogenetic context is 
obtaining correct values of information criteria that depend on sample size. 
Information criteria are necessary for e.g. finding the best evolutionary model, testing
evolutionary hypotheses, distinguishing
between competing phylogenies \citep{KBarPLio2014} or regime layouts \citep{MButAKin2004}.
If the evolutionary models/hypotheses are nested, then models can be compared by
a likelihood--ratio test. Such a test tells us whether the increase in the number of parameters
significantly improves the model fit. Alternatively when the models are not nested 
the Akaike information criterion
that penalizes for the number of extra parameters

$$
\mathrm{AIC}  = 2k - 2\log \mathcal{L}
$$
was proposed \citep{HAka1974}. In the above $k$ is the number of parameters and $\mathcal{L}$ the 
likelihood. The model with the lower AIC value is the better one. However, both the $\chi^{2}$
distribution of the likelihood ratio test and the AIC are asymptotic approaches. They
will be correct when the sample size is infinite (or large in practice). In
phylogenetic comparative studies the number of species is usually small.
Therefore two alternative
criteria that correct for small sample size were proposed to the phylogenetic
comparative methods community \citep{THanJPieSOrz2008}

$$
\mathrm{AIC}_{c} = \mathrm{AIC} + \frac{2k(k+1)}{n-k-1}
$$
and the Bayesian (or Schwarz) information criterion \citep{MButAKin2004}

$$
\mathrm{BIC} =  -2\log \mathcal{L} + k \log n.
$$
Of these two the AIC$_{c}$ seems to be the more used one (but AIC is also very popular). 

To see how much of a difference it makes, 
whether the observed or effective number of species is used,
I performed a simulation
study under various evolutionary scenarios. Under each scenario
I simulate data $N=1000$ times and from this obtain
histograms of the AIC$_{c}$ values under the true model
and an alternative using both the number of species and the effective sample size,
Figs. S.1---S.8  
in the supplementary material. 
I also plot in Fig. \ref{figphylAICcESS} how the average 
value of the small sample size correction
changes under the different evolutionary models and effective sample size value.
We consider the same evolutionary scenarios as in Fig. \ref{figphylESS}
and observe that for large $\alpha$ identifiability of the true model is easier.
The histograms of the AIC$_{c}$ are shown for small ($n=30$) and large ($n=205$)
phylogenies. We can see that in the large phylogeny case, all definitions 
of sample size result in the same distribution of AIC$_{c}$. However for 
the small phylogeny the mean and regression ESSs, $n_{e}^{R}$ and $n_{e}^{E}$, seem to be more effective
with the balanced phylogeny and fast adaptation.
The simulation results furthermore show that distinguishing different adapting OU models from each other and the BM one
can be difficult. This difficulty, especially with smaller $\alpha$s, is 
to be expected as the slowly adapting processes can take a lot
of time to reach stationarity and loose ancestral signal \citep{RAdaPMil2014,RAdaPMil2015,CAneLHoSRoc2014,KBarSSag2015}.
In fact our simulations confirm in this respect \citet{CCreMButAKin2015}'s recent study
--- ``Selection opportunity (\textit{i.e.} $\alpha$) is
substantially more difficult to estimate accurately: $\ldots$
relative errors exceeding $100\%$ are common, even when the correct model has been selected.''
\citep[especially for small $n$ and $\alpha$, see Fig. 6 of][]{CCreMButAKin2015}.
Hence, significantly larger sample sizes would be needed to identify slowly adapting models.
Figure \ref{figphylAICcESS} also tells us that even with smaller sized phylogenies
all pESS definitions should result in similar AIC$_{c}$ values. 
The observed agreement, between all tested sample size definitions, suggests that the likelihood dominates the AIC$_{c}$, which
is not surprising as the data is simulated under the BM or OU models. A similar
consistency is observed when working with real data (Section \ref{BiolpESS}).
The situation is different for the fully balanced tree which holds the most dependencies
between the species. In such a symmetric case, probably a much larger tree would be needed to 
obtain stability. 

\citet{CAne2008} noticed that for a Brownian motion model of evolution
effective sample sizes can be very small.
\citet{TGarADicCJanJJon1993}'s mammal phylogeny had $n_{e}^{\mathrm{E}}=6.111$ with $49$ tip species.
My simulations give very similar numbers (Fig. \ref{figphylESS}).
A Yule tree of $50$ tips has $\E{n_{e}^{\mathrm{E}}}=5.391$, $\E{n_{e}^{\mathrm{MI}}}=14.574$ 
and $\E{n_{e}^{\mathrm{R}}}=11.455$,
a fully unbalanced tree with $50$ tips has 
$\E{n_{e}^{\mathrm{E}}}=7.781$, $\E{n_{e}^{\mathrm{MI}}}=17.06$ and $\E{n_{e}^{\mathrm{R}}}=27.802$
and a fully balanced tree of $64$ tips has 
$\E{n_{e}^{\mathrm{E}}}=2.909$, $\E{n_{e}^{\mathrm{MI}}}=9.729$ and $\E{n_{e}^{\mathrm{R}}}=2.8$. 

The very low amount of independent information is evident. In Section \ref{BiolpESS}
I reanalyzed \citet{TGarADicCJanJJon1993}'s mammalian data \citep[from the ade4 R package][]{SDraADuf2007}.
Of course $n_{e}^{\mathrm{E}}=6.111$, as expected for the mammalian body size evolution
(the BM model was selected). The other pESSs were not much higher
$n_{e}^{\mathrm{MI}}=14.125$ and $n_{e}^{\mathrm{R}}=9.437$ (also BM model).
In Section \ref{BiolpESS} I discuss this data set in more detail. 

In most cases,
the mean effective sample size is the lowest because it measures
the information that the sample contains on the mean value. In the BM case, this is the ancestral
state and there is very little information on it. The other pESSs look more holistically
at what dependencies are in the data and hence are larger. If we move to more and 
more adaptive OU models (increase $\alpha$), then all, but especially $n_{e}^{E}$, increase.
The mean ESS is nearly always the smallest. However, if adaptation is fast and 
terminal branches are long (i.e. the contemporary sample is nearly independent), then
it can also be nearly $n$ (see Tab. \ref{tabBioDiv}).

Based on the simulation results alone, it is difficult to provide rules of thumb for the 
applied user. All methods essentially give the same results (as they should under simulated
data!). However the analyses of real data in Section \ref{BiolpESS} does provide 
some recommendations which are there discussed. 
One suggestion from the simulations is that it is not that important which information
criterion one uses --- all should result in the same conclusion. In the PCM field 
there is a tradition to prefer the AIC$_{c}$ and BIC over the AIC, but at least in this study
I did not notice significant differences. 

\begin{figure}
\begin{center}
\includegraphics[width=0.3\textwidth]{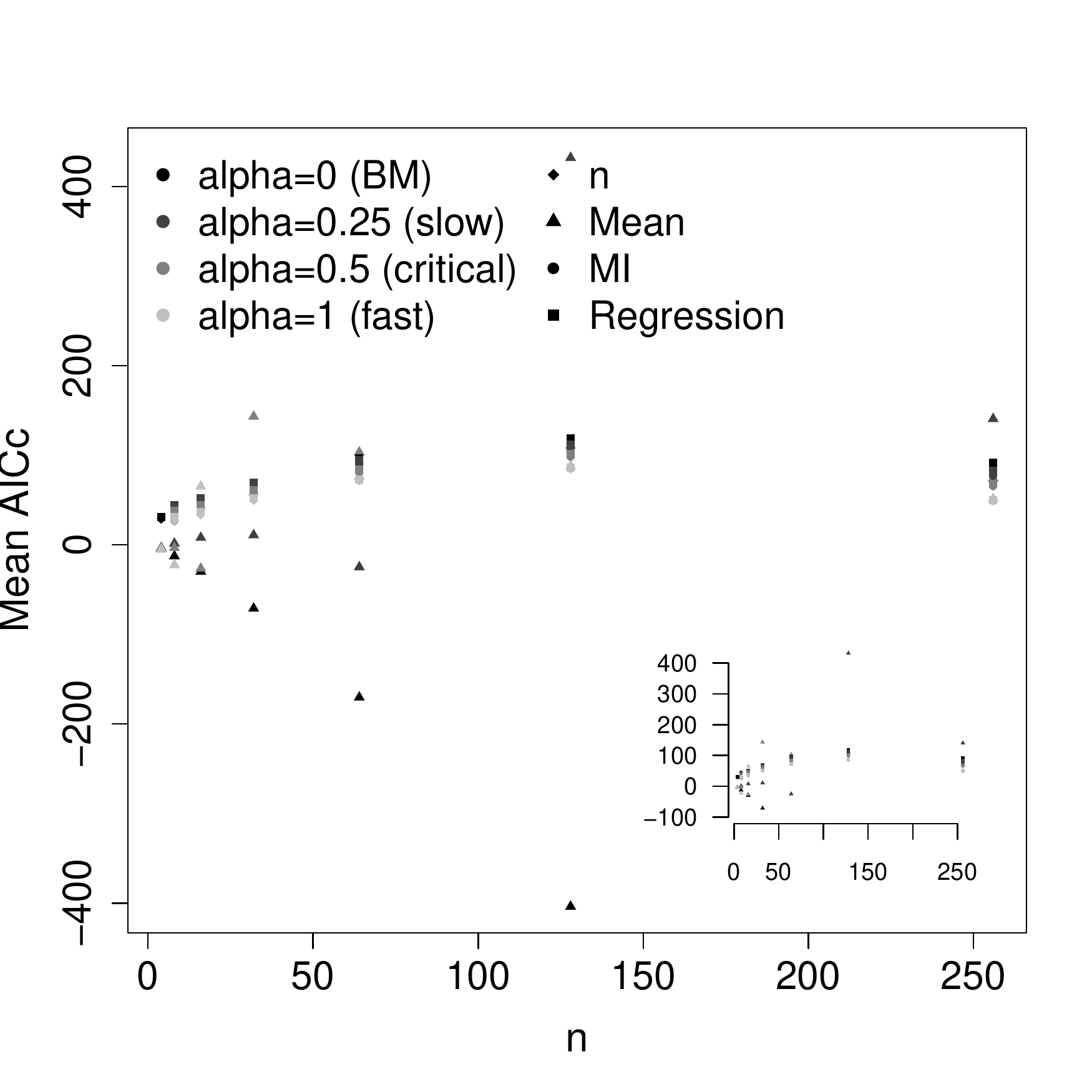}
\includegraphics[width=0.3\textwidth]{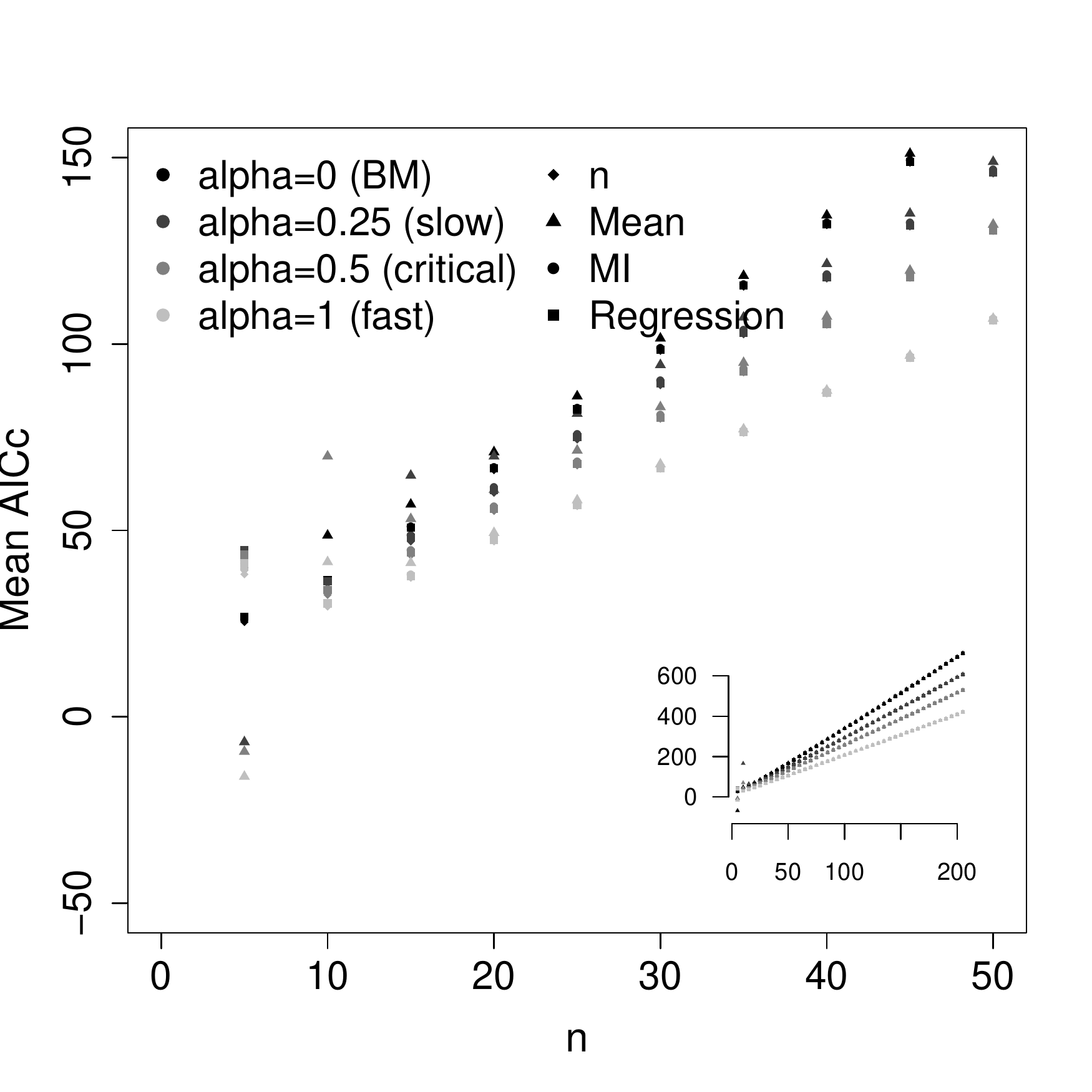}
\includegraphics[width=0.3\textwidth]{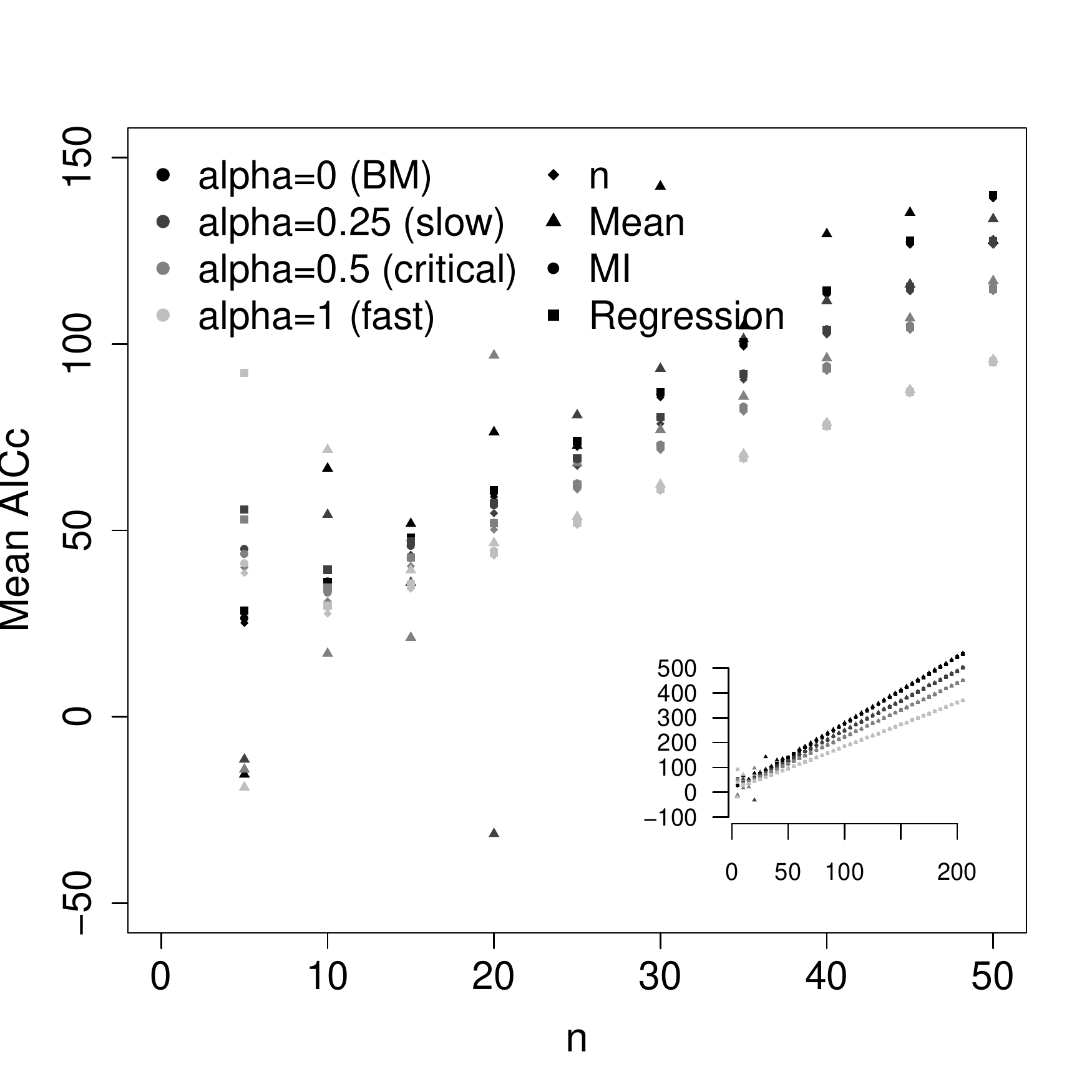}
\caption{
AIC$_{c}$ effective sample size correction for different types of trees and evolutionary processes.
Left: balanced tree, centre: left unbalanced tree, right: average of $1000$ pure--birth Yule trees ($\lambda=1$).
The balanced trees and unbalanced trees were generated using the function
stree() of the R ape package, the
Yule trees by the TreeSim R package.
The parameters of the processes are 
Brownian motion ($X_{0}=0$, $\sigma^{2}=1$),
second row: Ornstein--Uhlenbeck process ($\alpha=0.25$, $\sigma^{2}=1$, $X_{0}=0$, $\theta=0$),
third row: Ornstein--Uhlenbeck process ($\alpha=0.5$, $\sigma^{2}=1$, $X_{0}=0$,  $\theta=0$),
fourth row: Ornstein--Uhlenbeck process ($\alpha=1$, $\sigma^{2}=1$, $X_{0}=0$, $\theta=0$).\label{figphylAICcESS}}
\end{center}
\end{figure}

\section{Phenotypic diversity and conservation}
\label{SecPDcons}

An important application of phylogenetic methods is to quantify the biodiversity
of a group of species. 
Phylogenetic methods allow one to formulate
definitions of species that are useful from an evolutionary point of view \citep[Ch. 11][]{CNun2011}.
I will not be concerned with a definition of a species but assume that some phylogeny relating
predefined taxonomic units is available. The impact of a phylogenetic definition of
species was investigated by \citet{PAgaOBinKCraJGitGMacJMarAPur2004} and they noticed
that this caused an average increase of the number of species by about $49\%$ when compared
to alternative definitions. Such an influx can mean that a lot of species were ``split'',
resulting in species with smaller populations and geographic ranges. In turn as these
are variables contributing to classifying a species as endangered, it may lead to more labelled as such.
Therefore \citet{PAgaOBinKCraJGitGMacJMarAPur2004} postulated quantifying 
conservation value using alternative
variables, one of which was trait diversity. 

\citet{DFai1992} suggested to quantify biodiversity through phylogenetic information. 
The main idea is that one should concentrate on feature diversity --- how diverse are organisms.
Diversity is of course something difficult to quantify, we do not even know of all the variables to measure.
\citet{RCro1997} pointed out that one of the aims of conservation is to ``maximize the 
preserved information of the planet's biota best in terms of genetic information''.
He then points out that phylogenetic based measures which include branch lengths
will be better indicators than just counting the number of species.
Therefore as a proxy \citet{CNun2011} following
\citet{DFai1992}
proposes 
\citep[but also refers the reader to][]{DFai1994,DFai2002,RCro1997,APurJGitTBro2005} 
to quantify feature diversity with
phylogenetic diversity (PD) --- the sum of branch lengths of a tree/clade. 
The extinction of a clade (or species)
is therefore equivalent to subtracting the amount of branch lengths particular to this clade. 
Phylogenetic diversity has a rich literature \citep[e.g.][]{FCraMSuc2013,AMooetal2012,TStaMSte2012} 
and therefore
it is possible to make quantitative predictions about diversity loss/retention
under different models of tree growth, extinction and conservation. 

From a mathematical perspective PD quantifies the amount of feature diversity
as the amount of accumulated variance under the assumption that evolution follows Brownian motion.
One may say that this is sensible as an overall feature variable describing a species will be the sum  
of effects of many traits. Individually traits may be under selection but their sum is 
not necessarily adapting to anything --- providing an argument for Brownian drift.

An alternative approach that could be used to quantify the biodiversity (or feature diversity) of 
a clade of $n$ species is the effective amount of species in this clade $n_{e}$. 
This is done in a straightforward way. We prune the phylogeny to the subtree which 
contains only this clade, and use the methods described in this work to obtain
$n_{e}$ for this subtree.
Such an approach could be more appropriate for various reasons. 
For example
it could turn out that the traits important from a conservation point of view are quantified by
another process e.g. Ornstein--Uhlenbeck. In the OU case, the changes
along disjoint parts of the phylogeny are not independent and the 
variance is not a linear function of time.

The above trait based approach for quantifying biodiversity is closely related to the ideas
presented by \citet{SPavSOllADuf2005a}. They introduce the ``originality of a species within a
set'' concept based on \citet{CRao1982}'s quadratic entropy
that describes the ``average rarity of all the features belonging to this species''
\citep[see also][]{SPavSOllDPon2005b}.
In the discrete trait case, \citet{SPavSOllADuf2005a} find it equivalent to phylogenetic diversity.
They analyze the Carnivora data set \citep{JDinNTor2002,SPavSOllADuf2005a}
and plot (their Fig. $3$) how the PD changes with the amount of species dropped.
Interestingly the PD reaches a final plateau around $58$ (out of $70$) species 
--- the same amount that is the rESS for the range measurement.

Phylogenetic diversity measures can be overturned
if one uses the diversity of a (suite of) trait(s) as a proxy for biodiversity. In a very wide and recent shallow
radiation the diversity of a trait can be very small while the sum of branch lengths can be large. 
On the other hand if we have trees with few very old tips, then they may have much lower phylogenetic 
diversity. However, they might have diverged so far back in time and accumulated so much change in their phenotype
(without speciating), that the loss of even one tip results in a much more significant loss of phenotypic innovation 
than even of 
most of a recent shallow radiation. The latter is intuitively obvious --- in a recent shallow radiation the majority of information
about all species is coded inside essentially all of the species. It suffices for only one to survive for most of the information
to be retained. But making the radiation wider and wider one can imagine increasing 
the measure of phylogenetic diversity as much as desired.
Of course loosing tips is equivalent to loosing small innovations that set the species apart. 
All changes are naturally
a value in themselves but the majority of information is stored in any individual tip. However, in the many old species case,
every single species is a distinct entity not containing much information about the rest. Hence, any loss of a single
species leads to an irreplaceable loss of diversity
\citep{SNeeRMay1997}, while the phylogenetic diversity measure might not pick this up.
\citet[][p. 319]{CNun2011} points out that we are losing 
biological and cultural diversity at a faster rate than ever before.
Therefore, it is important to quantify how much of what we loose. 


Rather recently \citet{MVelWCorKMagForAMoo2011} compared various
phylogenetic based measures of biodiversity, including PD.
They found that mean PD (mPD, average over all pairs of species phylogenetic distance)
was more sensitive in detecting ``non--random community assembly'' in a clade. 
This is probably due to mPD taking advantage of more information, the branch lengths
and tree topology (averaging over pairs).


The pESS can be considered 
as a proposal of a new multi--omics currency of biodiversity. Instead of the standard 
currencies ``species'' or PD  I use diversity in traits. In other words, I sum up
innovations particular to species. 
Based on such a partition of the variance one can
identify ``innovative'' clades which
contain a lot of information. 
The proposed in this work approach can be
a step towards species--free methods postulated by \citet{PAga2005}. 
As yet, the pESS is not completely
species--free of course, it still includes the phylogeny. The tips of the tree are pre--defined
by experts taxonomic units. 
However, it is 
not an only--species methodology as e.g. counting species would be. It includes
evolutionary process information, that takes into account the topology of
the tree --- how much of one species is there in another. Also, \citet{PAga2005} discusses
that the problem with species methodologies is that depending on the definition
of species we can get wildly different counts. 
\citet{NIsaAPur2004} point out that correct identification of species 
numbers is important for understanding the diversity of our world.

Therefore, if one misidentifies a species, problems could occur ---
the species count will be wrong and hence the phylogenetic diversity. It will
be based on too few or too many branches.
And what if one missed a particular subpopulation that had something very special
attached to it? Can one still include its diversity even though it does not appear on the phylogeny?
The pESS can precisely do this through integrating data from different sources.
Assigning the effective clade size that takes into account the phylogeny
and trait variance and between--species covariance, should allow one to strike
a balance between expert knowledge concerning species and uncertainty attached 
to correct demarcation. For example evolutionary models can be easily extended
to include intra--species variance, often called ``measurement error''
\citep[e.g.][to name a few]{JFel2008,THanKBar2012,RRohPHarRNie2013}. Mathematically,
these methods boil down to adding to the matrix $\mathbf{V}$ a matrix $\mathbf{M}$
which is the intra--species variance (``measurement error''). Then this new
covariance matrix $\mathbf{V}+\mathbf{M}$ can be treated as the old $\mathbf{V}$
to obtain a value of effective species number. The intra--species variance
can be a representation of our uncertainty about species demarcation
and be used to correct for species miscall. If a species has many subpopulations,
that are very diverse, representing a species by only its mean over 
all (measured) individuals will not be the best option. Including
the variability of the trait inside the species can partially
alleviate the need to know the correct species structure. Such ``observational error'' 
can be thought of as averaging over all possible species demarcations that we are not sure about.

\citet{AMooSHeaEChr2005} discuss that one can look at conservation from an ethics point
of view should all species be considered equal and protected in the same way
or should one protect the features of evolution that are of some value for
us. Then phylogenetic diversity is a measure that quantifies a particular feature of evolution.
What I propose in my work is quantifying
a different feature of evolution. What sets it apart from PD is that it requires the researcher
to define traits --- exactly what features of evolution are valuable. To illustrate 
the statement, \citet{SNeeRMay1997} point out that the loss of \textit{Homo sapiens} would
result in a loss of a tiny fraction of evolutionary history, when one uses a measure that takes
into account only the tree. If we would choose a trait associated with e.g. civilization achievements and
then calculate the ESS of the human lineage ($1$ by definition) and non--human clades we
would obtain a completely different result.

In a way one could say that this is merely replacing counting species with counting the effective number
of species. However, the difference is in how we count. Counting just the number of species
means enumerating taxonomic units according to some definition. Counting the effective number
of species, in the way I propose, is really saying how much biodiversity we have in a clade, where
biodiversity is represented by some (suite of) trait(s). This measure can also be thought of
as calculating how much innovation we have in the clade. Of course my approach shifts the responsibility
to the biologist to identify what traits are important. 

\citet{WJetRFre2015} have very recently published an analysis that is distantly related to what I discuss.
They notice that on many species we have too little data, to say if they are endangered or not. 
On the one hand this would mean that we could assume that all data--deficient species 
are endangered, but as \citet{PAga2005} pointed out this would be far too costly. On the other hand
\citet{WJetRFre2015} point out, that \citet{SButJBir2010} observed that data--deficient birds are
at no greater extinction risk, than assessed birds. This suggests, that one could use, as \citet{WJetRFre2015}
do, e.g. body--mass, to predict threat status/threat probability. Of course, as species are dependent, in such an analysis
the phylogeny needs to be accounted for. Such an approach has the drawback, as \citet{WJetRFre2015} discuss, 
that a logistic regression, i.e. threatened/not threatened, will require a large dataset. Therefore, it might
be possible, but this of course requires further development and linkeage with phylogeographical models, that 
effective clade size could also be a proxy for threat status. In addition \citet{WJetRFre2015} point
out, that many species have missing measurements on phenotypes. The evolutionary models used to obtain 
pESS can handle unobserved data in a natural manner. There is no need to remove a species from an analysis
even if it has missing data.

\begin{figure}[!ht]
\begin{center}
\includegraphics[width=0.32\textwidth]{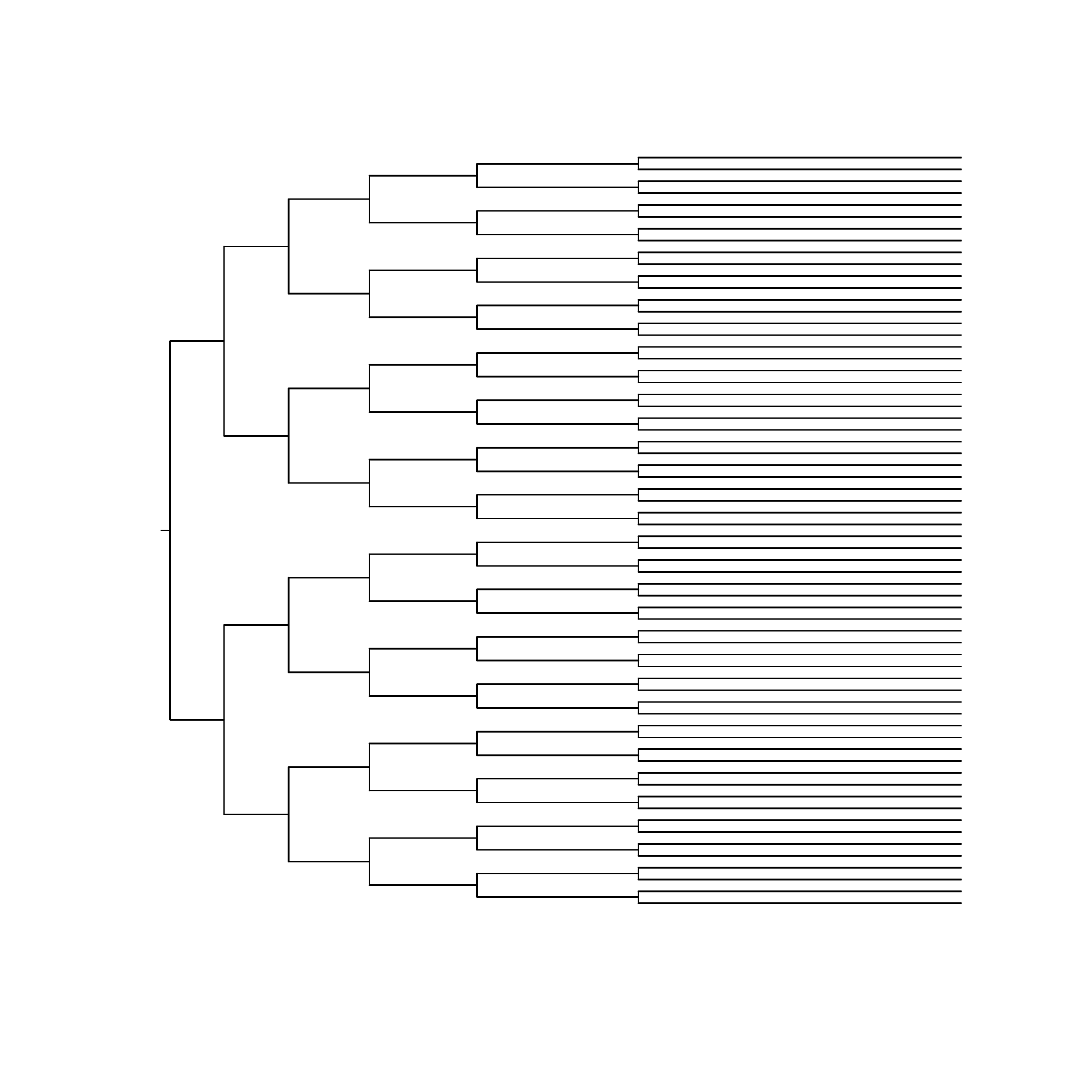}
\includegraphics[width=0.32\textwidth]{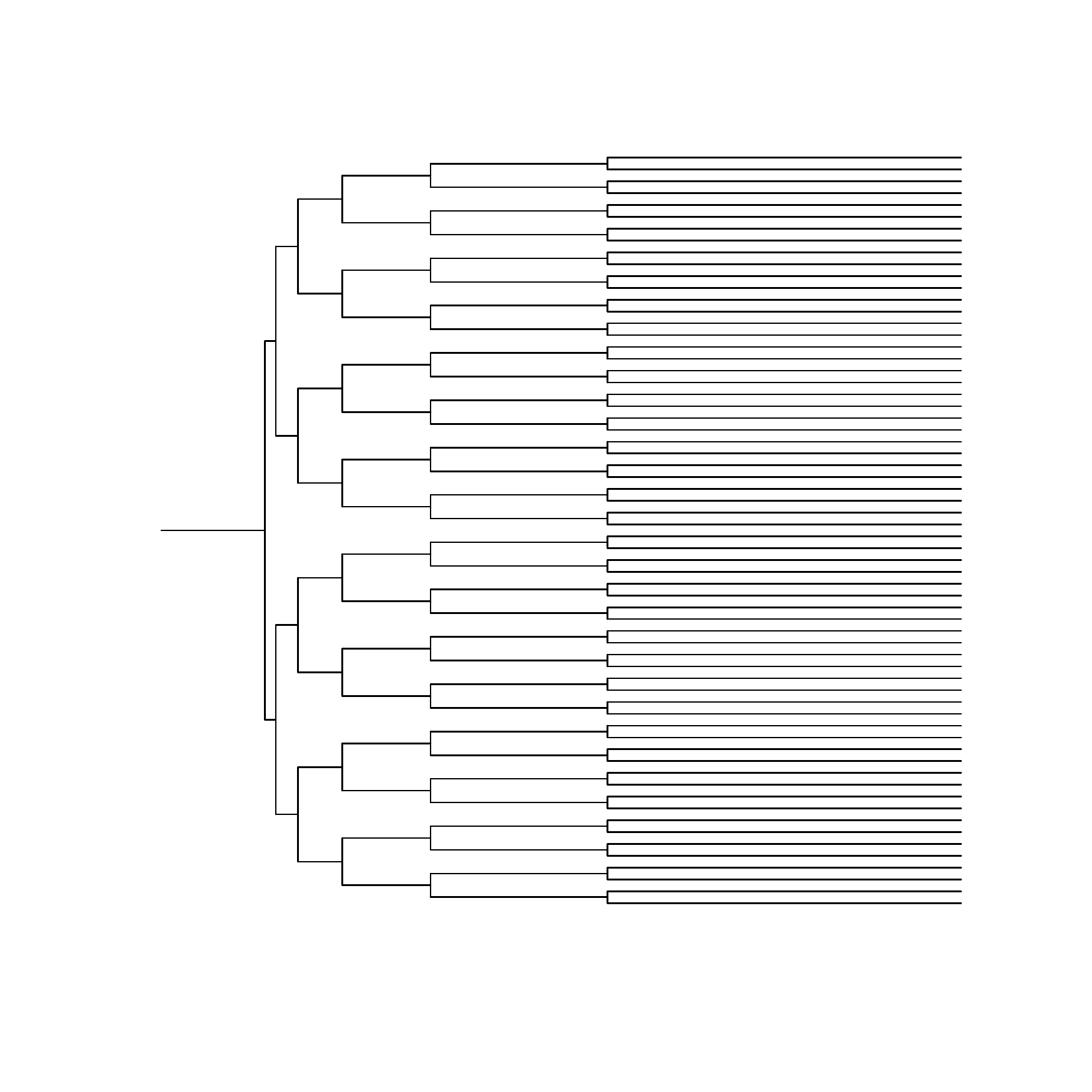}
\includegraphics[width=0.32\textwidth]{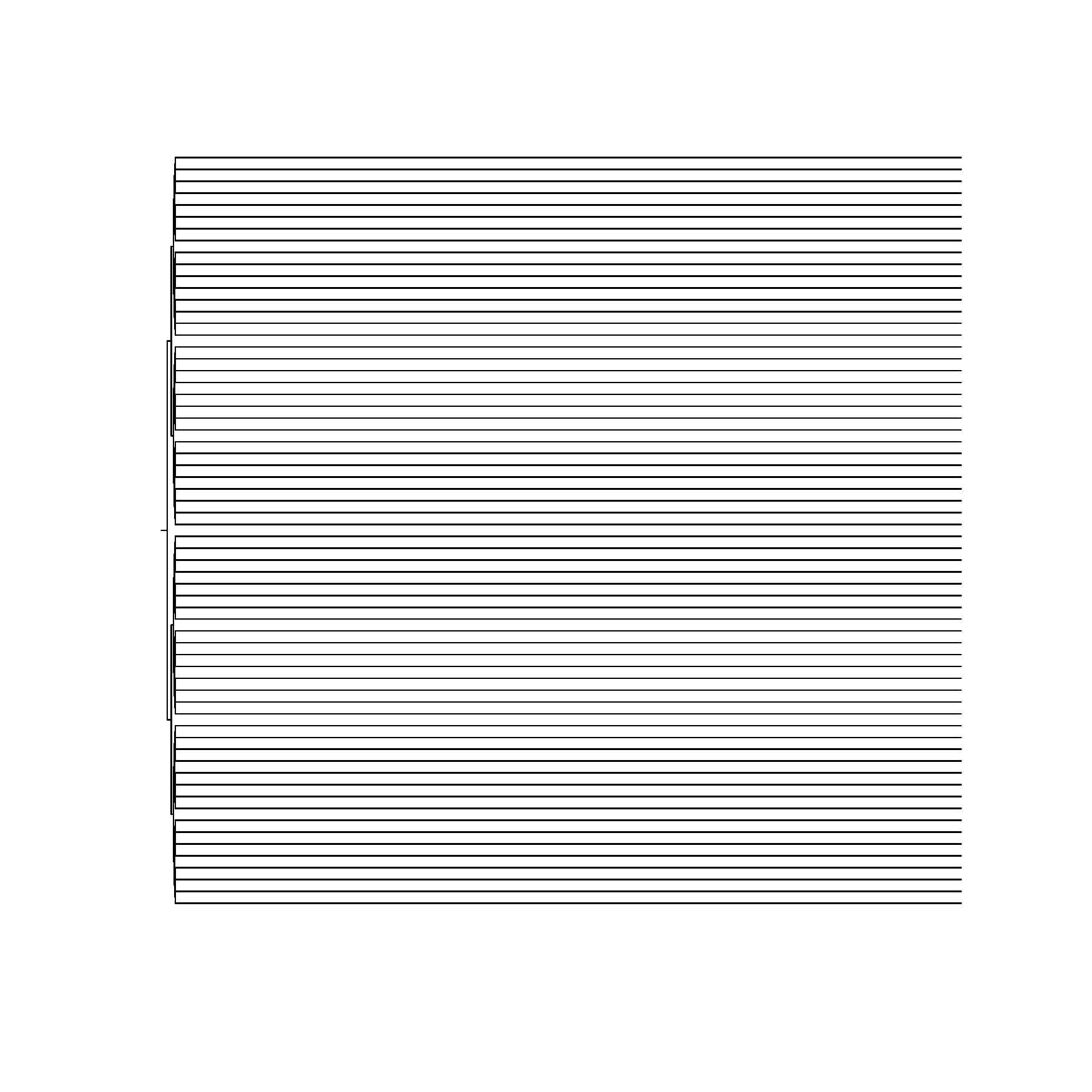} \\
\includegraphics[width=0.24\textwidth]{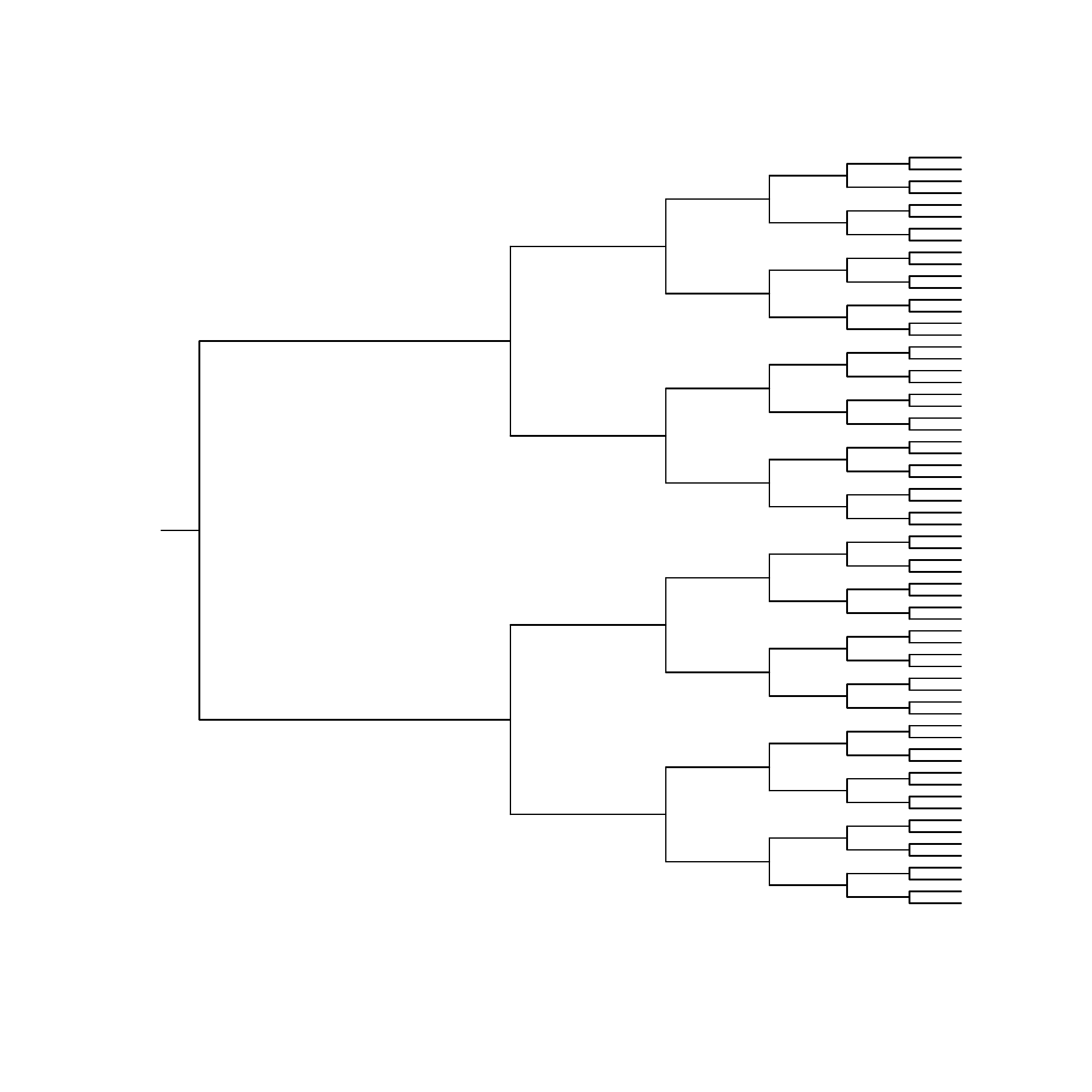} 
\includegraphics[width=0.24\textwidth]{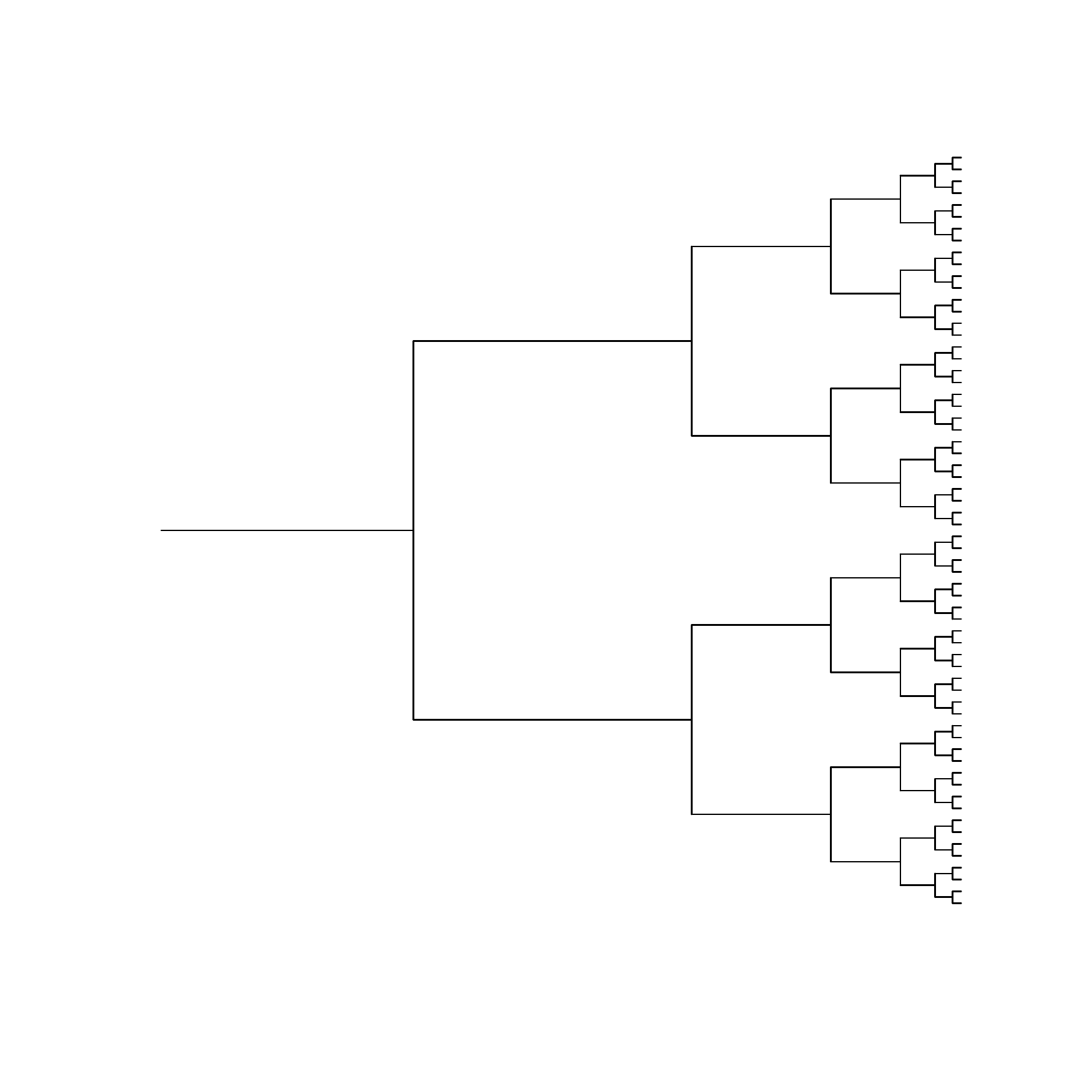}
\includegraphics[width=0.24\textwidth]{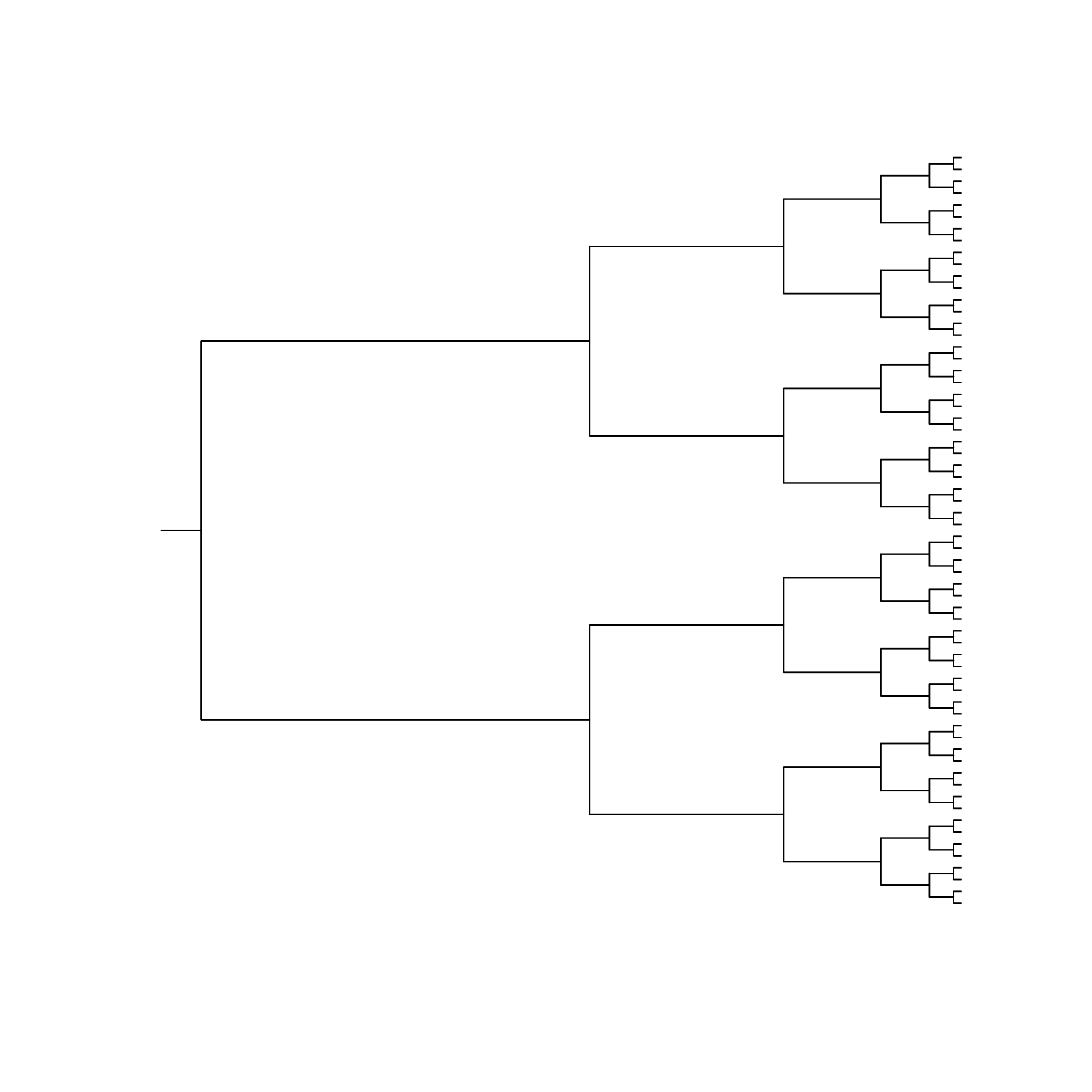}
\includegraphics[width=0.24\textwidth]{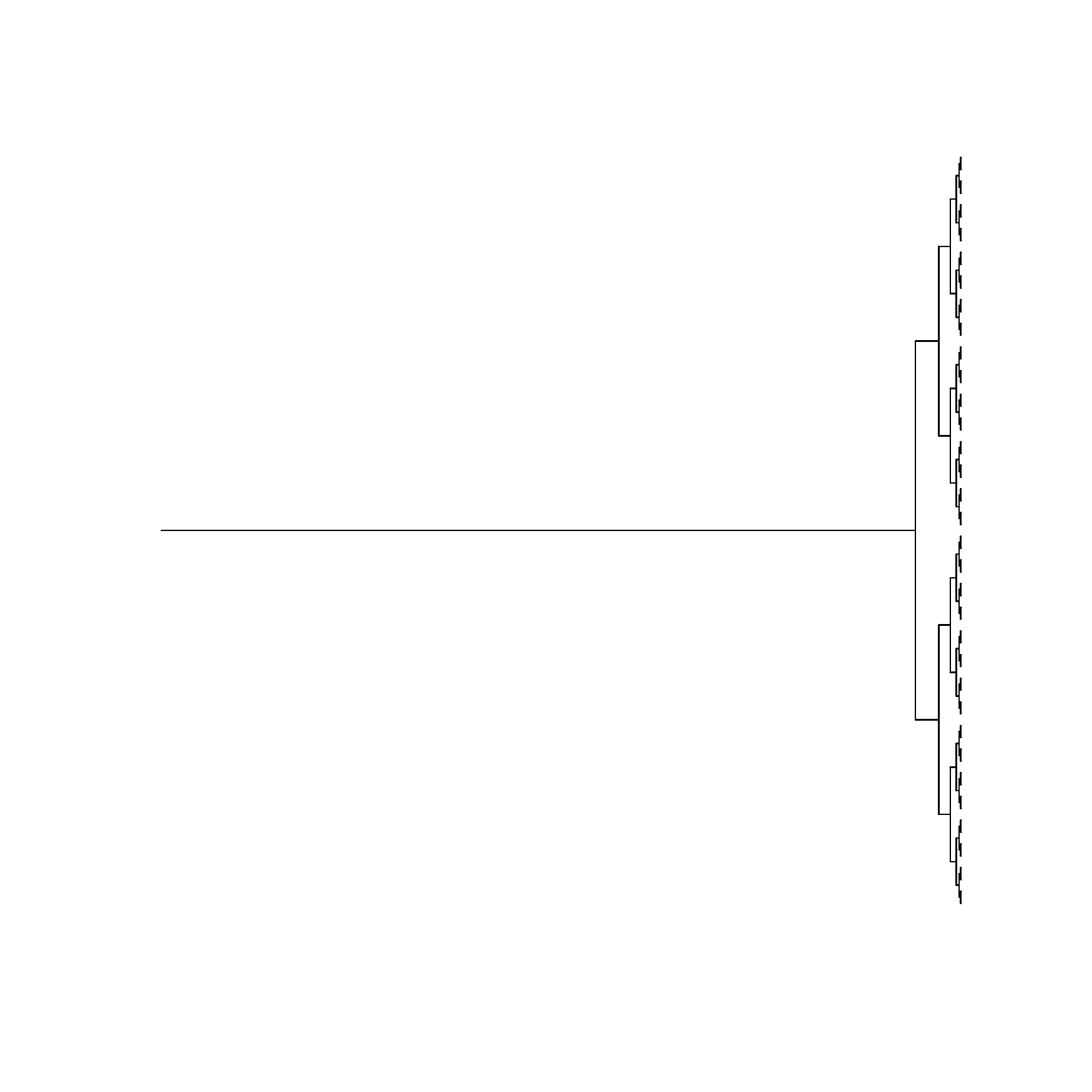}
\caption{Balanced phylogenies used in pESS for biodiversity simulations. 
Top row, left to right: branches increasing towards root in a harmonic fashion, 
branches increasing towards root in a geometric fashion,
terminal branches are $99\%$ of tree height. 
Bottom row, left to right: branches increasing towards root in a harmonic fashion, 
branches increasing towards root in a geometric fashion,
terminal branches are $1\%$ of tree height, root branch is $95\%$ of tree height.
The number of tips is $64$. 
For other types of phylogenies see Fig. \ref{figTreeTypes}. \label{figBioDivTrees}}
\end{center}
\end{figure}

In Tabs. \ref{tabBioDiv} and \ref{tabBioDivPD} I present situations where the pESS approaches produce results which are
in agreement and disagreement with phylogenetic diversity. I considered a number of different phylogenies 
(see Fig. \ref{figBioDivTrees}), with recent shallow radiations, with long tip branches, short tip branches
and Yule trees. 
Two considered types for balanced trees are geometrically or harmonically increasing or decreasing branch lengths.
In the geometric case, each level's branch is half of or twice as (decrease or increase)
the previous level's one. In the harmonic case, the branch length of the $i$--th level
(counting from the root --- decreasing or from the tips --- increasing) is $1/i$ of the tree's height.
On top of all trees I considered the BM process and the 
OU process with different parameter
values. All trees have an expected height of $\log n$. In deterministic trees (balanced and unbalanced, i.e. non--Yule)
some randomness to the topology is added by a root branch 
of length distributed exponentially with rate $1$. 
This is so that the models are more comparable --- that some variance is attached to the 
trait evolution and the OU model is allowed to approach stationarity/stasis before speciation effects begin.
For each setup $1000$ simulations were made.

The first thing that can strike us in Tabs. \ref{tabBioDiv} and \ref{tabBioDivPD} 
is that PD can be identical despite very different topologies, dependencies and
tip species numbers. For example the Yule and unbalanced trees have nearly identical
PDs for $n=16$ while the pESSs suggest that there is a difference between their 
information content. On the other hand when $n=125$ there is a large difference 
between the PDs, while not that much in the pESSs. 

If we compare the balanced short terminal tree
with $n=128$ and the $n=16$ balanced harmonic/geometric increase trees, then they have 
nearly identical PDs. Their pESSs are also similar but they explain what is going on,
in the first case, we have many very similar species in the second a few very distinct ones. 
In the latter situation, as discussed previously, the 
loss of a species means loosing a completely separate entity, in the former all species contain significant
information about all the others.  

Phylogenetic diversity's lack of explanatory power of
the dependency structure induced by the 
different topologies, is even more evident when considering relative
PDs and pESSs, i.e. PD$/n$, $n_{e}/n$, (Tab.  \ref{tabBioDivPD}). 
In the first example above (unbalanced and Yule)
the relative regression ESS seems stable (similar growth with $\alpha$) 
when comparing the small and large phylogenies
(both Yule and unbalanced). It clearly shows that there is more independence in the unbalanced tree ---
as expected there are more long terminal branches. The relative PD does not distinguish between
the small Yule and unbalanced phylogenies, and the large Yule phylogeny,
while $n_{e}^{R}/n=0.367,0.615,0.175$ for small Yule BM, small unbalanced BM and large Yule BM respectively.
The regression ESS clearly shows how the tree influences the dependency structure of the tips. 
Unfortunately the mutual information and mean ones do not describe these dependencies so clearly.
However, in their case this is explainable --- the mean measures only information on the expected value and 
the MI one needs further refinement with respect to the $e(\cdot)$ transformation.
\citet[][p. 208]{MVelWCorKMagForAMoo2011} comment that distance based metrics (e.g. PD) make it easy to detect phylogenetic 
clustering but not overdispersion on balanced trees, while the opposite is true for unbalanced
trees with accelerating diversification. Given an appropriate trait, 
the pESS should not have such topology dependent problems as uses both phylogenetic
and ``evolution on a lineage'' information. If the species are clustered, 
then this should be reflected in more dependencies between the observations and lower $n_{e}$.
On the other hand overdispersion should lead to more independence and hence higher $n_{e}$.

The general pattern from Tabs. \ref{tabBioDiv} and \ref{tabBioDivPD} is that if there is a lot of independence, 
then PD will
be large. But as said, the sum of branch lengths does not capture everything. For example, I look in more detail at the 
balanced long terminal, harmonic and geometric increase topologies. The PD measures (absolute and relative) do
not distinguish between these different situations. 
However, in Fig. \ref{figBioDivTrees} we can see that there is a substantial difference between the
long terminal one and the harmonic and geometric increases. The long terminal sample should essentially
be independent, while the other two should exhibit dependencies. The $n_{e}^{R}$ describes such a pattern 
perfectly. On the long terminal tree all processes generate a nearly independent sample with the rESS measure.
For the other two the process has to evolve quickly to loose ancestral dependencies. But on the other hand,
by the PD measure the long terminal branch tree carries less independence (diversity) than the harmonic
and geometric increase trees. Furthermore it is interesting to notice that the growth of the relative pESSs
with $\alpha$ is similar for all pESS definitions. The geometric increase has larger pESSs due to the 
longer terminal branches.

\begin{longtable}{cccccc}
 \caption{
 Comparison of phylogenetic diversity with the proposed pESS definitions
 for different evolutionary models and topologies.  The values are means from a $1000$ simulations.
 The value of $1$ for the regression ESS indicates that the calculated value was below $1$ 
 and hence the rounding up.\label{tabBioDiv} } \endlastfoot
\renewcommand*{\thefootnote}{\textit{\alph{footnote}}}
\setcounter{footnote}{0}
Model & $n$\footnote{number of tips} & $\E{\mathrm{PD}}$ \footnote{expected phylogenetic diversity} 
& $\E{n_{e}^{\mathrm{MI}}}$ \footnote{expected mutual information effective sample size}
& $\E{n_{e}^{\mathrm{E}}}$ \footnote{expected mean effective sample size}
& $\E{n_{e}^{\mathrm{R}}}$ \footnote{expected regression effective sample size}\\
\hline
\multicolumn{6}{l}{Yule} \\
 BM\footnote{Brownian motion  $\alpha=0$} & $15$ & $15.102$ & $7.119$ & $3.938$ & $5.505$ \\
 slow OU\footnote{Ornstein--Uhlenbeck $\alpha=0.25$} & $15$ & $15.102$ & $7.744$ & $4.661$ & $6.881$ \\
 medium OU\footnote{Ornstein--Uhlenbeck $\alpha=0.5$} & $15$ &  $15.102$ & $8.404$ & $5.562$ & $8.179$ \\
 fast OU\footnote{Ornstein--Uhlenbeck $\alpha=1$} & $15$ &  $15.102$ & $9.611$  &$7.507$ &$10.126$\\
\multicolumn{6}{l}{unbalanced} \\
 BM & $15$ & $15.461$ & $8.335$ & $4.55$ & $9.228$\\
slow OU & $15$ &  $15.461$ & $9.045$ & $5.208$ & $10.217$\\
 medium OU & $15$ &  $15.461$ & $9.808$ & $6.006$ & $11.117$ \\
 fast OU & $15$ &  $15.461$ & $11.292$ & $7.885$ &  $12.506$  \\
\multicolumn{6}{l}{balanced} \\
 BM & $16$ & $8.584$ & $6.735$ & $2.824$ & $2.793$ \\
 slow OU & $16$ & $8.584$ & $7.146$ &$3.172$ & $3.6$  \\
 medium OU & $16$ & $8.584$ & $7.598$ & $3.608$ & $4.532$ \\
 fast OU & $16$ & $8.584$ & $8.568$ & $4.698$ & $6.544$\\
\multicolumn{6}{l}{balanced short terminal}\\
 BM & $16$ & $10.829$ & $5.892$ &  $2.689$ & $1.294$ \\
 slow OU & $16$ & $10.829$ & $6.2$ & $3.159$ & $1.539$ \\
 medium OU & $16$ & $10.829$ & $6.497$ & $3.762$ & $1.853$  \\
 fast OU & $16$ & $10.829$ & $7.003$ & $5.078$ & $2.57$
\\
\multicolumn{6}{l}{balanced long terminal}  \\
 BM & $16$ & $44.023$ & $15.991$ & $15.238$ & $15.994$\\
 slow OU & $16$& $44.023$ & $15.998$ &  $15.637$ &  $15.999$ \\
 medium OU & $16$ & $44.023$ & $16$ & $15.852$ &  $16$\\ 
 fast OU & $16$ & $44.023$ & $16$ &  $15.982$ & $16$\\
\multicolumn{6}{l}{balanced harmonic decrease} \\
 BM & $16$ & $19.908$ & $7.292$ & $3.053$ & $4.094$\\
 slow OU & $16$ & $19.908$ & $8.578$ & $4.13$ & $6.839$\\
 medium OU & $16$ & $19.908$ & $10.107$ & $5.694$ & $9.634$ \\
 fast OU & $16$ &  $19.908$ & $12.896$ & $9.158$ & $13.333$ \\
\multicolumn{6}{l}{balanced harmonic increase} \\
 BM & $16$ & $39.702$ & $11.242$ & $6.25$ &  $11.196$ \\
 slow OU & $16$ & $39.702$ & $13.921$ & $9.187$ &  $14.418$ \\
 medium OU & $16$ & $39.702$ & $15.454$ & $12.325$ & $15.622$ \\
fast OU & $16$ & $39.702$ & $15.986$ & $15.379$ &  $15.987$\\
\multicolumn{6}{l}{balanced geometric decrease}  \\
 BM & $16$ & $15.171$ & $6.735$ & $2.824$ & $2.794$\\
slow OU & $16$ & $15.171$ & $7.54$ & $3.549$ & $4.41$\\
 medium OU & $16$ & $15.171$ & $8.445$ & $4.55$ & $6.294$\\
 fast OU & $16$ & $15.171$ & $10.247$ & $6.827$ &  $9.644$\\
\multicolumn{6}{l}{balanced geometric increase} \\
 BM & $16$ & $38.661$ & $12.072$ & $7.5$ & $12.463$ \\
 slow OU & $16$ & $38.661$ & $14.335$ & $10.198$ & $14.75$ \\
 medium OU & $16$ & $38.661$ & $15.544$ & $12.836$ & $15.684$\\
 fast OU & $16$ & $38.661$ & $15.983$ & $15.418$ & $15.988$  \\
\multicolumn{6}{l}{balanced long root branch}\\
 BM & $16$ & $3.598$ & $7.045$ & $2.824$ & $1.914$\\
 slow OU & $16$ & $3.598$ & $7.116$ & $2.868$ & $2.025$\\
 medium OU & $16$ & $3.598$ & $7.189$ & $2.914$ & $2.141$\\
 fast OU & $16$ & $3.598$ & $7.34$ & $3.011$ & $2.382$\\
\multicolumn{6}{l}{Yule} \\
 BM & $125$ & $124.839$ & $27.745$ &    $6.636$ &  $21.864$ \\
 slow OU & $125$ & $124.839$ &  $30.748$ &  $12.229$ &  $40.549$ \\
 medium OU & $125$ & $124.839$ & $33.829$ &   $22.896$ &  $58.002$ \\
 fast OU & $125$ & $124.839$ &  $39.212$ &  $51.921$&  $80.243$ \\
\multicolumn{6}{l}{unbalanced} \\
 BM & $125$ & $244.124$ & $32$ &  $11.905$ &    $66.687$ \\
slow OU & $125$ & $244.124$ &$36.518$ &  $18.825$ & $85.375$ \\
 medium OU & $125$ & $244.124$ & $41.85$ &    $30.96$ &    $98.777$ \\
 fast OU & $125$ & $244.124$ & $52.44$ &   $64.035$ &   $111.456$ \\
\multicolumn{6}{l}{balanced} \\
 BM & $128$ & $28.19$ & $24.599$ &    $2.977$ &    $2.796$ \\
 slow OU & $128$ &  $28.19$  & $25.746$ &    $4.073$ &     $5.008$  \\
 medium OU & $128$ &  $28.19$ & $26.798$ &  $5.765$ &   $7.797$ \\
 fast OU & $128$ &  $28.19$ & $28.526$ &    $10.189$  &  $14.288$  \\
\multicolumn{6}{l}{balanced short terminal}\\
 BM & $128$ & $36.485$ &   $24.776$ &  $2.983$ & $3.232$  \\
 slow OU & $128$ & $36.485$ & $26.3$ &  $4.478$ &  $6.846$\\
 medium OU & $128$ &  $36.485$  &$27.662$ &   $6.886$ &    $11.571$  \\
 fast OU & $128$ & $36.485$ & $29.87$ & $12.816$ & $21.3$      \\
\multicolumn{6}{l}{balanced long terminal}  \\
 BM & $128$ & $615.155$ &  $124.514$ &   $89.930$ &    $127.722$\\
 slow OU & $128$& $615.155$  &  $127.759$ &  $116.316$ &   $127.980$  \\
 medium OU & $128$ & $615.155$ & $127.993$ &   $125.927$&  $127.999$\\ 
 fast OU & $128$ & $615.155$ & $128$ & $127.967$ & $128$ \\
\multicolumn{6}{l}{balanced harmonic decrease} \\
 BM & $128$ & $178.258$  & $27.567$ &  $3.745$ &  $12.813$ \\
 slow OU & $128$ & $178.258$  & $35.106$ &  $13.792$ &   $50.429$ \\
 medium OU & $128$ & $178.258$ & $44.221$ &   $36.794$ & $82.256$  \\
 fast OU & $128$ & $178.258$ & $67.237$ & $78.888$ & $113.169$   \\
\multicolumn{6}{l}{balanced harmonic increase} \\
 BM & $128$ & $683.621$ & $39.878$ &  $14.424$ &  $73.063$  \\
 slow OU & $128$ &  $683.621$ & $101.417$ &   $81.108$ &    $124.902$ \\
 medium OU & $128$ &   $683.621$  & $127.303$ &   $123.606$ &    $127.94$  \\
fast OU & $128$ & $683.621$ & $128$ & $127.939$ & $128$  \\
\multicolumn{6}{l}{balanced geometric decrease}  \\
 BM & $128$ & $54.926$ & $24.599$ &  $2.977$ &  $2.796$ \\
slow OU & $128$ & $54.926$ & $26.783$ &  $5.735$ & $7.878$ \\
 medium OU & $128$ & $54.926$&  $28.502$ &   $10.116$ &  $14.189$ \\
 fast OU & $128$ & $54.926$ & $31.144$ & $19.056$ & $26.283$    \\
\multicolumn{6}{l}{balanced geometric increase} \\
 BM & $128$ & $658.375$  & $48.870$ &  $36.286$&  $93.302$ \\
 slow OU & $128$ & $658.375$  & $106.934$ &    $95.213$ &  $125.693$ \\
 medium OU & $128$ & $658.375$  & $127.340$ &    $123.876$ &   $127.943$ \\
 fast OU & $128$ & $658.375$ & $128$ & $127.939$ & $128$      \\
\multicolumn{6}{l}{balanced long root branch}\\
 BM & $128$ & $7.259$ & $24.599$ & $2.977$ & $1.81$ \\
 slow OU & $128$ & $7.637$ & $24.768$ & $3.072$ & $2.015$ \\
 medium OU & $128$ & $7.637$ & $24.902$ & $3.175$ & $2.233$  \\
 fast OU & $128$ & $7.637$  & $25.171$ & $3.404$ & $2.712$ 
\end{longtable}

\begin{longtable}{cccccc}
\caption{
Comparison of relative phylogenetic diversity with the proposed relative pESSs. The values are means from the same $1000$ simulations from Tab. \ref{tabBioDiv}. 
\label{tabBioDivPD}}
\endlastfoot
\renewcommand*{\thefootnote}{\textit{\alph{footnote}}}
\setcounter{footnote}{0}
Model & $n$ & $\E{\mathrm{PD}/n}$ & $\E{n_{e}^{\mathrm{MI}}/n}$ & $\E{n_{e}^{\mathrm{E}}/n}$ & $\E{n_{e}^{\mathrm{R}}/n}$ \\
\hline
\multicolumn{6}{l}{Yule} \\
 BM & $15$ & \pgfmathparse{15.102/15} \MyNum{\pgfmathresult} & \pgfmathparse{7.119/15} \MyNum{\pgfmathresult}  & \pgfmathparse{3.938/15} \MyNum{\pgfmathresult} & \pgfmathparse{5.505/15} \MyNum{\pgfmathresult} \\
 slow OU & $15$ & \pgfmathparse{15.102/15} \MyNum{\pgfmathresult} & \pgfmathparse{7.744/15} \MyNum{\pgfmathresult} & \pgfmathparse{4.661/15} \MyNum{\pgfmathresult} & \pgfmathparse{6.881/15} \MyNum{\pgfmathresult} \\
 medium OU & $15$ &  \pgfmathparse{15.102/15} \MyNum{\pgfmathresult} & \pgfmathparse{8.404/15} \MyNum{\pgfmathresult} & \pgfmathparse{5.562/15} \MyNum{\pgfmathresult} & \pgfmathparse{8.719/15} \MyNum{\pgfmathresult} \\
 fast OU & $15$ &  \pgfmathparse{15.102/15} \MyNum{\pgfmathresult} & \pgfmathparse{9.611/15} \MyNum{\pgfmathresult}  & \pgfmathparse{7.507/15} \MyNum{\pgfmathresult} & \pgfmathparse{10.126/15} \MyNum{\pgfmathresult}\\
\multicolumn{6}{l}{unbalanced} \\
 BM & $15$ & \pgfmathparse{15.461/15} \MyNum{\pgfmathresult} & \pgfmathparse{8.335/15} \MyNum{\pgfmathresult} & \pgfmathparse{4.55/15} \MyNum{\pgfmathresult} & \pgfmathparse{9.228/15} \MyNum{\pgfmathresult}\\
slow OU & $15$ &  \pgfmathparse{15.461/15} \MyNum{\pgfmathresult} & \pgfmathparse{9.045/15} \MyNum{\pgfmathresult} & \pgfmathparse{5.208/15} \MyNum{\pgfmathresult} & \pgfmathparse{10.217/15} \MyNum{\pgfmathresult}\\
 medium OU & $15$ &  \pgfmathparse{15.461/15} \MyNum{\pgfmathresult} & \pgfmathparse{9.808/15} \MyNum{\pgfmathresult} & \pgfmathparse{6.006/15} \MyNum{\pgfmathresult} & \pgfmathparse{11.117/15} \MyNum{\pgfmathresult} \\
 fast OU & $15$ &  \pgfmathparse{15.461/15} \MyNum{\pgfmathresult} & \pgfmathparse{11.292/15} \MyNum{\pgfmathresult} & \pgfmathparse{7.885/15} \MyNum{\pgfmathresult} &  \pgfmathparse{12.506/15} \MyNum{\pgfmathresult}  \\
\multicolumn{6}{l}{balanced} \\
 BM & $16$ & \pgfmathparse{8.584/16} \MyNum{\pgfmathresult} & \pgfmathparse{6.735/16} \MyNum{\pgfmathresult} & \pgfmathparse{2.824/16} \MyNum{\pgfmathresult} & \pgfmathparse{2.793/16} \MyNum{\pgfmathresult} \\
 slow OU & $16$ & \pgfmathparse{8.584/16} \MyNum{\pgfmathresult} & \pgfmathparse{7.146/16} \MyNum{\pgfmathresult} & \pgfmathparse{3.172/16} \MyNum{\pgfmathresult} & \pgfmathparse{3.6/16} \MyNum{\pgfmathresult}  \\
 medium OU & $16$ & \pgfmathparse{8.584/16} \MyNum{\pgfmathresult} & \pgfmathparse{7.598/16} \MyNum{\pgfmathresult} & \pgfmathparse{3.608/16} \MyNum{\pgfmathresult} & \pgfmathparse{4.532/16} \MyNum{\pgfmathresult} \\
 fast OU & $16$ & \pgfmathparse{8.584/16} \MyNum{\pgfmathresult} & \pgfmathparse{8.568/16} \MyNum{\pgfmathresult} & \pgfmathparse{4.698/16} \MyNum{\pgfmathresult} & \pgfmathparse{6.544/16} \MyNum{\pgfmathresult}\\
\multicolumn{6}{l}{balanced short terminal}\\
 BM & $16$ & \pgfmathparse{10.829/16} \MyNum{\pgfmathresult} & \pgfmathparse{5.892/16} \MyNum{\pgfmathresult} & \pgfmathparse{2.689/16} \MyNum{\pgfmathresult} & \pgfmathparse{1.294/16} \MyNum{\pgfmathresult} \\
 slow OU & $16$ & \pgfmathparse{10.829/16} \MyNum{\pgfmathresult} & \pgfmathparse{6.2/16} \MyNum{\pgfmathresult} & \pgfmathparse{3.159/16} \MyNum{\pgfmathresult} & \pgfmathparse{1.539/16} \MyNum{\pgfmathresult} \\
 medium OU & $16$ & \pgfmathparse{10.829/16} \MyNum{\pgfmathresult} & \pgfmathparse{6.497/16} \MyNum{\pgfmathresult} & \pgfmathparse{3.762/16} \MyNum{\pgfmathresult} & \pgfmathparse{1.853/16} \MyNum{\pgfmathresult}  \\
 fast OU & $16$ & \pgfmathparse{10.829/16} \MyNum{\pgfmathresult} & \pgfmathparse{7.003/16} \MyNum{\pgfmathresult} & \pgfmathparse{5.078/16} \MyNum{\pgfmathresult} & \pgfmathparse{2.57/16} \MyNum{\pgfmathresult}
\\
\multicolumn{6}{l}{balanced long terminal}  \\
 BM & $16$ & \pgfmathparse{44.023/16} \MyNum{\pgfmathresult} & \pgfmathparse{15.991/16} \MyNum{\pgfmathresult} & \pgfmathparse{15.238/16} \MyNum{\pgfmathresult} & \pgfmathparse{15.994/16} \MyNum{\pgfmathresult}\\
 slow OU & $16$ & \pgfmathparse{44.023/16} \MyNum{\pgfmathresult} & \pgfmathparse{15.998/16} \MyNum{\pgfmathresult} &  \pgfmathparse{15.637/16} \MyNum{\pgfmathresult} &  \pgfmathparse{15.999/16} \MyNum{\pgfmathresult} \\
 medium OU & $16$ & \pgfmathparse{44.023/16} \MyNum{\pgfmathresult} & \pgfmathparse{16/16} \MyNum{\pgfmathresult} & \pgfmathparse{15.852/16} \MyNum{\pgfmathresult} &  \pgfmathparse{16/16} \MyNum{\pgfmathresult}\\ 
 fast OU & $16$ &  \pgfmathparse{44.023/16} \MyNum{\pgfmathresult}  & \pgfmathparse{16/16} \MyNum{\pgfmathresult} &  \pgfmathparse{15.982/16} \MyNum{\pgfmathresult} & \pgfmathparse{16/16} \MyNum{\pgfmathresult} \\
\multicolumn{6}{l}{balanced harmonic decrease} \\
 BM & $16$ & \pgfmathparse{19.908/16} \MyNum{\pgfmathresult} & \pgfmathparse{7.292/16} \MyNum{\pgfmathresult} & \pgfmathparse{3.053/16} \MyNum{\pgfmathresult} & \pgfmathparse{4.094/16} \MyNum{\pgfmathresult}\\
 slow OU & $16$ & \pgfmathparse{19.908/16} \MyNum{\pgfmathresult} & \pgfmathparse{8.578/16} \MyNum{\pgfmathresult} & \pgfmathparse{4.13/16} \MyNum{\pgfmathresult} & \pgfmathparse{6.839/16} \MyNum{\pgfmathresult}\\
 medium OU & $16$ & \pgfmathparse{19.908/16} \MyNum{\pgfmathresult} & \pgfmathparse{10.107/16} \MyNum{\pgfmathresult} & \pgfmathparse{5.694/16} \MyNum{\pgfmathresult} & \pgfmathparse{9.634/16} \MyNum{\pgfmathresult} \\
 fast OU & $16$ &  \pgfmathparse{19.908/16} \MyNum{\pgfmathresult} & \pgfmathparse{12.896/16} \MyNum{\pgfmathresult} & \pgfmathparse{9.158/16} \MyNum{\pgfmathresult} & \pgfmathparse{13.333/16} \MyNum{\pgfmathresult} \\
\multicolumn{6}{l}{balanced harmonic increase} \\
 BM & $16$ & \pgfmathparse{39.702/16} \MyNum{\pgfmathresult} & \pgfmathparse{11.242/16} \MyNum{\pgfmathresult} & \pgfmathparse{6.25/16} \MyNum{\pgfmathresult} &  \pgfmathparse{11.196/16} \MyNum{\pgfmathresult} \\
 slow OU & $16$ & \pgfmathparse{39.702/16} \MyNum{\pgfmathresult} & \pgfmathparse{13.921/16} \MyNum{\pgfmathresult} & \pgfmathparse{9.187/16} \MyNum{\pgfmathresult} &  \pgfmathparse{14.418/16} \MyNum{\pgfmathresult} \\
 medium OU & $16$ & \pgfmathparse{39.702/16} \MyNum{\pgfmathresult} & \pgfmathparse{15.454/16} \MyNum{\pgfmathresult} & \pgfmathparse{12.325/16} \MyNum{\pgfmathresult} & \pgfmathparse{15.622/16} \MyNum{\pgfmathresult} \\
fast OU & $16$ & \pgfmathparse{39.702/16} \MyNum{\pgfmathresult} & \pgfmathparse{15.986/16} \MyNum{\pgfmathresult} & \pgfmathparse{15.379/16} \MyNum{\pgfmathresult} &  \pgfmathparse{15.987/16} \MyNum{\pgfmathresult}\\
\multicolumn{6}{l}{balanced geometric decrease}  \\
 BM & $16$ & \pgfmathparse{15.171/16} \MyNum{\pgfmathresult} & \pgfmathparse{6.735/16} \MyNum{\pgfmathresult} & \pgfmathparse{2.824/16} \MyNum{\pgfmathresult} & \pgfmathparse{2.824/16} \MyNum{\pgfmathresult}\\
slow OU & $16$ & \pgfmathparse{15.171/16} \MyNum{\pgfmathresult} & \pgfmathparse{7.54/16} \MyNum{\pgfmathresult} & \pgfmathparse{3.549/16} \MyNum{\pgfmathresult} & \pgfmathparse{4.41/16} \MyNum{\pgfmathresult}\\
 medium OU & $16$ & \pgfmathparse{15.171/16} \MyNum{\pgfmathresult} & \pgfmathparse{8.445/16} \MyNum{\pgfmathresult} & \pgfmathparse{4.55/16} \MyNum{\pgfmathresult} & \pgfmathparse{6.294/16} \MyNum{\pgfmathresult}\\
 fast OU & $16$ & \pgfmathparse{15.171/16} \MyNum{\pgfmathresult} & \pgfmathparse{10.247/16} \MyNum{\pgfmathresult} & \pgfmathparse{6.827/16} \MyNum{\pgfmathresult} &  \pgfmathparse{9.644/16} \MyNum{\pgfmathresult}\\
\multicolumn{6}{l}{balanced geometric increase} \\
 BM & $16$ & \pgfmathparse{38.661/16} \MyNum{\pgfmathresult} & \pgfmathparse{12.072/16} \MyNum{\pgfmathresult} & \pgfmathparse{7.5/16} \MyNum{\pgfmathresult} & \pgfmathparse{12.463/16} \MyNum{\pgfmathresult} \\
 slow OU & $16$ & \pgfmathparse{38.661/16} \MyNum{\pgfmathresult} & \pgfmathparse{14.335/16} \MyNum{\pgfmathresult} & \pgfmathparse{10.198/16} \MyNum{\pgfmathresult} & \pgfmathparse{14.75/16} \MyNum{\pgfmathresult} \\
 medium OU & $16$ & \pgfmathparse{38.661/16} \MyNum{\pgfmathresult} & \pgfmathparse{15.544/16} \MyNum{\pgfmathresult} & \pgfmathparse{12.836/16} \MyNum{\pgfmathresult} & \pgfmathparse{15.684/16} \MyNum{\pgfmathresult}\\
 fast OU & $16$ & \pgfmathparse{38.661/16} \MyNum{\pgfmathresult} & \pgfmathparse{15.983/16} \MyNum{\pgfmathresult} & \pgfmathparse{15.418/16} \MyNum{\pgfmathresult} & \pgfmathparse{15.988/16} \MyNum{\pgfmathresult}  \\
\multicolumn{6}{l}{balanced long root branch}\\
 BM & $16$ & \pgfmathparse{3.598/16} \MyNum{\pgfmathresult} & \pgfmathparse{7.045/16} \MyNum{\pgfmathresult} & \pgfmathparse{2.824/16} \MyNum{\pgfmathresult} & \pgfmathparse{1.914/16} \MyNum{\pgfmathresult}\\
 slow OU & $16$ & \pgfmathparse{3.598/16} \MyNum{\pgfmathresult} & \pgfmathparse{7.116/16} \MyNum{\pgfmathresult} & \pgfmathparse{2.868/16} \MyNum{\pgfmathresult} & \pgfmathparse{2.025/16} \MyNum{\pgfmathresult}\\
 medium OU & $16$ & \pgfmathparse{3.598/16} \MyNum{\pgfmathresult} & \pgfmathparse{7.189/16} \MyNum{\pgfmathresult} & \pgfmathparse{2.914/16} \MyNum{\pgfmathresult} & \pgfmathparse{2.141/16} \MyNum{\pgfmathresult}\\
 fast OU & $16$ & \pgfmathparse{3.598/16} \MyNum{\pgfmathresult} & \pgfmathparse{7.34/16} \MyNum{\pgfmathresult} & \pgfmathparse{3.011/16} \MyNum{\pgfmathresult} & \pgfmathparse{2.382/16} \MyNum{\pgfmathresult}\\
\multicolumn{6}{l}{Yule} \\
 BM & $125$ & \pgfmathparse{124.839/125} \MyNum{\pgfmathresult} & \pgfmathparse{27.745/125} \MyNum{\pgfmathresult} &    \pgfmathparse{6.636/125} \MyNum{\pgfmathresult} &  \pgfmathparse{21.864/125} \MyNum{\pgfmathresult} \\
 slow OU & $125$ & \pgfmathparse{124.839/125} \MyNum{\pgfmathresult} &  \pgfmathparse{30.748/125} \MyNum{\pgfmathresult} &  \pgfmathparse{12.229/125} \MyNum{\pgfmathresult} &  \pgfmathparse{40.549/125} \MyNum{\pgfmathresult} \\
 medium OU & $125$ & \pgfmathparse{124.839/125} \MyNum{\pgfmathresult} & \pgfmathparse{33.829/125} \MyNum{\pgfmathresult} & \pgfmathparse{22.896/125} \MyNum{\pgfmathresult} &  \pgfmathparse{58.002/125} \MyNum{\pgfmathresult} \\
 fast OU & $125$ & \pgfmathparse{124.839/125} \MyNum{\pgfmathresult} &  \pgfmathparse{39.212/125} \MyNum{\pgfmathresult} & \pgfmathparse{51.921/125} \MyNum{\pgfmathresult}&  \pgfmathparse{80.243/125} \MyNum{\pgfmathresult} \\
\multicolumn{6}{l}{unbalanced} \\
 BM & $125$ & \pgfmathparse{244.124/125} \MyNum{\pgfmathresult} & \pgfmathparse{32/125} \MyNum{\pgfmathresult} &  \pgfmathparse{11.905/125} \MyNum{\pgfmathresult} & \pgfmathparse{66.687/125} \MyNum{\pgfmathresult}\\
slow OU & $125$ & \pgfmathparse{244.124/125} \MyNum{\pgfmathresult} & \pgfmathparse{36.518/125} \MyNum{\pgfmathresult} & \pgfmathparse{18.825/125} \MyNum{\pgfmathresult} & \pgfmathparse{85.375/125} \MyNum{\pgfmathresult} \\
 medium OU & $125$ & \pgfmathparse{244.124/125} \MyNum{\pgfmathresult} & \pgfmathparse{41.85/125} \MyNum{\pgfmathresult} & \pgfmathparse{30.96/125} \MyNum{\pgfmathresult} & \pgfmathparse{98.777/125} \MyNum{\pgfmathresult} \\
 fast OU & $125$ & \pgfmathparse{244.124/125} \MyNum{\pgfmathresult} & \pgfmathparse{52.44/125} \MyNum{\pgfmathresult} & \pgfmathparse{64.035/125} \MyNum{\pgfmathresult} &   \pgfmathparse{111.456/125} \MyNum{\pgfmathresult} \\
\multicolumn{6}{l}{balanced} \\
 BM & $128$ & \pgfmathparse{ 28.19/128} \MyNum{\pgfmathresult} & \pgfmathparse{ 24.599/128} \MyNum{\pgfmathresult} & \pgfmathparse{2.977/128} \MyNum{\pgfmathresult} &    \pgfmathparse{2.976/128} \MyNum{\pgfmathresult} \\
 slow OU & $128$ &  \pgfmathparse{28.19/128} \MyNum{\pgfmathresult}  & \pgfmathparse{25.746/128} \MyNum{\pgfmathresult} &    \pgfmathparse{4.073/128} \MyNum{\pgfmathresult} &     \pgfmathparse{5.008/128} \MyNum{\pgfmathresult}  \\
 medium OU & $128$ &  \pgfmathparse{28.19/128} \MyNum{\pgfmathresult} & \pgfmathparse{26.798/128} \MyNum{\pgfmathresult} & \pgfmathparse{5.765/128} \MyNum{\pgfmathresult} &   \pgfmathparse{7.797/128} \MyNum{\pgfmathresult} \\
 fast OU & $128$ &  \pgfmathparse{28.19/128} \MyNum{\pgfmathresult} & \pgfmathparse{28.526/128} \MyNum{\pgfmathresult} & \pgfmathparse{10.189/128} \MyNum{\pgfmathresult}  &  \pgfmathparse{14.288/128} \MyNum{\pgfmathresult}  \\
\multicolumn{6}{l}{balanced short terminal}\\
 BM & $128$ & \pgfmathparse{ 36.485/128} \MyNum{\pgfmathresult} &   \pgfmathparse{24.776/128} \MyNum{\pgfmathresult} &  \pgfmathparse{2.983/128} \MyNum{\pgfmathresult} & \pgfmathparse{ 3.232/128} \MyNum{\pgfmathresult}  \\
 slow OU & $128$ & \pgfmathparse{ 36.485/128} \MyNum{\pgfmathresult} & \pgfmathparse{ 26.3/128} \MyNum{\pgfmathresult} &  \pgfmathparse{4.478/128} \MyNum{\pgfmathresult} &  \pgfmathparse{6.846/128} \MyNum{\pgfmathresult}\\
 medium OU & $128$ &  \pgfmathparse{36.485/128} \MyNum{\pgfmathresult}  & \pgfmathparse{27.662/128} \MyNum{\pgfmathresult} &   \pgfmathparse{6.886/128} \MyNum{\pgfmathresult} &    \pgfmathparse{11.571/128} \MyNum{\pgfmathresult}  \\
 fast OU & $128$ & \pgfmathparse{ 36.485/128} \MyNum{\pgfmathresult} & \pgfmathparse{ 29.87/128} \MyNum{\pgfmathresult} & \pgfmathparse{ 12.816/128} \MyNum{\pgfmathresult} & \pgfmathparse{ 21.3/128} \MyNum{\pgfmathresult}      \\
\multicolumn{6}{l}{balanced long terminal}  \\
 BM & $128$ & \pgfmathparse{615.155/128} \MyNum{\pgfmathresult} &  \pgfmathparse{124.514/128} \MyNum{\pgfmathresult} &   \pgfmathparse{89.930/128} \MyNum{\pgfmathresult} &    \pgfmathparse{127.722/128} \MyNum{\pgfmathresult}\\
 slow OU & $128$ & \pgfmathparse{615.155/128} \MyNum{\pgfmathresult}  &  \pgfmathparse{127.759/128} \MyNum{\pgfmathresult} &  \pgfmathparse{116.316/128} \MyNum{\pgfmathresult} &   \pgfmathparse{127.980/128} \MyNum{\pgfmathresult}  \\
 medium OU & $128$ & \pgfmathparse{615.155/128} \MyNum{\pgfmathresult} & \pgfmathparse{127.993/128} \MyNum{\pgfmathresult} &  \pgfmathparse{125.927/128} \MyNum{\pgfmathresult}&  \pgfmathparse{127.999/128} \MyNum{\pgfmathresult}\\ 
 fast OU & $128$ & \pgfmathparse{615.155/128} \MyNum{\pgfmathresult} & \pgfmathparse{128/128} \MyNum{\pgfmathresult} & \pgfmathparse{ 127.967/128} \MyNum{\pgfmathresult} & \pgfmathparse{128/128} \MyNum{\pgfmathresult} \\
\multicolumn{6}{l}{balanced harmonic decrease} \\
 BM & $128$ & \pgfmathparse{178.258/128} \MyNum{\pgfmathresult}  & \pgfmathparse{27.567/128} \MyNum{\pgfmathresult} &  \pgfmathparse{3.745/128} \MyNum{\pgfmathresult} &  \pgfmathparse{12.813/128} \MyNum{\pgfmathresult} \\
 slow OU & $128$ & \pgfmathparse{178.258/128} \MyNum{\pgfmathresult}  & \pgfmathparse{35.106/128} \MyNum{\pgfmathresult} &  \pgfmathparse{13.792/128} \MyNum{\pgfmathresult} &   \pgfmathparse{50.429/128} \MyNum{\pgfmathresult} \\
 medium OU & $128$ & \pgfmathparse{178.258/128} \MyNum{\pgfmathresult} & \pgfmathparse{44.221/128} \MyNum{\pgfmathresult} &   \pgfmathparse{36.794/128} \MyNum{\pgfmathresult} & \pgfmathparse{82.256/128} \MyNum{\pgfmathresult}  \\
 fast OU & $128$ & \pgfmathparse{178.258/128} \MyNum{\pgfmathresult} & \pgfmathparse{67.237/128} \MyNum{\pgfmathresult} & \pgfmathparse{78.888/128} \MyNum{\pgfmathresult} & \pgfmathparse{113.169/128} \MyNum{\pgfmathresult}   \\
\multicolumn{6}{l}{balanced harmonic increase} \\
 BM & $128$ & \pgfmathparse{683.621/128} \MyNum{\pgfmathresult} & \pgfmathparse{39.878/128} \MyNum{\pgfmathresult} &  \pgfmathparse{14.424/128} \MyNum{\pgfmathresult} &  \pgfmathparse{73.063/128} \MyNum{\pgfmathresult}  \\
 slow OU & $128$ &  \pgfmathparse{683.621/128} \MyNum{\pgfmathresult} & \pgfmathparse{101.417/128} \MyNum{\pgfmathresult} &   \pgfmathparse{81.108/128} \MyNum{\pgfmathresult} &    \pgfmathparse{124.902/128} \MyNum{\pgfmathresult} \\
 medium OU & $128$ &   \pgfmathparse{683.621/128} \MyNum{\pgfmathresult}  & \pgfmathparse{127.303/128} \MyNum{\pgfmathresult} &   \pgfmathparse{123.606/128} \MyNum{\pgfmathresult} &    \pgfmathparse{127.94/128} \MyNum{\pgfmathresult}  \\
fast OU & $128$ & \pgfmathparse{683.621/128} \MyNum{\pgfmathresult} & $128$ & \pgfmathparse{127.939/128} \MyNum{\pgfmathresult} & \pgfmathparse{128/128} \MyNum{\pgfmathresult}  \\
\multicolumn{6}{l}{balanced geometric decrease}  \\
 BM & $128$ & \pgfmathparse{54.926/128} \MyNum{\pgfmathresult} & \pgfmathparse{24.599/128} \MyNum{\pgfmathresult} &  \pgfmathparse{2.977/128} \MyNum{\pgfmathresult} &  \pgfmathparse{2.796/128} \MyNum{\pgfmathresult} \\
slow OU & $128$ & \pgfmathparse{54.926/128} \MyNum{\pgfmathresult} & \pgfmathparse{26.783/128} \MyNum{\pgfmathresult} &  \pgfmathparse{5.735/128} \MyNum{\pgfmathresult} & \pgfmathparse{7.878/128} \MyNum{\pgfmathresult} \\
 medium OU & $128$ & \pgfmathparse{54.926/128} \MyNum{\pgfmathresult} & \pgfmathparse{28.502/128} \MyNum{\pgfmathresult} &   \pgfmathparse{10.116/128} \MyNum{\pgfmathresult} &  \pgfmathparse{14.189/128} \MyNum{\pgfmathresult} \\
 fast OU & $128$ & \pgfmathparse{54.926/128} \MyNum{\pgfmathresult} & \pgfmathparse{31.144/128} \MyNum{\pgfmathresult} & \pgfmathparse{19.056/128} \MyNum{\pgfmathresult} & \pgfmathparse{26.283/128} \MyNum{\pgfmathresult}    \\
\multicolumn{6}{l}{balanced geometric increase} \\
 BM & $128$ & \pgfmathparse{658.375/128} \MyNum{\pgfmathresult} & \pgfmathparse{48.870/128} \MyNum{\pgfmathresult} &  \pgfmathparse{36.286/128} \MyNum{\pgfmathresult}&  \pgfmathparse{93.302/128} \MyNum{\pgfmathresult} \\
 slow OU & $128$ & \pgfmathparse{658.375/128} \MyNum{\pgfmathresult} & \pgfmathparse{106.934/128} \MyNum{\pgfmathresult} & \pgfmathparse{95.213/128} \MyNum{\pgfmathresult} &  \pgfmathparse{125.693/128} \MyNum{\pgfmathresult} \\
 medium OU & $128$ & \pgfmathparse{658.375/128} \MyNum{\pgfmathresult} & \pgfmathparse{127.340/128} \MyNum{\pgfmathresult} & \pgfmathparse{123.876/128} \MyNum{\pgfmathresult} &   \pgfmathparse{127.943/128} \MyNum{\pgfmathresult} \\
 fast OU & $128$ & \pgfmathparse{658.375/128} \MyNum{\pgfmathresult} & \pgfmathparse{128/128} \MyNum{\pgfmathresult} & \pgfmathparse{127.939/128} \MyNum{\pgfmathresult} & \pgfmathparse{128/128} \MyNum{\pgfmathresult}      \\
\multicolumn{6}{l}{balanced long root branch}\\
 BM & $128$ & \pgfmathparse{7.259/128} \MyNum{\pgfmathresult} & \pgfmathparse{24.599/128} \MyNum{\pgfmathresult} & \pgfmathparse{2.977/128} \MyNum{\pgfmathresult} & \pgfmathparse{1.81/128} \MyNum{\pgfmathresult} \\
 slow OU & $128$ & \pgfmathparse{7.637/128} \MyNum{\pgfmathresult} & \pgfmathparse{24.768/128} \MyNum{\pgfmathresult} & \pgfmathparse{3.072/128} \MyNum{\pgfmathresult} & \pgfmathparse{2.015/128} \MyNum{\pgfmathresult} \\
 medium OU & $128$ & \pgfmathparse{7.637/128} \MyNum{\pgfmathresult} & \pgfmathparse{24.902/128} \MyNum{\pgfmathresult} & \pgfmathparse{3.175/128} \MyNum{\pgfmathresult} & \pgfmathparse{2.233/128} \MyNum{\pgfmathresult}  \\
 fast OU & $128$ & \pgfmathparse{7.637/128} \MyNum{\pgfmathresult}  & \pgfmathparse{25.171/128} \MyNum{\pgfmathresult} & \pgfmathparse{3.404/128} \MyNum{\pgfmathresult} & \pgfmathparse{2.712/128} \MyNum{\pgfmathresult} \\
\end{longtable}

\section{pESS in biological data sets}\label{BiolpESS}

Using the new version of mvSLOUCH 
I analyzed a number of data sets to see what effects using different definitions 
of pESS would have on inference. The data sets are a collection from various
sources. All but one are vertebrates.
The sole exception is the fruit length, a fitness related trait,
data for $33$ \textit{Chaerophyllum} species \citep{MPiwRPucKSpa2015}.
Ten datasets from the animal kingdom are looked into.
I consider
Madagascar Mantellidae male snout--vent length and range 
measurements for $40$ species \citep{MPabKWolMVen2012},
Carnivora body size (natural and log scale) and range data for $70$ species 
from the \texttt{carni70} data set, ade4 R package
\citep{JDinNTor2002,SPavSOllADuf2005a}.
I look into the same data that \citet{CAne2008} used
to introduce what I called the mESS:
body mass, running speed and hind limb length of 
$49$ mammalian (both carnivores, herbivores) species
\citep{SDraADuf2007,TGarADicCJanJJon1993,TGarCJan1993}.
I also consider the data sets attached to the GEIGER R package 
\citep{LHaretalGEIGER}: log body size of $16$ Carnivores species,
log body size of $197$ salamanders species,
log body size of $226$ turtles species,
log body size of $233$ primates species,
and 
log  wing, tarsus, culmen  lengths, log beak diameter and log gonys width 
$13$ Darwin's finches species.
I also look into log brightness, hue and spacing
of $38$ Duck species \citep{CEliRMaiMSha2014}.
Lastly I also use log sexual size dimorphism of $23$
\textit{Anolis} species \citep{MButTSchJLos2000,MButAKin2004}.

Data are analyzed on the natural scale unless mentioned above.
The results of this analysis are presented in
Tab. \ref{tabDataRes}. 
In all datasets the phylogenetic trees are ultrametric.
All trees were rescaled to a height of $\log(n) -1$ to be comparable 
with other results here. I take the $-1$ as there is no root branch in these trees.
In all the analysis,  except the mammalian hind limb length,
the OU processes were assumed to have a single constant optimum over the phylogeny.

As common in many comparative studies BM was selected for the body size/mass 
variables. There was one exception to this, the logarithm of body size for
the Carnivora data gave more support to an OU process. On the other hand measurements
on the natural scale are in favour of a BM. 

All definitions of pESS, except for the mESS, lead to the
same conclusions. Using the mESS can lead, at first sight, to dramatically
different conclusions --- an Ornstein--Uhlenbeck process with disruptive selection
(i.e. $\alpha<0$). However when looking into the estimate of $\alpha$ in all
cases, it was negative but very close to $0$ --- hence resembling a Brownian motion.
Also mESS is not, as explained in the beginning, designed to measure 
how much independent signal there is in the data. It measures how 
much information there is in the data to make inference about the mean value
parameters. The quantification of the independent signal depends rather on
the covariance between the data points --- hence the regression and mutual information
ESSs seem to make more sense. A reader might ask how is it possible
that a more complex model (nOU) is chosen when the mESS is significantly smaller
than $n$. 
But the mESS for BM models in these situations is even lower hence the disruptive OU model is favoured.
However, the $\alpha$ parameter is estimated at the magnitude of $-10^{9}$ so effectively 
this is a Brownian motion.
A similar phenomena can be observed in the \textit{Anolis} SSD analysis.
With the mean ESS the more complex OU is chosen as $n_{e}^{\mathrm{E}}\approx n$.
Such a choice is made by the model selection procedure, 
as under the BM model, $n_{e}^{\mathrm{E}}\approx 5.199$.

From these results one can draw the conclusion that even with noisy ``real--world''
data the likelihood should still be expected to dominate. However, the mESS
will not be a fortunate choice to use especially if the data seem to follow
a BM. There is a very good explanation for low values of $n_{e}^{\mathrm{E}}$. 
Under the 
phylogenetic BM model inference about the ancestral state are next to impossible
from only the contemporary sample. Due to the noise level
one cannot obtain consistent estimators of it \citep{CAne2008,KBarSSag2014,SSagKBar2012}.
As the mESS measures the amount of information available to estimate
mean parameters and the ancestral state equals the mean in the BM model,
then $n_{e}^{\mathrm{E}}$ will be small. Hence, AIC$_{c}$ will be high in this
case, and this model will not be favoured. However, with other definitions 
of pESS, Brownian motion is not discriminated in this way.
When the true model is the OU one,
the mESS does not seem to lead to wrong conclusions. This is, as in the OU
model there is a lot of information about $\theta$ ---
approximately the mean value \citep{KBarSSag2015}.

If we look at the turtles and primates results, then we can again see 
that the PD does not tell the full story of diversity. Both have similar
relative (and absolute) PDs but their $n_{e}^{\mathrm{R}}$ are very different.
The primates body size follows a Brownian motion and the 
phylogeny highly correlates contemporary species. The turtles' body size on the other
hand follows an OU process and there is much more independence in the data set. 
This is despite the fact that when investigating the phylogenies the primates' one
has clades diverging further back in the past.

An interesting data set to look at are the 
$49$ mammalian (Carnivores, Herbivores) measurements 
\citep{TGarADicCJanJJon1993,TGarCJan1993}
that can be found in the ade4 R package
\citep[as the \texttt{carni70} data set, ][]{SDraADuf2007}. 
This was the data that \citet{CAne2008} used
to illustrate her work. In line with her conclusions, as mentioned before, 
I found that the body size variable
has a very low amount of independent information. However, the two other considered
variables running speed and hind limb length are more informative, with the 
rESS being more than half $n$ for the latter variable. 
Hind limb length is also interesting that the best supported OU process
has different optima for the carnivores and herbivores. On the other
hand running speed supported a common optimum for the two groups of mammals.

I ended Section \ref{secphilIC} by writing that the analyses
of real biological data sets would be better for providing
rules of thumb for what how best to use information criteria 
in a phylogenetic context. Essentially all performed analyses
indicate that the choice of criterion and whether to use
the observed or effective sample size does not have much
effect on model selection. However, unlike in the simulation
study, $n_{e}^{\mathrm{E}}$ can give very different results.
Seeing as this pESS often points to a disruptive OU process, 
while the other definitions to a more biologically realistic BM
or adaptive OU process, indicates that $n_{e}^{\mathrm{E}}$ 
is probably not a good choice for model selection purposes.
This seemingly undesirable behaviour, alongside tiny effective sample sizes,
of the mESS occurs for samples as large as $70$.

A proposed rule of thumb is that if one has a very small
sample (like the $13$ species for Darwin's finches), then it is worth
trying out the different definitions of ESS for model selection. 
Of course with such a small sample drawing any conclusions is risky.
However, sometimes it is impossible to collect more measurements.
The pESS approach might allow the user to look at the observations 
from different angles in such a data deficient situation.
When the sample size is moderately large all methods (bar the mESS based ones) 
seem to be robust and lead to similar conclusions. Of the three proposed
definitions of pESS I found that the rESS performed best. It
has furthermore the advantage of a solid mathematical explanation
on how it quantifies independence in a phylogenetic data set.
However, I only tested it for Gaussian processes.
In a non--Gaussian setting the miESS could work better ---
a topic for further investigation.

\begin{longtable}{lccccc}
\caption{Results of analysis on real data with different definitions of pESS. 
In the situation where the OU model with disruptive selection $(\alpha<0)$ the value of $\alpha$ was
tiny, about $10^{-9}$. Hence these dynamics on the scale of the phylogeny are indistinguishable
from a BM.\label{tabDataRes}}
\endlastfoot
\renewcommand*{\thefootnote}{\textit{\alph{footnote}}}
\setcounter{footnote}{0}
Data set & $n$ & PD$/n$ & $n_{e}^{\mathrm{MI}}/n$ & $n_{e}^{\mathrm{E}}/n$ &  $n_{e}^{\mathrm{R}}/n$  \\
\hline 
Mantellidae male snout--vent length \footnote{\label{ftPab}\citep{MPabKWolMVen2012}} &
$40$  BM\footnote{Brownian motion} & \pgfmathparse{54.548/40} \MyNum{\pgfmathresult}&  \pgfmathparse{16.431/40} \MyNum{\pgfmathresult} BM & \pgfmathparse{3.873/40} \MyNum{\pgfmathresult} nOU\footnote{Ornstein--Uhlenbeck with $\alpha<0$} & \pgfmathparse{20.968/40} \MyNum{\pgfmathresult}  BM
\\
Mantellidae Range \addtocounter{footnote}{-3}\footnotemark{} & 
$40$  OU \addtocounter{footnote}{+2}\footnote{Ornstein--Uhlenbeck with $\alpha>0$} 
& \pgfmathparse{54.548/40} \MyNum{\pgfmathresult}& \pgfmathparse{40/40} \MyNum{\pgfmathresult}  OU & \pgfmathparse{39.993/40} \MyNum{\pgfmathresult}  OU & \pgfmathparse{40/40} \MyNum{\pgfmathresult}  OU
\\
\textit{Chaerophyllum} fruit length \footnote{\citep{MPiwRPucKSpa2015}} &
$33$  OU & \pgfmathparse{13.829/33} \MyNum{\pgfmathresult}  & \pgfmathparse{20.938/33} \MyNum{\pgfmathresult} OU & \pgfmathparse{3.215/33} \MyNum{\pgfmathresult}  nOU & \pgfmathparse{27.702/33} \MyNum{\pgfmathresult}  OU 
\\
Carnivora body size \footnote{\citep{JDinNTor2002,SPavSOllADuf2005a}} & 
$70$ BM  & \pgfmathparse{701.6/70} \MyNum{\pgfmathresult}  & \pgfmathparse{17.366/70} \MyNum{\pgfmathresult} BM & \pgfmathparse{3.669/70} \MyNum{\pgfmathresult} OU & \pgfmathparse{9.902/70} \MyNum{\pgfmathresult}  BM
\\
Carnivora log(body size) \addtocounter{footnote}{-1}\footnotemark{} &
$70$ OU  & \pgfmathparse{701.6/70} \MyNum{\pgfmathresult}  & \pgfmathparse{21.252/70} \MyNum{\pgfmathresult} OU & \pgfmathparse{3.669/70} \MyNum{\pgfmathresult} nOU & \pgfmathparse{28.231/70} \MyNum{\pgfmathresult} OU 
\\
Carnivora range \addtocounter{footnote}{-1}\footnotemark{} &
$70$ OU & \pgfmathparse{701.6/70} \MyNum{\pgfmathresult}  & \pgfmathparse{34.892/70} \MyNum{\pgfmathresult} OU  & \pgfmathparse{3.669/70} \MyNum{\pgfmathresult} nOU & \pgfmathparse{57.729/70} \MyNum{\pgfmathresult}  OU
\\
Mammalian body mass \footnote{\citep{SDraADuf2007,TGarADicCJanJJon1993,TGarCJan1993}} & 
$49$ BM & \pgfmathparse{905.5/49} \MyNum{\pgfmathresult}  & \pgfmathparse{6.111/49} \MyNum{\pgfmathresult} BM & \pgfmathparse{14.125/49} \MyNum{\pgfmathresult} BM  & \pgfmathparse{9.437/49} \MyNum{\pgfmathresult} BM 
\\
Mammalian running speed \addtocounter{footnote}{-1}\footnotemark{} &
$49$ OU  & \pgfmathparse{905.5/49} \MyNum{\pgfmathresult}  & \pgfmathparse{11.675/49} \MyNum{\pgfmathresult} OU  & \pgfmathparse{17.212/49} \MyNum{\pgfmathresult} OU & \pgfmathparse{21.15/49} \MyNum{\pgfmathresult}  OU
\\
Carnivores, Herbivores hind limb length \addtocounter{footnote}{-1}\footnotemark{} &
$49$ OU & \pgfmathparse{905.5/49} \MyNum{\pgfmathresult}  & \pgfmathparse{15.635/49} \MyNum{\pgfmathresult} OU  & \pgfmathparse{19.3/49} \MyNum{\pgfmathresult} OU  & \pgfmathparse{27.173/49} \MyNum{\pgfmathresult}  OU
\\
Carnivores log(body size) \footnote{\citep[datasets in GEIGER R package][]{LHaretalGEIGER}} & 
$16$  BM & \pgfmathparse{17.663/16} \MyNum{\pgfmathresult}  & \pgfmathparse{11.387/16} \MyNum{\pgfmathresult} BM& \pgfmathparse{5.8/16} \MyNum{\pgfmathresult}  BM & \pgfmathparse{12.264/16} \MyNum{\pgfmathresult}  BM
\\
Salamanders log(body size) \addtocounter{footnote}{-1}\footnotemark& 
$197$  OU & \pgfmathparse{171.176/197} \MyNum{\pgfmathresult}  & \pgfmathparse{44.69/197} \MyNum{\pgfmathresult} OU & \pgfmathparse{11.569/197} \MyNum{\pgfmathresult}  OU & \pgfmathparse{73.927/197} \MyNum{\pgfmathresult}  OU 
\\
Turtles log(body size) \addtocounter{footnote}{-1}\footnotemark{} & 
$226$  OU & \pgfmathparse{122.989/226} \MyNum{\pgfmathresult}  & \pgfmathparse{51.635/226} \MyNum{\pgfmathresult} OU & \pgfmathparse{52.721/226} \MyNum{\pgfmathresult}  OU & \pgfmathparse{125.386/226} \MyNum{\pgfmathresult}  OU 
\\
Primates log(body size) \addtocounter{footnote}{-1}\footnotemark{} & 
$233$  BM & \pgfmathparse{119.546/233} \MyNum{\pgfmathresult} &\pgfmathparse{42.238/233} \MyNum{\pgfmathresult} BM & \pgfmathparse{5.908/233} \MyNum{\pgfmathresult}  BM & \pgfmathparse{19.822/233} \MyNum{\pgfmathresult}  BM 
\\
Darwin's finches log(wing length) \addtocounter{footnote}{-1}\footnotemark{} & 
$13$  BM & \pgfmathparse{9.143/13} \MyNum{\pgfmathresult}  &\pgfmathparse{11.987/13} \MyNum{\pgfmathresult} OU & \pgfmathparse{3.798/13} \MyNum{\pgfmathresult}  nOU & \pgfmathparse{12.229/13} \MyNum{\pgfmathresult}  OU
\\
Darwin's finches log(tarsus length) \addtocounter{footnote}{-1}\footnotemark{} & 
$13$  BM & \pgfmathparse{9.143/13} \MyNum{\pgfmathresult}  & \pgfmathparse{6.556/13} \MyNum{\pgfmathresult} BM & \pgfmathparse{3.798/13} \MyNum{\pgfmathresult}  nOU & \pgfmathparse{5.885/13} \MyNum{\pgfmathresult}  BM
\\
Darwin's finches log(culmen length) \addtocounter{footnote}{-1}\footnotemark{} & 
$13$  OU & \pgfmathparse{9.143/13} \MyNum{\pgfmathresult} & \pgfmathparse{12.376/13} \MyNum{\pgfmathresult} OU & \pgfmathparse{3.803/13} \MyNum{\pgfmathresult}  nOU & \pgfmathparse{12.518/13} \MyNum{\pgfmathresult}  OU 
\\
Darwin's finches log(beak diameter) \addtocounter{footnote}{-1}\footnotemark{} & 
$13$  BM & \pgfmathparse{9.143/13} \MyNum{\pgfmathresult} & \pgfmathparse{6.556/13} \MyNum{\pgfmathresult} BM & \pgfmathparse{3.798/13} \MyNum{\pgfmathresult}  nOU & \pgfmathparse{5.885/13} \MyNum{\pgfmathresult}  BM
\\
Darwin's finches log(gonys width) \addtocounter{footnote}{-1}\footnotemark{} & 
$13$  BM & \pgfmathparse{9.143/13} \MyNum{\pgfmathresult} & \pgfmathparse{9.783/13} \MyNum{\pgfmathresult} OU & \pgfmathparse{3.798/13} \MyNum{\pgfmathresult}  nOU & \pgfmathparse{10.381/13} \MyNum{\pgfmathresult}  OU
\\
Ducks log(brightness) \footnote{\citep{CEliRMaiMSha2014}} &
$38$  OU & \pgfmathparse{32.784/38} \MyNum{\pgfmathresult} & \pgfmathparse{17.524/38} \MyNum{\pgfmathresult} OU & \pgfmathparse{22.488/38} \MyNum{\pgfmathresult}  OU & \pgfmathparse{25.997/38} \MyNum{\pgfmathresult}  OU
\\
Ducks log(hue) \addtocounter{footnote}{-1}\footnotemark{} &
$38$  OU & \pgfmathparse{32.784/38} \MyNum{\pgfmathresult} & \pgfmathparse{18.686/38} \MyNum{\pgfmathresult} OU & \pgfmathparse{24.289/38} \MyNum{\pgfmathresult}  OU & \pgfmathparse{27.531/38} \MyNum{\pgfmathresult}  OU
\\
Ducks log(spacing) \addtocounter{footnote}{-1}\footnotemark{} & 
$38$  OU & \pgfmathparse{32.784/38} \MyNum{\pgfmathresult} &  \pgfmathparse{26.292/38} \MyNum{\pgfmathresult} OU & \pgfmathparse{31.92/38} \MyNum{\pgfmathresult}  OU & \pgfmathparse{32.535/38} \MyNum{\pgfmathresult}  OU 
\\
\textit{Anolis} log(SSD) \footnote{\citep{MButTSchJLos2000,MButAKin2004}} &
$23$  BM & \pgfmathparse{22.31/23} \MyNum{\pgfmathresult} & \pgfmathparse{14.023/23} \MyNum{\pgfmathresult} BM & \pgfmathparse{23/23} \MyNum{\pgfmathresult} OU & \pgfmathparse{16.545/23} \MyNum{\pgfmathresult}  BM
\end{longtable}

\section{Discussion}

In this study I approached the question of quantifying the amount
of independent signal in a phylogenetic data set. I  
proposed two definitions of an effective sample size
and compared it to the one considered by \citet{CAne2008}.
My work is mainly heuristic --- to see how do these proposed
definitions behave on real and simulated datasets. 

The most important goal of my paper is --- does it make sense to
use information criteria for model selection with phylogenetically correlated data.
The most popular information criterion, Akaike's, is an asymptotic one with infinite sample size.
Because phylogenetic samples are usually small this was not satisfactory --- e.g. more realistic
but parameter richer models are rejected in favour of simpler ones. Therefore
small sample size corrected criteria were implemented, e.g. the considered here
AIC$_{c}$ (BIC an alternative one). However, these corrections were derived under 
the assumption of independence. One of the aims of this paper is to propose a formula
that allows for replacing the sample size with the amount of independent observations
and then see if this changes the models indicated by the criterion.
In most cases, it seems that the likelihood part of the information criterion
dominates and all definitions of pESS lead to similar conclusions
especially with many tip species. One can assume therefore, that for model selection,
dependencies in the data do not cause serious problems. However for 
small phylogenies
it seems reasonable to compare the conclusions from different 
pESS definitions (Tab. \ref{tabDataRes} Darwin's finches OU conclusion for $n_{e}^{R}$
and BM for $n$). 

The second goal of the paper is to quantify the amount of dependency in a phylogenetic
sample and to understand patterns associated with it.
Obtaining the pESS of clades, can indicate clades where more sampling or research effort is needed.
For example, is a low pESS due to there being really few species or should we expect more or possibly
a reclassification of species is needed? 
Of course, all of this is with respect to a specific trait(s). This specificity allows for identification
of interesting clades. Considering a trait like body size we obtain the distribution
of relative (for comparability between clades) pESSs across a set of clades. 
In the next step one may identify outlier clades --- extremely high or low
pESSs for further research. Low relative pESSs could indicate recent radiations or other factors
not allowing different species to evolve independently. High relative pESS, especially close to $1$,
would mean that the species are under completely independent evolutionary pressures. 
Phylogenetic ESSs of a clade can indicate undersampling of species. If we have 
high relative pESS with a low absolute number of species, then perhaps the
very recently evolved species are missing. This can be helpful to indicate 
where biologists and taxonomists should put efforts to fill in the gaps 
\citep{NIsaAPur2004}. 

A possibly appealing application of this measurement of independence is the quantification of 
biodiversity. The most commonly used evolutionary measurement tool is 
phylogenetic diversity --- the sum of branch lengths. It seems however that this number
does not say much (even when scaled by the number of tip species) about the
``value'' of an individual species and comparison between clades is difficult
(very different ones can have identical values, cf. Tabs. \ref{tabBioDiv} and \ref{tabBioDivPD}
long terminal with geometric and harmonic increases, or
Tab. \ref{tabDataRes} primates and turtles). Therefore, to give the ``value'' 
of a single species I propose to use the relative pESS (i.e. $n_{e}/n$). If the value is low,
then the loss of a single species does not result in much biodiversity loss --- as the other
species contain information on it. On the other hand loosing a species when the number is close 
to $1$ results in the loss of a unique entity. 

The pESS approach also forces one to 
define biodiversity in terms of a specific trait --- the one described by the stochastic 
process. Using a particular trait has the advantage of precision --- biodiversity is expressed by the variability
of specific entities directly linked to species. In a sense, the pESS links
the concept of a species as both a pattern and process \citep{MLidBOxe1989}.
The process is the evolving trait, an entity that can be directly observed
and measured. The patterns are the pre--identified entities on the phylogenetic tree.
On the other hand it has the disadvantage of being specific --- one looks only at one 
(or a couple if it is a suite of traits) dimension of the species. 

Quantifying the number of species by the pESS of a clade has the advantage of being
objective and not subject to potentially arbitrary calls.
Not splitting 
a group is compensated by intra--species variability which can be accommodated by the pESS concept.
The need to identify exceptional 
lineages and possibly novel traits associated with them is discussed by
\citet{JBeaBOMea2015}, in the context of clade specific increased/decreased
speciation rates. The phylogenetic effective sample size allows for 
direct comparison between clades with respect to traits, e.g. ones suspected/known of 
contributing to speciation. Outlier values of pESSs will indicate ``interesting'' groups of species.
Such a methodology combines
data from multiple sources, morphological (the traits) and genetic (the phylogeny)  --- a truly multi--omics approach.
With the availability of more and more data from diverse sources mathematical methods that integrate them 
are being developed more and more in the evolutionary biology world
\citep[e.g.][]{KBarPLio2014,CSolLKnoCAne2014}. 

\citet{EMarTHan1996AB} point out that one should expect comparative data sets to
contain phylogenetic correlations. It is their absence that should be proved.
To prove dependence or independence is a difficult problem in general.
One way would be to use information criteria, but it is not clear how many
degrees of freedom does the tree have. The relative pESS is an alternative
way of showing that phylogenetic correlations are not important. If the value of the relative pESS is
close to $1$, then the data set is essentially independent. 

\citet{WMadRFit2015} regret the lack of a method to quantify the number
of pseudoreplicates in a phylogenetically correlated dataset. They point
out, that the case of discrete traits is even more complicated, 
as 
it is the unobserved number of independent origins 
that matters.
Power and p--values,
unless one derives model specific tests or uses simulation
methods, of e.g. association tests should depend on this number and not on the 
observed number of species. However, as this number is unknown there is ``no quantitative correction
to apply to these methods'' \citep{WMadRFit2015}, e.g. a contingency table test. 
The concept of the pESS is what \citet{WMadRFit2015} seem to be looking for, but
I considered it here in the continuous trait case. Further work is needed to carry
the ideas over to the discrete case. However, there is a potential heuristic way
of applying the pESS to categorical traits. If one is able to identify continuous
traits, that are reasonably related to the discrete one and their pESSs are
similar, then their average can be used, as a plug--in for the pESS of the discrete trait
in a further downstream analysis/test i.e. an estimator of the number of shifts. 
The fact that these pESSs are correlated,
the traits are dependent through the categorical one and probably between themselves,
is actually an advantage. We want the pESS to be nearly the same for each trait 
and their similarity would indicate sensibility of the described ``proxy'' approach. 
If the pESSs for the different traits 
are dissimilar, then this indicates the need for further investigation, especially
choice of traits. The described approach is of course only a suggestion 
for dealing with discrete, evolutionary correlated data. 
Further study is needed alongside the development of models where continuous
and categorical traits can jointly co--evolve. Another alternative approach to develop
in the discrete case, as already mentioned, is the phylogenetic informativeness
\citep{WMulFCra2015,JTow2007}.

The phylogenetic ESS definitions are also interesting from a statistical point of view. The
mESS measures the amount of information on the mean value and hence often results
in a small pESS, especially in the BM case, where there is limited information
on the ancestral state. From all the simulations presented, it seems that the regression
ESS captures 
the amount of independent observations in the data for BM and OU evolution.
The good behaviour of the rESS is not surprising as, by construction, it adds up the variance
of the independent residuals. Both of these definitions can be used for non--normal 
processes but we should not expect the regression ESS to be so effective. Rather it would
only measure the amount of linearly independent observations. In a general case, I suggest
the mutual information ESS, but here work still needs to be done on defining an appropriate
$e(\cdot)$ transformation in order for $n_{e}^{MI} \in [1,n]$ to be in agreement with
$n_{e}^{R}$ for normal samples.

It could be possible that the proposed pESS approach
a step in solving a problem indicated by \citet{LFayRMatDorAMoo2015}:
``Unfortunately, not a single of these metrics (providing isolation scores for species -- KB)
has a strong empirical connection to things we might actually value about biodiversity
--- trait diversity or trait rarity, evolutionary potential, improved ecosystem function
and/or overall genetic information.'' The phylogenetic effective sample size forces 
one to work with a specific trait --- if that trait is interesting for biodiversity,
then we could have an index that is interesting from 
\citet{LFayRMatDorAMoo2015} point of view. What is more important, pESSs are cheap to obtain.

\section*{Acknowledgments}
I was supported by the Knut and Alice Wallenberg Foundation.
I would like to like to acknowledge two anonymous reviewers, whose comments greatly
improved the manuscript. 
I would like to thank Chad Eliason for providing me with the ducks data and Marcin Piwczy\'nski
for the Mantellidae and \textit{Chaerophyllum} data. I would like to thank Bengt Oxelman
and Tanja Stadler for encouragement and numerous comments on this work.

\bibliography{Bartoszek_pESS}

\begin{thebibliography}{74}
\providecommand{\natexlab}[1]{#1}
\providecommand{\url}[1]{\texttt{#1}}
\expandafter\ifx\csname urlstyle\endcsname\relax
  \providecommand{\doi}[1]{doi: #1}\else
  \providecommand{\doi}{doi: \begingroup \urlstyle{rm}\Url}\fi

\bibitem[Adamczak and {Mi\l o\'s}(2014)]{RAdaPMil2014}
R.~Adamczak and P.~{Mi\l o\'s}.
\newblock U--statistics of {O}rnstein--{U}hlenbeck branching particle system.
\newblock \emph{J. Th. Probab.}, 27\penalty0 (4):\penalty0 1071--1111, 2014.

\bibitem[Adamczak and {Mi\l o\'s}(2015)]{RAdaPMil2015}
R.~Adamczak and P.~{Mi\l o\'s}.
\newblock {CLT} for {O}rnstein--{U}hlenbeck branching particle system.
\newblock \emph{Elect. J. Probab.}, 20\penalty0 (42):\penalty0 1--35, 2015.

\bibitem[Agapow(2005)]{PAga2005}
P.~Agapow.
\newblock Species: demarcation and diversity.
\newblock In A.~Purvis, J.~L. Gittleman, and T.~Brooks, editors,
  \emph{Phylogeny and Conservation}, pages 57--75. Cambridge University Press,
  Cambridge, 2005.

\bibitem[Agapow et~al.(2004)Agapow, Crandall, Gittleman, Mace, Marshall, and
  Purvis]{PAgaOBinKCraJGitGMacJMarAPur2004}
P.~Agapow, K.~A. Crandall, J.~L. Gittleman, G.~M. Mace, J.~C. Marshall, and
  A.~Purvis.
\newblock The impact of species concept on biodiversity studies.
\newblock \emph{Quart. Rev. Biol.}, 79\penalty0 (2):\penalty0 161--179, 2004.

\bibitem[Akaike(1974)]{HAka1974}
H.~Akaike.
\newblock A new look at the statistical model identification.
\newblock \emph{IEEE T. Automat. Contr.}, 19\penalty0 (6):\penalty0 716--723,
  1974.

\bibitem[An\'e(2008)]{CAne2008}
C.~An\'e.
\newblock Analysis of comparative data with hierarchical autocorrelation.
\newblock \emph{Ann. Appl. Stat}, 2\penalty0 (3):\penalty0 1078--1102, 2008.

\bibitem[An\'e et~al.(2014)An\'e, Ho, and Roch]{CAneLHoSRoc2014}
C.~An\'e, L.~S.~T. Ho, and S.~Roch.
\newblock Phase transition on the convergence rate of parameter estimation
  under an {O}rnstein--{U}hlenbeck diffusion on a tree.
\newblock \emph{ArXiv e-prints}, 2014.

\bibitem[Bartoszek(2014)]{KBar2014}
K.~Bartoszek.
\newblock Quantifying the effects of anagenetic and cladogenetic evolution.
\newblock \emph{Math. Biosci.}, 254:\penalty0 42--57, 2014.

\bibitem[Bartoszek and Li\'o(2014)]{KBarPLio2014}
K.~Bartoszek and P.~Li\'o.
\newblock A novel algorithm to reconstruct phylogenies using gene sequences and
  expression data.
\newblock In \emph{Int. Proc. Chem. Biol. Env. Eng.; Env. Energy and Biotech.
  III}, volume~70, pages 8--12, 2014.

\bibitem[Bartoszek and Sagitov(2015{\natexlab{a}})]{KBarSSag2014}
K.~Bartoszek and S.~Sagitov.
\newblock A consistent estimator of the evolutionary rate.
\newblock \emph{J. Theor. Biol.}, 371:\penalty0 69--78, 2015{\natexlab{a}}.

\bibitem[Bartoszek and Sagitov(2015{\natexlab{b}})]{KBarSSag2015}
K.~Bartoszek and S.~Sagitov.
\newblock Phylogenetic confidence intervals for the optimal trait value.
\newblock \emph{J. App. Prob.}, 52\penalty0 (4):\penalty0 1115--1132,
  2015{\natexlab{b}}.

\bibitem[Bartoszek et~al.(2012)Bartoszek, Pienaar, Mostad, Andersson, and
  Hansen]{KBaretal2012}
K.~Bartoszek, J.~Pienaar, P.~Mostad, S.~Andersson, and T.~F. Hansen.
\newblock A phylogenetic comparative method for studying multivariate
  adaptation.
\newblock \emph{J. Theor. Biol.}, 314:\penalty0 204--215, 2012.

\bibitem[Beaulieu et~al.(2012)Beaulieu, Jhwueng, Boettiger, and
  O'Meara]{JBeauetal2012}
J.~M. Beaulieu, D.-C. Jhwueng, C.~Boettiger, and B.~C. O'Meara.
\newblock Modeling stabilizing selection: Expanding the {O}rnstein--{U}hlenbeck
  model of adaptive evolution.
\newblock \emph{Evolution}, 66:\penalty0 2369--2389, 2012.

\bibitem[Beaulieu and O'Meara(2016)]{JBeaBOMea2015}
J.M. Beaulieu and B.C. O'Meara.
\newblock Detecting hidden diversification shifts in models of trait-dependent
  speciation and extinction.
\newblock \emph{Syst. Biol.}, 2016.

\bibitem[Butchart and Bird(2010)]{SButJBir2010}
S.~H.~M. Butchart and J.~P. Bird.
\newblock Data deficient birds on the {IUCN} {R}ed {L}ist: what we don't know
  and why does it matter?
\newblock \emph{Biol. Conserv.}, 143:\penalty0 239--247, 2010.

\bibitem[Butler and King(2004)]{MButAKin2004}
M.~A. Butler and A.~A. King.
\newblock Phylogenetic comparative analysis: a modelling approach for adaptive
  evolution.
\newblock \emph{Am. Nat.}, 164\penalty0 (6):\penalty0 683--695, 2004.

\bibitem[Butler et~al.(2000)Butler, Schoener, and Losos]{MButTSchJLos2000}
M.~A. Butler, T.~W. Schoener, and J.~B. Losos.
\newblock The relationship between sexual size dimorphism and habitat use in
  {G}reater {A}ntillean {A}nolis lizards.
\newblock \emph{Evolution}, 54:\penalty0 259--272, 2000.

\bibitem[Clavel et~al.(2015)Clavel, Escarguel, and Merceron]{JClaGEscGMer2015}
J.~Clavel, G.~Escarguel, and G.~Merceron.
\newblock {mvMORPH}: an {R} package for fitting multivariate evolutionary
  models to morphometric data.
\newblock \emph{Meth. Ecol. Evol.}, 11\penalty0 (6):\penalty0 1311--1319, 2015.

\bibitem[Crawford and Suchard(2013)]{FCraMSuc2013}
F.~W. Crawford and M.~A. Suchard.
\newblock Diversity, disparity, and evolutionary rate estimation for unresolved
  {Y}ule trees.
\newblock \emph{Syst. Biol.}, 62\penalty0 (3):\penalty0 439--455, 2013.

\bibitem[Cressler et~al.(2015)Cressler, Butler, and King]{CCreMButAKin2015}
C.~E. Cressler, M.~A. Butler, and A.~A. King.
\newblock Detecting adaptive evolution in phylogenetic comparative analysis
  using the {O}rnstein--{U}hlenbeck model.
\newblock \emph{Syst. Biol.}, 64\penalty0 (6):\penalty0 953--968, 2015.

\bibitem[Crozier(1997)]{RCro1997}
R.~H. Crozier.
\newblock Preserving the information content of species: {G}enetic diversity,
  phylogeny and concervation worth.
\newblock \emph{Annu. Rev. Ecol. Syst.}, 28:\penalty0 243--268, 1997.

\bibitem[Diniz-Filho and T\^orres(2002)]{JDinNTor2002}
J.~A.~F. Diniz-Filho and N.~M. T\^orres.
\newblock Phylogenetic comparative methods and the geographic range size ---
  body size relationship in {N}ew {W}orld terrestrial {C}arnivora.
\newblock \emph{Evol. Ecol.}, 16:\penalty0 351--367, 2002.

\bibitem[Dray and Durfor(2007)]{SDraADuf2007}
S.~Dray and A.~B. Durfor.
\newblock The ade4 package: implementing the duality diagram for ecologists.
\newblock \emph{J. Stat. Soft.}, 22\penalty0 (4):\penalty0 1--20, 2007.

\bibitem[Eliason et~al.(2014)Eliason, Maia, and Shawkey]{CEliRMaiMSha2014}
C.~M. Eliason, R.~Maia, and M.~D. Shawkey.
\newblock Modular color evolution facilitated by a complex nanostructure in
  birds.
\newblock \emph{Evolution}, 69\penalty0 (2):\penalty0 357--367, 2014.

\bibitem[Elliot and Mooers(2014)]{MEllAMoo2014}
M.~G. Elliot and A.~{\O}. Mooers.
\newblock Inferring ancestral states without assuming neutrality or gradualism
  using a stable model of continuous character evolution.
\newblock \emph{BMC Evol. Biol.}, 14:\penalty0 226, 2014.

\bibitem[Faes et~al.(2009)Faes, Molenberghs, Aerts, Verbeke, and
  Kenward]{CFaeetal2009}
C.~Faes, G.~Molenberghs, M.~Aerts, G.~Verbeke, and M.~G. Kenward.
\newblock The effective sample size and an alternative small--sample
  degrees--of--freedom method.
\newblock \emph{Am. Stat.}, 63\penalty0 (4):\penalty0 389--399, 2009.

\bibitem[Faith(1992)]{DFai1992}
D.~P. Faith.
\newblock Conservation evaluation and phylogenetic diversity.
\newblock \emph{Biol. Conserv.}, 61:\penalty0 1--10, 1992.

\bibitem[Faith(1994)]{DFai1994}
D.~P. Faith.
\newblock Phylogenetic pattern and the quantification of organismal
  biodiversity.
\newblock \emph{Philos. Trans. Roy. Soc. B}, 345:\penalty0 45--58, 1994.

\bibitem[Faith(2002)]{DFai2002}
D.~P. Faith.
\newblock Quantifying biodiversity: {A} phylogenetic perspective.
\newblock \emph{Conserv. Biol.}, 16:\penalty0 248--252, 2002.

\bibitem[Faye et~al.(2015)Faye, Matthey-Doret, and Mooers]{LFayRMatDorAMoo2015}
L.~Faye, R.~Matthey-Doret, and A.~{\O}. Mooers.
\newblock Valuing species on the cheap.
\newblock \emph{Anim. Conserv.}, 18:\penalty0 313--314, 2015.

\bibitem[Felsenstein(2008)]{JFel2008}
J.~Felsenstein.
\newblock Comparative methods with sampling error and within-species variation:
  contrasts revisited and revised.
\newblock \emph{Am. Nat.}, 171:\penalty0 713--725, 2008.

\bibitem[Garland and Janis(1993)]{TGarCJan1993}
T.~J. Garland and C.~M. Janis.
\newblock Does metatarsal-femur ratio predict maximal running speed in
  cursorial mammals?
\newblock \emph{J. Zool.}, 229:\penalty0 133--151, 1993.

\bibitem[{Garland, T., Jr.} et~al.(1993){Garland, T., Jr.}, Dickerman, Janis,
  and Jones]{TGarADicCJanJJon1993}
{Garland, T., Jr.}, A.~W. Dickerman, C.~M. Janis, and J.~A. Jones.
\newblock Phylogenetic analysis of covariance by computer simulation.
\newblock \emph{Syst. Biol.}, 42:\penalty0 265--292, 1993.

\bibitem[Gernhard(2008{\natexlab{a}})]{TGer2008a}
T.~Gernhard.
\newblock The conditioned reconstructed process.
\newblock \emph{J. Theor. Biol.}, 253:\penalty0 769--778, 2008{\natexlab{a}}.

\bibitem[Gernhard(2008{\natexlab{b}})]{TGer2008b}
T.~Gernhard.
\newblock New analytic results for speciation times in neutral models.
\newblock \emph{B. Math. Biol.}, 70:\penalty0 1082--1097, 2008{\natexlab{b}}.

\bibitem[Hansen and Bartoszek(2012)]{THanKBar2012}
T.~F. Hansen and K.~Bartoszek.
\newblock Interpreting the evolutionary regression: the interplay between
  observational and biological errors in phylogenetic comparative studies.
\newblock \emph{Syst. Biol.}, 61\penalty0 (3):\penalty0 413--425, 2012.

\bibitem[Hansen and Orzack(2005)]{THanSOrz2005}
T.~F. Hansen and S.~H. Orzack.
\newblock Assessing current adaptation and phylogenetic inertia as explanations
  of trait evolution: the need for controlled comparisons.
\newblock \emph{Evolution}, 59\penalty0 (10):\penalty0 2063--2072, 2005.

\bibitem[Hansen et~al.(2008)Hansen, Pienaar, and Orzack]{THanJPieSOrz2008}
T.~F. Hansen, J.~Pienaar, and S.~H. Orzack.
\newblock A comparative method for studying adaptation to a randomly evolving
  environment.
\newblock \emph{Evolution}, 62:\penalty0 1965--1977, 2008.

\bibitem[Harmon et~al.(2008)Harmon, Weir, Brock, Glor, and
  Challenger]{LHaretalGEIGER}
L.~J. Harmon, J.~T. Weir, C.~D. Brock, R.~E. Glor, and W.~Challenger.
\newblock {GEIGER}: investigating evolutionary radiations.
\newblock \emph{Bioinformatics}, 24:\penalty0 129--131, 2008.

\bibitem[Hurvich and Tsai(1989)]{CHurCTsa1989}
C.~M. Hurvich and C.~L. Tsai.
\newblock Regression and time series model selection in small samples.
\newblock \emph{Biometrika}, 76\penalty0 (2):\penalty0 297--307, 1989.

\bibitem[Ingram and Mahler(2013)]{TIngDMah2013}
T.~Ingram and D.~L. Mahler.
\newblock {SURFACE}: detecting convergent evolution from comparative data by
  fitting {O}rnstein--{U}hlenbeck models with stepwise {A}kaike {I}nformation
  {C}riterion.
\newblock \emph{Meth. Ecol. Evol.}, 4\penalty0 (5):\penalty0 416--425, 2013.

\bibitem[Isaac and Purvis(2004)]{NIsaAPur2004}
N.~J.~B. Isaac and A.~Purvis.
\newblock The `species problem' and testing macroevolutionary hypotheses.
\newblock \emph{Diversity Distrib.}, 10:\penalty0 275--281, 2004.

\bibitem[Jetz and Freckleton(2015)]{WJetRFre2015}
W.~Jetz and R.~P. Freckleton.
\newblock Towards a general framework for predicting threat status of
  data--deficient species from phylogenetic, spatial and environmental
  information.
\newblock \emph{Phil. Trans. R. Soc. B.}, 370:\penalty0 20140016, 2015.

\bibitem[Jones and Moriarty(2013)]{NJonJMor2013}
N.~S. Jones and J.~Moriarty.
\newblock Evolutionary inference for function--valued traits: {G}aussian
  process regression on phylogenies.
\newblock \emph{J. R. Soc. Interface}, 10:\penalty0 20120616, 2013.

\bibitem[Koch(2014)]{IKoc2014}
I.~Koch.
\newblock \emph{Analysis of Multivariate and High--Dimensional Data}.
\newblock Cambridge University Press, Cambridge, 2014.

\bibitem[Lid\'en and Oxelman(1989)]{MLidBOxe1989}
M.~Lid\'en and B.~Oxelman.
\newblock Species---pattern or process?
\newblock \emph{Taxon}, 38\penalty0 (2):\penalty0 228--232, 1989.

\bibitem[Lin et~al.(2007)Lin, Steel, Pittman, and Clarke]{XLinJPitBCla2007}
X.~Lin, M.~Steel, J.~Pittman, and B.~Clarke.
\newblock Information conversion, effective samples, and parameter size.
\newblock \emph{IEEE. Trans. Inf. Theory}, 53\penalty0 (12):\penalty0
  4438--4456, 2007.

\bibitem[Maddison and Fitz{J}ohn(2015)]{WMadRFit2015}
W.~P. Maddison and R.~G. Fitz{J}ohn.
\newblock The unsolved challenge to phylogenetic correlation tests for
  categorical characters.
\newblock \emph{Syst. Biol.}, 64\penalty0 (1):\penalty0 127--136, 2015.

\bibitem[Martins and Hansen(1996)]{EMarTHan1996AB}
E.~P. Martins and T.~F. Hansen.
\newblock The statistical analysis of interspecific data: {A} review and
  evaluation of phylogenetic comparative methods.
\newblock In E.~P. Martins, editor, \emph{Phylogenies and the Comparative
  Method in Animal Behaviour}, pages 22--75. Oxford University Press, 1996.

\bibitem[Mooers et~al.(2005)Mooers, Heard, and Chrostowski]{AMooSHeaEChr2005}
A.~{\O}. Mooers, S.~B. Heard, and E.~Chrostowski.
\newblock Evolutionary heritage as a metric for conservation.
\newblock In A.~Purvis, J.~L. Gittleman, and T.~Brooks, editors,
  \emph{Phylogeny and Conservation}, pages 120--138. Cambridge University
  Press, Cambridge, 2005.

\bibitem[Mooers et~al.(2012)Mooers, Gascuel, Stadler, Li, and
  Steel]{AMooetal2012}
A.~{\O}. Mooers, O.~Gascuel, T.~Stadler, H.~Li, and M.~Steel.
\newblock Branch lengths on birth–-death trees and the expected loss of
  phylogenetic diversity.
\newblock \emph{Syst. Biol.}, 61\penalty0 (2):\penalty0 195--203, 2012.

\bibitem[Mulder and Crawford(2015)]{WMulFCra2015}
W.~H. Mulder and F.~W. Crawford.
\newblock On the distribution of interspecies correlation for {M}arkov models
  of character evolution on {Y}ule trees.
\newblock \emph{J. Theor. Biol.}, 364:\penalty0 275--283, 2015.

\bibitem[Nee and May(1997)]{SNeeRMay1997}
S.~Nee and R.~M. May.
\newblock Extinction and loss of evolutionary history.
\newblock \emph{Science}, 278:\penalty0 692--694, 1997.

\bibitem[Nunn(2011)]{CNun2011}
C.~L. Nunn.
\newblock \emph{The Comparative Approach in Evolutionary Anthropology and
  Biology}.
\newblock The University of Chicago Press, Chicago, 2011.

\bibitem[Pabijan et~al.(2012)Pabijan, Wollenberg, and Vences]{MPabKWolMVen2012}
M.~Pabijan, K.~C. Wollenberg, and M.~Vences.
\newblock Small body size increases the regional differentiation of populations
  of tropical mantellid frogs ({A}nura: {M}antellidae).
\newblock \emph{J. Evol. Biol.}, 25:\penalty0 2310--2324, 2012.

\bibitem[Pagel(1993)]{MPag1993}
M.~D. Pagel.
\newblock Seeking the evolutionary regression coefficent: {A}n analysis of what
  compaarative methods measure.
\newblock \emph{J. Theor. Biol.}, 164:\penalty0 191--205, 1993.

\bibitem[Paradis(2012)]{EPar2012}
E.~Paradis.
\newblock \emph{Analysis of Phylogenetics and Evolution with R}.
\newblock Springer, New York, 2012.

\bibitem[Pavoine et~al.(2005{\natexlab{a}})Pavoine, Ollier, and
  Dufour]{SPavSOllADuf2005a}
S.~Pavoine, S.~Ollier, and A.-B. Dufour.
\newblock Is the originality of a species measurable?
\newblock \emph{Ecol. Lett.}, 8:\penalty0 579--586, 2005{\natexlab{a}}.

\bibitem[Pavoine et~al.(2005{\natexlab{b}})Pavoine, Ollier, and
  Pontier]{SPavSOllDPon2005b}
S.~Pavoine, S.~Ollier, and D.~Pontier.
\newblock Measuring diversity from dissimilarities with {R}ao's quadratic
  entropy: Are any dissimilarities suitable?
\newblock \emph{Theor. Pop. Biol.}, 67:\penalty0 231--239, 2005{\natexlab{b}}.

\bibitem[Piwczy{\'n}ski et~al.(2015)Piwczy{\'n}ski, Pucha{\l }ka, and
  Spalik]{MPiwRPucKSpa2015}
M.~Piwczy{\'n}ski, R.~Pucha{\l }ka, and K.~Spalik.
\newblock The infrageneric taxonomy of {C}haerophyllum ({A}piaceae) revisited:
  new evidence from {nrDNA} {ITS} sequences and fruit anatomy.
\newblock \emph{Bot. J. Linn. Soc.}, 178:\penalty0 298--313, 2015.

\bibitem[Purvis et~al.(2005)Purvis, Gittleman, and Brooks]{APurJGitTBro2005}
A.~Purvis, J.~L. Gittleman, and T.~Brooks, editors.
\newblock \emph{Phylogeny and Conservation}.
\newblock Cambridge University Press, Cambridge, 2005.

\bibitem[{R Core Team}(2013)]{R}
{R Core Team}.
\newblock \emph{{R}: A Language and Environment for Statistical Computing}.
\newblock {R} Foundation for Statistical Computing, Vienna, Austria, 2013.
\newblock URL \url{http://www.R-project.org}.

\bibitem[Rao(1982)]{CRao1982}
C.~R. Rao.
\newblock Diversity: its measurements, decomposition, apportionment and
  analysis.
\newblock \emph{Sankhya. Ind. J. Stat.}, A44:\penalty0 1--22, 1982.

\bibitem[Rohlfs et~al.(2013)Rohlfs, Harrigan, and Nielsen]{RRohPHarRNie2013}
R.~V. Rohlfs, P.~Harrigan, and R.~Nielsen.
\newblock Modeling gene expression evolution with an extended
  {O}rnstein--{U}hlenbeck process accounting for within-species variation.
\newblock \emph{Mol. Biol. Evol.}, 31\penalty0 (1):\penalty0 201--211, 2013.

\bibitem[{Sagitov} and {Bartoszek}(2012)]{SSagKBar2012}
S.~{Sagitov} and K.~{Bartoszek}.
\newblock Interspecies correlation for neutrally evolving traits.
\newblock \emph{J. Theor. Biol.}, 309:\penalty0 11--19, 2012.

\bibitem[Schwarz(1978)]{GSch1978}
G.~Schwarz.
\newblock Estimating the dimension of a model.
\newblock \emph{Ann. Stat.}, 5\penalty0 (2):\penalty0 461--461, 1978.

\bibitem[Smith(1994)]{RSmi1994}
R.~J. Smith.
\newblock Degrees of freedom in interspecific allometry: An adjustment for the
  effects of phylogenetic constraint.
\newblock \emph{Am. J. Phys. Anthropol.}, 93:\penalty0 95--107, 1994.

\bibitem[Sol\'is-Lemus et~al.(2014)Sol\'is-Lemus, Knowles, and
  An\'e]{CSolLKnoCAne2014}
C.~Sol\'is-Lemus, L.~L. Knowles, and C.~An\'e.
\newblock {B}ayesian species delimitation combining multiple genes and traits
  in a unified framework.
\newblock \emph{Evolution}, 69\penalty0 (2):\penalty0 492--507, 2014.

\bibitem[Stadler(2009)]{TreeSim1}
T.~Stadler.
\newblock On incomplete sampling under birth-death models and connections to
  the sampling-based coalescent.
\newblock \emph{J. Theor. Biol.}, 261\penalty0 (1):\penalty0 58--68, 2009.

\bibitem[Stadler(2011)]{TreeSim2}
T.~Stadler.
\newblock Simulating trees with a fixed number of extant species.
\newblock \emph{Syst. Biol.}, 60\penalty0 (5):\penalty0 676--684, 2011.

\bibitem[Stadler and Steel(2012)]{TStaMSte2012}
T.~Stadler and M.~Steel.
\newblock Distribution of branch lengths and phylogenetic diversity under
  homogeneous speciation models.
\newblock \emph{J. Theor. Biol.}, 297:\penalty0 33--40, 2012.

\bibitem[Townsend(2007)]{JTow2007}
J.~P. Townsend.
\newblock Profiling phylogenetic informativeness.
\newblock \emph{Syst. Biol.}, 56\penalty0 (2):\penalty0 222--231, 2007.

\bibitem[Uyeda and Harmon(2014)]{JUyeLHar2014}
J.~C. Uyeda and L.~J. Harmon.
\newblock A novel {B}ayesian method for inferring and interpreting the dynamics
  of adaptive landscapes from phylogenetic comparative data.
\newblock \emph{Syst. Biol.}, 63\penalty0 (6):\penalty0 902--918, 2014.

\bibitem[Vellend et~al.(2011)Vellend, Cornwell, Magnuson-Ford, and
  Mooers]{MVelWCorKMagForAMoo2011}
M.~Vellend, W.~K. Cornwell, K.~Magnuson-Ford, and A.~{\O}. Mooers.
\newblock Measuring phylogenetic diversity.
\newblock In A.~E. Magurran and B.~J. McGill, editors, \emph{Biological
  Diversity: Frontiers in Measurement and Assessment}, pages 194--207. Oxford
  University Press, Oxford, 2011.

\end{thebibliography}
\bibliographystyle{plainnat}

\includepdf[pages=-]{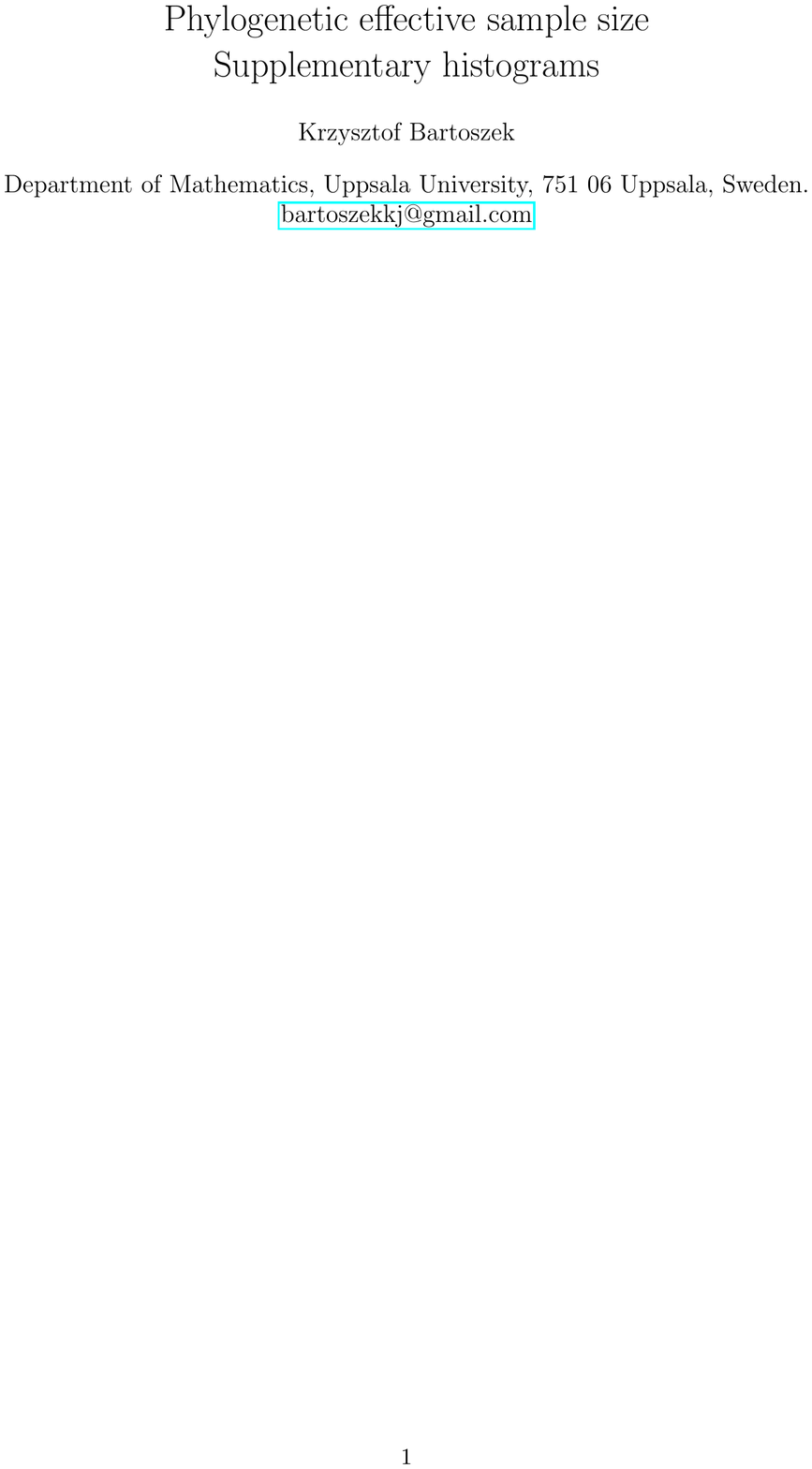}

\end{document}